\documentclass[]{aa} % 
\usepackage{natbib}
\bibpunct{(}{)}{;}{a}{}{,}
\usepackage{graphicx}
\usepackage{txfonts}
\begin{document}
   \title{Transit timing variations in eccentric hierarchical triple exoplanetary systems}
   \subtitle{I. Perturbations on the time-scale of the orbital period of the perturber}

   \author{T. Borkovits\inst{1} \and Sz. Csizmadia\inst{2} \and E. Forg\'acs-Dajka\inst{3} \and T. Heged\"us\inst{1}}

   \offprints{T. Borkovits}

   \institute{\inst{1}Baja Astronomical Observatory of B\'acs-Kiskun County,
              H-6500 Baja, Szegedi \'ut, Kt. 766, Hungary, \\
              \email{borko@electra.bajaobs.hu}; \email{hege@electra.bajaobs.hu} \\
	      \inst{2}Deutsches Zentrum f\"ur Luft- und Raumfahrt (DLR), Institut f\"ur Planetenforschung, 
	      Rutherfordstr. 2, 12489 Berlin, Germany, 	\\
	      \email{szilard.csizmadia@dlr.de} \\
	      \inst{3}E\"otv\"os University, Department of Astronomy, H-1518 Budapest, Pf. 32, Hungary, \\
              \email{E.Forgacs-Dajka@astro.elte.hu}}         

   \date{Received <date> / Accepted <date>}

   \titlerunning{}

\abstract{}
{We study the long-term time-scale (i.e. period comparable to the orbital period of the outer perturber object) 
transit timing variations in transiting exoplanetary systems which contain a further, more distant ($a_2>>a_1$)
either planetary, or stellar companion.}
{ We give an analytical form of the $O-C$ diagram (which describes such TTV-s) in trigonometric
series, valid for arbitrary mutual inclinations, up to the sixth order in the inner eccentricity.}
{We show that the dependence of the $O-C$ on the orbital and physical
parameters can be separated into three parts. Two of these are independent of the
real physical parameters (i.e. masses, separations, periods) of a concrete system,
and depend only on dimensionless orbital elements, and so, can be analysed in general.
We find, that for a specific transiting system, where eccentricity ($e_1$) and 
the observable argument of periastron ($\omega_1$) are known e.g. from spectroscopy, 
the main characteristics of any, caused by a possible third-body, transit timing variations can be mapped simply.
Moreover, as the physical attributes of a given system occur only as scaling
parameters, the real amplitude of the $O-C$ can also be estimated for a given
system, simply as a function of the $m_3/P_2$ ratio.
We analyse the above-mentioned dimensionless amplitudes for different arbitrary
initial parameters, as well as for two particular systems \object{CoRoT-9b} and
\object{HD 80606b}. We find in general, that while the shape of the $O-C$
strongly varies with the angular orbital elements, the net amplitude (departing from
some specific configurations) depends only weakly on these elements, but strongly
on the eccentricities. 
As an application, we illustrate how the formulae work for
the weakly eccentric \object{CoRoT-9b}, and the highly eccentric \object{HD 80606b}.
We consider also the question of detection, as well as the correct identification of such perturbations.
Finally, we illustrate the operation and effectiveness of Kozai cycles
with tidal friction (KCTF) in the case of \object{HD 80606b}.}
{}

\maketitle

%
%________________________________________________________________

\section{Introduction}
The rapidly increasing number of exoplanetary systems, as well as the 
lengthening time interval of the observations naturally leads to
the search for perturbations in the motion
of the known planets which can provide the possibility to detect further
planetary (or stellar) components in a given system, and/or can produce
further information about the oblateness of the host star (or the planet),
or might even refer to evolutionary effects.

The detection and the interpretation of such perturbations in the orbital
revolution of the exoplanets usually depends on such methods and theoretical
formulae which are well-known and have been applied for a long time in the field of
the close eclipsing and spectroscopic binaries. We mainly refer to the methods
developed in connection with the observed period variations in eclipsing binaries.
These period variations manifests themselve in the departure of the occurrence 
of an eclipse event (either transit, or occultation) from its predicted time.
Applying the nomenclature of exoplanet studies this phenomenon is called transit timing
variation (TTV). Plotting the observed minus calculated mid-transit times with respect to
the cycle numbers we get the $O-C$ diagram which has been the main tool for period studies
by variable star observers (not only for eclipsing binaries) for more than a century.
Consequently, the effect of the various types of period variations (both real and
apparent) for the $O-C$ diagram were already widely studied in the last one hundred years.
Some of these are less relevant in the case of transiting exoplanets, but others
are important. For example, the two classical cases are the simple geometrical 
light-time effect (LITE) (due to a further, distant companion), and the apsidal motion effect (AME)
(due to both the stellar oblateness in eccentric binaries and the relativistic effect).
Due to its small amplitude however, LITE, which has been widely used for identification
further stellar companions of many variable stars (not only in case of eclipsing binaries)
since the papers of \citet{chandler892,hertzsprung22,woltjer22,irwin52}, is less significant in the case of a planetary-mass wide component.
Nevertheless in the recent years there have been some efforts to discover exoplanets
on this manner (\citealp[see e.g.][\object{V391 Pegasi}]{silvottietal07}, \citealp[][\object{CM Draconis}]{deegetal08}, \citealp[][\object{QS Virginis}]{qianetal09}, \citealp[][\object{HW Virginis}]{leeetal09}).
The importance of AME in close exoplanet systems
has been investigated in several papers \citep[see e.g.][and further references therein]{miralda-escude02,heylgladman07,jordanbakos08}. 
Note, that this latter effect
has been studied since the pioneering works of \citet{cowling38,sterne39}, and
besides its evident importance in the checking of the general relativity
theory, it plays also an important role in studying the inner mass distributions of
stars (via their quadrupol moment), i.e. it provides observational verification of stellar modells.

Nevertheless, besides their similarities, there are also several differences between the period variations
of close binary and multiple stars, and planetary systems. For example, most of the known
stellar multiple systems form hierarchic subsystems \citep[see e.g.][]{tokovinin97}. 
This is mainly the consequence of the
dynamical stability of such configurations, or put another way, the dynamical instability
of the non-hierarchical stellar multiple systems. In contrast, a multiple planetary
system may form a stable (or at least long-time quasi-stable) non-hierarchical configuration
as we can see in our solar system, and furthermore, the same result was shown by numeric integrations
for several known exoplanet systems \citep[see e. g][]{sandoretal07}. Due to this fact
we can expect several different configurations, the examination of wich has, until now not been considered in the field of the period variations of multiple stellar systems. Examples of these are
as follows, the perturbations of an inner planet, as well as of a companion on a resonant
orbit \citep{agoletal05}, or as a special case of the latter, the possibility of
Trojan exoplanets \citep{schwarzetal09}. Recently, the detectability of exomoons
has also been studied \citep{simonetal07,kipping09a,kipping09b}.

Furthermore, due to the enhanced activities related to extrasolar planetary searches,
which led to missions like CoRoT and Kepler which produce long-term, extraordinarily
accurate data, we can expect in the close future such kinds of observations
which give the possibility of detecting and studying further phenomena which was never observed earlier.
For example, it is well-known, that neither the close binary stars, nor the hot-Jupiter-type
exoplanets could have been formed in their present positions. Different orbital shrinking mechanisms are
described in the literature. Instead of listing them, we refer to the short summaries by 
\citet{tokovininetal06,tokovinin08} and \citet{fabryckytremaine07}. Here we note only, that 
one of the most preferred theories for the formation of close binary stars, which
also might have produced at least a portion of hot-Jupiters, as well, is 
the combination of the Kozai cycles with tidal friction (KCTF). The Kozai resonance 
(recently frequently referred to as Kozai cycle[s]) was first described 
by \citet{kozai62} investigating secular perturbations of asteroids. The first 
(theoretical) investigation of this phenomenon with respect to multiple stellar 
systems can be found in the studies by \citet{harrington68,harrington69,mazehsaham79,soderhjelm82}.
A higher, third order theory of Kozai cycles was given by \citet{fordetal00},
while the first application of KCTF to explain the present configuration of
a close, hierarchical triple system (the emblematic \object{Algol}, itself)
was presented by \citet{kiselevaetal98}.
According to this theory the close binaries (as well as hot-Jupiter systems) 
should have originally formed as significantly wider primordial binaries 
having a distant, inclined third companion. Due to the third object induced 
Kozai oscillation, the inner eccentricity becomes cyclically so large 
that around the periastron passages, the two stars (or the host star and its planet) 
approach each other so closely that tidal friction may be effective, which, 
during one or more Kozai cycles, shrinks the orbit remarkably. Due to
the smaller separation, the tidal forces remain effective on a larger and larger portion 
of the whole revolution, and finally, they will switch off the Kozai cycles,
producing a highly eccentric, moderately inclined, small-separation intermediate orbit. 
After the last Kozai cycle,  some additional tidally
forced circularization may then form close systems in their recent configurations.

Nevertheless, independently from the question, as to which mechanism(s) is (are) the really effective ones, up to now we were not able
to study these mechanisms in operation, only the end-results were observed.
This is mainly the consequence of selection effects. In order to study these phenomena when
they are effective, one should observe the variation of the orbital elements of
such extrasolar planets, as well as binary systems which are in the period
range of months to years. The easisest way to carry out such observations is
the monitoring of the transit timing variations of these systems. 
However, there are only a few known transiting extrasolar planets in this period regime. 
Furthermore, although binary stars are also known with such separations,
similarly, they are not appropriate subjects for this investigation because of their non-eclipsing nature.

The continuous long-term monitoring of several hundreds of stars with the CoRoT and
Kepler satellites, as well as the long-term systematic terrestrial surveys provide an
excellent opportunity to discover transiting exoplanets (or as by-products: eclipsing binaries)
with the period of months. Then continuous, long-term
transit monitoring of such systems (combining the data with spectroscopy) may allow the tracing of dynamical evolutionary effects (i. e. orbital shrinking)
already on the timescale of a few decades. Furthermore, the larger the characteristic 
size of a multiple planetary, stellar (or mixed) system the greater the amplitude of
even the shorter period perturbations in the transit timing variations, as was shown
in detail in the discussion of \citet{borkovitsetal03}. 

In the last few years several papers have been published on transit timing variations,
both from theoretical aspects \citep[e.g. (non-complete)][and see further references therein]{agoletal05,holmanmurray05,nesvornybeauge10,holman10,cabrera10,fabrycky10},
and large numbers of papers on observational aspects for individual transiting exoplanetary systems. Nevertheless, most (but not all) of the theoretical
papers above, mainly concentrate simply on the detectibility of further companions 
(especially super-Earths) from the transit timing variations.  

In this paper we consider this question in greater detail. We calculate the analytical form
of the long-period\footnote{\bf In this paper we follow the original classification of \citet{brown36} who divided
the periodic perturbations occuring in hierarchical systems into the following three groups:
\begin{itemize}
\item[--] Short period perturbations. The typical period is equal to the orbital period $P_1$ of
the close pair, while the amplitude is of the order $(P_1/P_2)^2$;
\item[--] Long period perturbations. This group has a typical period of $P_2$, and magnitude of the order $(P_1/P_2)$;
\item[--] Apse-node terms. In this group the typical period is anout $P_2^2/P_1$, and the order of the amplitude reaches unity.
\end{itemize}
This classification differs from what is used in the classical planetary perturbation theories.
There, the first two groups termed together ``short period'' perturbations, while the ''apse-node terms"
are called ``long period ones''. Nevertheless, in the hierarchical
scenario, the first two groups differ from each other both in period and amplitude,
consequently, we feel this nomenclature more appropriate in the present situation.} (i. e. with a period on the order of the orbital period of the ternary component $P_2$)
time-scale  perturbations of the $O-C$ diagram for hierarchical (i.e. $P_2>>P_1$) triple systems. 
(Note, as we mainly concentrated on transiting systems with a period
of weeks to months, we omitted the possible tidal forces. However, our formulae
can be practically applied even for the closest exoplanetary systems, because
the tidal perturbations become effective usually on a notably longer time-scale.)
This work is a continuation and extension of the previous paper of 
\citet{borkovitsetal03}. In that paper we formulated the long-period
perturbations of an (arbitrarily eccentric and inclined) distant companion to the $O-C$ diagram for a circular inner orbit. 
(Note, that our formula is a generalized variation (in the relative inclination) of the one of \citealp{agoletal05}.
For the coplanar case the two results become identical.) Now we extend the results to the case
of an eccentric inner binary (formed either a host-star with its planet, or two stars). 
As it will be shown, our formulae have a satisfactory
accuracy even for a high eccentricity, such as $e_1=0.9$. Note, that in the period regime of a few months the
tidal forces are ineffective, so we expect eccentric orbits. This is especially valid
in the case of the predecessor systems of hot-Jupiters, in which case the 
above mentioned theory predicts very high eccentricities. 
%In the second paper we examined 
%the combined effect of stellar oblateness and the perturbations of a third component for
%the orbital elements of a close binary. We mainly concentrated on the deduction of the
%apsidal motion period from an $O-C$ curve distorted such a way. Tehrefore, we considered
%only the apse-node-scale variations, not combining with the long-period ones. Furthermore,
%as the presence of a significant stellar oblateness significantly changes the 
%analytical solution of the perturbation equations in the high mutual inclination regime
%(e. g. the stellar oblateness may prevent the Kozai cycles),
%in the present paper besides the combinations of the formulae of the two types
%of perturbations we also give new results for the high-mutual inclination case.

In the next section we give a very brief summary of our calculations. (A somewhat
detailed description can be found in \citealp{borkovitsetal03,borkovitsetal07},
nevertheless for self-consistency of the present paper we provide here a brief overview .) In Sect. 3. we discuss our results, while in Sect. 4
we illustrate the results with both analytical and numerical calculations on two individual systems,
\object{CoRoT-9b} and \object{HD 80606b}. {\bf Finally in Sect. 5 we conclude our results,
and, furthermore, we compare our method and results with that of \citet{nesvornymorbidelli08,nesvorny09}.}

\section{Analytical investigations}

\subsection{General considerations and equations of the problem}
As is well-known, at the moment of the mid-transit (which in case of an eccentric orbit 
usually does not coincide with the half-time of the whole transit event)
\begin{equation}
u\approx\pm\frac{\pi}{2}+2k\pi,
\label{eq:fundamental}
\end{equation}
where $u$ is the true longitude measured from the intersection of the orbital
plane and the plane of sky, and $k$ is an integer. (Note, since, traditionally the positive $z$-axis 
is directed away from the observer, the primary transit occurs at $u=-\pi/2$.)
An exact equality is valid only if the binary has a circular orbit, or if the orbit is seen edge-on exactly.
(The correct inclination dependence of the occurrence of the mid-eclipses can be
found in \citealp{gimenezgarcia-pelayo83}.)
Nevertheless, the observable inclination of a potentially month-long period transiting
extrasolar planet should be close to $i=90\degr$, if the latter condition 
is to be satisfied.
Due to its key role in the occurrence of the transits, instead of the usual variables,
we use $u$ as our independent time-like variable.
It is known from the textbooks of celestial mechanics, that
\begin{eqnarray}
\dot{u}&=&\frac{c_1}{\rho_1^2}-\dot\Omega\cos{i}, \nonumber \\
&=&\mu^{1/2}a^{-3/2}(1-e^2)^{-3/2}(1+e\cos{v})^2-\dot\Omega\cos{i},
\label{eq:upont}
\end{eqnarray}
consequently, the moment of the $N$-th primary minimum (transit) after an epoch $t_0$ can be
calculated as
\begin{eqnarray}
\int^{t_N}_{t_0}\mathrm{d}t&=&\int_{-\pi/2}^{2N\pi-\pi/2}\frac{a^{3/2}}{\mu^{1/2}}\frac{(1-e^2)^{3/2}}{[1+e\cos(u-\omega)]^2}\frac{\mathrm{d}u}{1-\frac{\rho_1^2}{c_1}\dot\Omega\cos{i}}, \nonumber\\
&\approx&\int\frac{a^{3/2}}{\mu^{1/2}}\frac{(1-e^2)^{3/2}}{[1+e\cos(u-\omega)]^2}\left(1+\frac{\rho_1^2}{c_1}\dot\Omega\cos{i}\right)\mathrm{d}u.
\label{eq:TN}
\end{eqnarray}
In the equations above $c_1$ denotes the specific angular momentum of the inner binary, $\rho_1$ is the radius vector
of the planet with respect to its host-star, while the orbital elements have their usual meanings. Furthermore, in Eq.~(\ref{eq:TN}) we 
applied that the true anomaly can be written as $v=u-\omega$.
In order to evaluate Eq.~(\ref{eq:TN}) first we have to express the perturbations in the orbital elements with respect to $u$. 
Assuming that the orbital elements (except of $u$) are constant, the first
term of the right hand side yields the following closed solution
\begin{eqnarray}
\overline{P}_{I,II}\!&\!=\!&\!\frac{P}{2\pi}\!\left[2\!\arctan\!\left(\!\sqrt{\frac{1-e}{1+e}}\frac{\mp\cos\omega}{1\mp\sin\omega}\right)\!\pm(1-e^2)^{1/2}\!\frac{e\cos\omega}{1\mp e\sin\omega}\right], \nonumber \\
\label{eq:apsisclosed}
\end{eqnarray}
for the two types of minima, respectively. (Here $P$ denotes the anomalistic or Keplerian period which is considered to be constant.)
Note, that instead of the exact forms above, widely used is the expansion (as in this paper),
which, up to the fifth order in $e$, is as follows:
\begin{eqnarray}
\overline{P}_{I,II}&=&P_\mathrm{s}E+\frac{P}{2\pi}\left[\mp\frac{1}{2}\pi\pm2e\cos\omega+\left(\frac{3}{4}e^2+\frac{1}{8}e^4\right)\sin2\omega\right. \nonumber \\
&&\left.\mp\left(\frac{1}{3}e^3+\frac{1}{8}e^5\right)\cos3\omega-\frac{5}{32}e^4\sin4\omega\pm\frac{3}{40}e^5\cos5\omega\right], \nonumber \\
\label{eq:apsisexp}
\end{eqnarray}
where $P_\mathrm{s}$ is the sidereal (or eclipsing) period of, for example, the first cycle, and $E$ is the cycle-number.
{\bf In the present case the orbital elements cease to remain constant. Nevertheless,
as it can be seen from Eqs.~(\ref{Eq:da1dv2})--(\ref{Eq:di1dv2}), their variation on $P_2$ timescale
is related to $P_1/P_2<<1$, which allows the linearization of the problem, i.e.
in such a case Eq.~(\ref{eq:apsisexp}) is formally valid in the same form, but $e$, $\omega$
and $P_\mathrm{s}$ are no longer constant. Then a further integration of Eq.~(\ref{eq:apsisexp})
with respect to $v_2$ gives the analytical form of the perturbed $O-C$ on $P_2$ time-scale.
To this, as} a next step, we have to calculate the long-period and apse-node perturbations in $u$.
Some of these arise simply from the similar perturbations  of the orbital elements (or directly of the $e\cos\omega$, $e\sin\omega$ functions),
while others (we will refer to them as direct perturbations in $u$) come from the
variations of the mean motion \citep[for more details see][]{borkovitsetal07}
\footnote{The minus signs on the rhs of Eqs. (10) and (11) of \citet{borkovitsetal07} should be replaced by plus.}. 
(In other words this means that in such a case $P$ will no longer be constant.)

At this point, to avoid any confusion, we emphasize that during our calculations we use different sets
of the angular orbital elements. As we are interested in such a phenomenon which primarily
depends on the relative positions of the orbiting celestial bodies with respect to the observer, the 
angular elements (i.e. $u$, $\omega$, $i$, $\Omega$) should be expressed in the ``observational'' frame of
reference having the plane of the sky as the fundamental plane, and $u$, as well as $\omega$ is measured 
from the intersection of the binary's orbital plane with that plane, while $\Omega$ is measured along 
the plane of the sky from an arbitrary origin. On the other hand, the physical
variations of the motion of the bodies depend on their relative positions
to each other, and consequently, it is more beneficial (and convenient) to express the equations of perturbations of 
the orbital elements in a different frame of reference, (we shall refer to it as the
dynamical frame) which depends on the relative positions of the bodies, independently
from any outer observer. The fundamental plane of this latter frame of reference is
the invariable plane of the triple system, i.e. the plane perpendicular to the net
angular momentum vector of the complete triple system. In this frame of reference the longitude of the ascending node ($h$)
gives the arc between the sky and the corresponding intersection of the two orbital
planes measured along the invariable plane, while the true longitude ($w$) and
the argument of periastron ($g$) is measured from that ascending node, along the
respected orbital plane. In order to avoid a further confusion, the relative inclination
of the orbits to the invariable plane is denoted by $j$. The meaning and the relation
between the different elements can be seen in Fig.~\ref{fig:krsz1}, and
listed also in Appendix~\ref{App:elemrel}.
\begin{figure}
\centering
\includegraphics[width=\hsize]{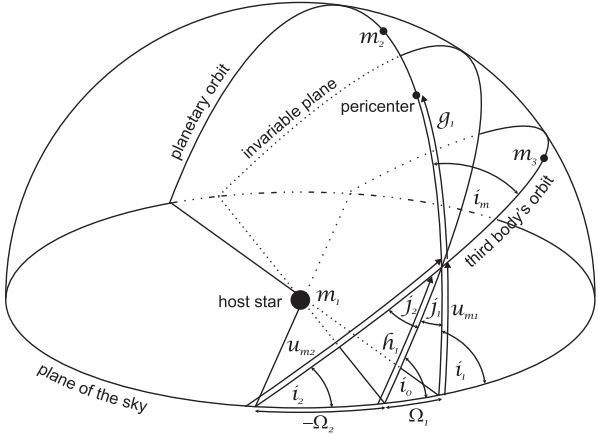}
\caption{The spatial configuration of the system.}
\label{fig:krsz1}
\end{figure}

\subsection{Long period perturbations}
From now on we refer to the orbital elements of the inner binary (i.e. the pair formed by
either a host star and the inner planet, or two stellar mass objects) by subscript $_1$,
while those of the wider binary (i.e. the orbit of the third component around the centre
of mass of the inner system) by subscript $_2$.
The differential perturbation equations of the orbital elements are listed
in \citet{borkovitsetal07}\footnote{Note, in the denominator of the Eqs. (14) and (15)
of that paper for $e$ and $\omega$ a closing bracket is evidently missing, and furthermore, 
the equation for $\Omega$ should be divided by $e$ for the correct result.}. {\bf (In that
paper, from practical considerations, we did not restrict ourselves to the most usual representation of the perturbation
equations, i.e. the Lagrange equations with the perturbing function, rather we used the somewhat more general form, expressing
the perturbations with the three orthogonal components of the perturbing force. Nevertheless,
as far as only conservative, three-body terms are considered, the two representations are
perfectly equivalent.)}
In order to get the long-period terms of the perturbation equations, the usual method involves averaging the equations for the short-period ($\approx P_1$) variables, which
is usually the mean anomaly ($l_1$) or the true anomaly ($v_1$) of the inner
binary, but in our special case it is the true longitude $u_1$. 
This means that we get the variation of the orbital elements by averaged for an eclipsing
period. Furthermore, in the case of the averaged equations we change the independent variable
from $u_1$ to (the averaged) $v_2$, by the use of
\begin{eqnarray}
\frac{\mathrm{d}u_1}{\mathrm{d}v_2}&\approx&\frac{\rho_2^2}{c_2}\frac{c_1}{\rho_1^2}\left(1-\frac{\rho_1^2}{c_1}\dot\Omega_1\cos i_1\right),
\end{eqnarray}
from which, after averaging we get:
\begin{eqnarray}
\frac{\mathrm{d}\overline{u_1}}{\mathrm{d}\overline{v_2}}&\approx&\frac{\mu_1^{1/2}}{a_1^{3/2}}\frac{\rho_2^2}{\sqrt{\mu_2a_2(1-e_2^2)}}-\frac{\mathrm{d}\overline{\Omega_1}}{\mathrm{d}\overline{v_2}}\cos i_1 \nonumber \\
&\approx&\frac{P_2}{P_1}\frac{(1-e_2^2)^{3/2}}{(1+e_2\cos v_2)^2}-\frac{\mathrm{d}\Omega_1}{\mathrm{d}v_2}\cos i_1. 
\label{Eq:dudvaverage}
\end{eqnarray}
In the following we omit the overlining. Furthermore, as one can see from Eq.~(\ref{Eq:dOm1dv2}) below,
the second term in the r.h.s of (\ref{Eq:dudvaverage}) is of the order $P_1/P_2$,
and consequently can be neglected.
So for the long-period perturbations of the orbital elements of the close orbit we get as follows:
\begin{eqnarray}
\frac{\mathrm{d}a_1}{\mathrm{d}v_2}&=&0, \label{Eq:da1dv2}\\
\nonumber \\
\frac{\mathrm{d}e_1}{\mathrm{d}v_2}&=&A_\mathrm{L}(1-e_1^2)^{1/2}e_1(1+e_2\cos v_2)\nonumber \\
&&\times\left[\left(1-I^2\right)\sin2g_1\right. \nonumber \\
&&-\frac{1}{2}(1+I)^2\sin(2v_2+2g_2-2g_1) \nonumber \\
&&\left.+\frac{1}{2}(1-I)^2\sin(2v_2+2g_2+2g_1)\right], \label{Eq:de1dv2}\\
\nonumber \\
\frac{\mathrm{d}g_1}{\mathrm{d}v_2}&=&A_\mathrm{L}(1-e_1^2)^{1/2}(1+e_2\cos v_2)\nonumber \\
&&\times\left\{\frac{3}{5}\left(I^2-\frac{1}{3}\right)\right. \nonumber \\
&&+\left(1-I^2\right)\left[\cos2g_1+\frac{3}{5}\cos(2v_2+2g_2)\right] \nonumber \\
&&+\frac{1}{2}(1+I)^2\cos(2v_2+2g_2-2g_1)\nonumber \\
&&\left.+\frac{1}{2}(1-I)^2\cos(2v_2+2g_2+2g_1)\right\} \nonumber \\
&&-\frac{\mathrm{d}h_1}{\mathrm{d}v_2}\cos{j_1}, \label{Eq:dg1dv2}\\
\nonumber \\
\frac{\mathrm{d}h_1}{\mathrm{d}v_2}&=&-A_\mathrm{L}(1-e_1^2)^{-1/2}\frac{\sin i_\mathrm{m}}{\sin{j_1}}(1+e_2\cos v_2)\nonumber \\
&&\times\left[\frac{2}{5}\left(1+\frac{3}{2}e_1^2\right)I\right. \nonumber \\
&&-e_1^2I\cos2g_1 \nonumber \\
&&-\frac{2}{5}\left(1+\frac{3}{2}e_1^2\right)I\cos(2v_2+2g_2) \nonumber \\
&&+\frac{1}{2}e_1^2(1+I)\cos(2v_2+2g_2-2g_1) \nonumber \\
&&\left.-\frac{1}{2}e_1^2(1-I)\cos(2v_2+2g_2+2g_1)\right], \label{Eq:dh1dv2}
\nonumber \\
\frac{\mathrm{d}j_1}{\mathrm{d}v_2}&=&A_\mathrm{L}(1-e_1^2)^{-1/2}\sin i_\mathrm{m}(1+e_2\cos v_2) \nonumber \\
&&\times\left[\frac{2}{5}\left(1+\frac{3}{2}e_1^2\right)\sin(2v_2+2g_2)\right. \nonumber \\
&&+e_1^2I\sin2g_1 \nonumber \\
&&-\frac{1}{2}e_1^2(1+I)\sin(2v_2+2g_2-2g_1) \nonumber \\
&&\left.+\frac{1}{2}e_1^2(1-I)\sin(2v_2+2g_2+2g_1)\right], \label{Eq:dj1dv2}\\
\nonumber \\
\frac{\mathrm{d}\omega_1}{\mathrm{d}v_2}&=&A_\mathrm{L}(1-e_1^2)^{1/2}(1+e_2\cos v_2)\nonumber \\
&&\times\left\{\frac{3}{5}\left(I^2-\frac{1}{3}\right)\right. \nonumber \\
&&+\left(1-I^2\right)\left[\cos2g_1+\frac{3}{5}\cos(2v_2+2g_2)\right] \nonumber \\
&&+\frac{1}{2}(1+I)^2\cos(2v_2+2g_2-2g_1)\nonumber \\
&&\left.+\frac{1}{2}(1-I)^2\cos(2v_2+2g_2+2g_1)\right\} \nonumber \\
&&-\frac{\mathrm{d}\Omega_1}{\mathrm{d}v_2}\cos{i_1}, \label{Eq:domega1dv2}\\
\nonumber  \\
\frac{\mathrm{d}\Omega_1}{\mathrm{d}v_2}&=&A_\mathrm{L}(1-e_1^2)^{-1/2}\frac{\sin i_\mathrm{m}}{\sin{i_1}}(1+e_2\cos v_2)\nonumber \\
&&\times\left[\frac{2}{5}\left(1+\frac{3}{2}e_1^2\right)I\cos(\omega_1-g_1)\right. \nonumber \\
&&+\frac{1}{5}\left(1+\frac{3}{2}e_1^2\right)(1-I)\cos(2v_2+2g_2-\omega_1+g_1) \nonumber \\
&&-\frac{1}{5}\left(1+\frac{3}{2}e_1^2\right)(1+I)\cos(2v_2+2g_2+\omega_1-g_1) \nonumber \\
&&-e_1^2I\cos(\omega_1+g_1) \nonumber \\
&&+\frac{1}{2}e_1^2(1+I)\cos(2v_2+2g_2-\omega_1-g_1) \nonumber \\
&&\left.-\frac{1}{2}e_1^2(1-I)\cos(2v_2+2g_2+\omega_1+g_1)\right], \label{Eq:dOm1dv2}\\
\nonumber \\
\frac{\mathrm{d}i_1}{\mathrm{d}v_2}&=&A_\mathrm{L}(1-e_1^2)^{-1/2}\sin i_\mathrm{m}(1+e_2\cos v_2)\nonumber \\
&&\times\left[-\frac{2}{5}\left(1+\frac{3}{2}e_1^2\right)I\sin(\omega_1-g_1)\right. \nonumber \\
&&+\frac{1}{5}\left(1+\frac{3}{2}e_1^2\right)(1-I)\sin(2v_2+2g_2-\omega_1+g_1) \nonumber \\
&&+\frac{1}{5}\left(1+\frac{3}{2}e_1^2\right)(1+I)\sin(2v_2+2g_2+\omega_1-g_1) \nonumber \\
&&+e^2I\sin(\omega_1+g_1) \nonumber \\
&&-\frac{1}{2}e_1^2(1+I)\sin(2v_2+2g_2-\omega_1-g_1) \nonumber \\
&&\left.+\frac{1}{2}e_1^2(1-I)\sin(2v_2+2g_2+\omega_1+g_1)\right], 
\label{Eq:di1dv2}
\end{eqnarray}
and, finally, the direct term is as follows:
\begin{eqnarray}
\frac{\mathrm{d}\lambda_1}{\mathrm{d}v_2}&=&A_\mathrm{L}(1-e_1^2)^{1/2}(1+e_2\cos v_2)\nonumber \\
&&\times\left\{\frac{4}{5}\left(I^2-\frac{1}{3}\right)f_1(e_1)\right. \nonumber \\
&&+\frac{51}{20}\left(1-I^2\right)e_1^2f_2(e_1)\cos2g_1 \nonumber \\
&&+\frac{4}{5}\left(1-I^2\right)f_1(e_1)\cos(2v_2+2g_2) \nonumber \\
&&+\frac{51}{40}(1+I)^2e_1^2f_2(e_1)\cos(2v_2+2g_2-2g_1) \nonumber \\
&&\left.+\frac{51}{40}(1-I)^2e_1^2f_2(e_1)\cos(2v_2+2g_2+2g_1)\right\}, \nonumber \\
\label{Eq:dlambda1dv2}
\end{eqnarray}
where
\begin{equation}
A_\mathrm{L}=\frac{15}{8}\frac{m_3}{m_{123}}\frac{P_1}{P_2}(1-e_2^2)^{-3/2},
\end{equation}
and
\begin{eqnarray}
f_1(e)&=&1+\frac{25}{8}e^2+\frac{15}{8}e^4+\frac{95}{64}e^6, \\
f_2(e)&=&1+\frac{31}{51}e^2+\frac{23}{48}e^4.
\end{eqnarray}
Furthermore, $i_\mathrm{m}$ denotes the mutual inclination of the two orbital planes, while
\begin{equation}
I=\cos i_\mathrm{m},
\end{equation}
and $m_{123}$ stands for the total mass of the system. Note, formal integration of the first five of 
the equations above (i.e. those which refer to the orbital elements in the dynamical 
system, Eqs.~[\ref{Eq:da1dv2}]-[\ref{Eq:dj1dv2}]) reproduce the results of \citet{soderhjelm82}.

Strictly speaking, some of the equations above are valid only in that case when both orbits
are non-circular, and the orbital planes are inclined to each other, and/or to the plane of the sky.
For example, if the outer orbit is circular, neither $v_2$, nor $g_2$ has any meaning. 
Nevertheless, their sum i. e. $v_2+g_2$ is meaningful as before. Furthermore,
although in the case of circular inner orbits the derivative of $g_1$ has no meaning,
yet the derivative of $e_1\cos g_1$, $e_1\sin g_1$ (or $e_1\cos\omega_1$, $e_1\sin\omega_1$),
i. e. the so-called Lagrangian elements can be calculated correctly. (Note, instead
of the ``pure'' derivatives of $e_1$ and $g_1$ (or $\omega_1$) these latter occur directly
in the $O-C$.) Similar redefenitions can be done in the case of coplanarity. So, 
for practical reasons, and for the sake of clarity, we retain the original formulations even in such cases, when
it is formally not valid.

As one can see, there are some terms on the r.h.s. of these equations which do not
depend on $v_2$. Primarily, these terms give the so-called apse-node time-scale contribution to
the variation of the orbital elements and the transit timing variations. Nevertheless, 
in order to get a correct result for the long-term
behaviour of the orbital elements these terms must nevertheless be retained in our long-term formulae. 
Note, these terms were calculated for tidally distorted triplets in \citet{borkovitsetal07}. 
Such formulae (after an omission of the tidal terms) are also valid for the present case 
in the low mutual inclination (i.e. approx. $I^2>\frac{3}{5}$) domain. 
(Formulae valid for arbitrary mutual inclinations will be presented in a subsequent paper.) 

Carrying out the integrations, all the orbital elements on the r.h.s of these equations 
with the exception of $v_2$ are considered as constants. This can be justified for two reasons.
First, as one can see, for $P_2>>P_1$ the amplitudes of the long-period perturbative terms ($A_\mathrm{L}$)
remain small (which is especially valid for the case where the host star is orbited by two planets, when
$m_3<<m_{123}$)\footnote{\bf This assumption is analogous to that of the classical
low order perturbation theory where the squares of the orbital element changes are neglected,
as they should be proportional to $(m_\mathrm{perturber}/m_\mathrm{star})^2$, but with the
difference that here the small parameter is the ratio of the semi-major axes $a_1/a_2$ instead of
the mass ratio. Consequently, this assumption remains valid even if the perturber
would be a sufficiently distant stellar-mass object.}, 
and second, although the amplitude of the apse-node perturbative terms can reach
unity, the period is usually so long ($\sim P_2^2/P_1$, \citealp[see e.g.][]{brown36}), 
that its contribution can be safely ignored during one revolution
of the outer object.\footnote{Note, this is not necessarily true in the presence of other perturbative effects.
Nevertheless, in such cases one can assume, that if there is a physical effect producing
apse-node time-scale perturbations having a period comparable to the period of the long-period
perturbations arising from some other sources, then the amplitudes of these long-period perturbations
are usually so small with respect to the amplitudes of the apse-node perturbations that their effect can be neglected.}
The final result of such an analysis is an analytical form of the transit timing variations, i.e. the long-period
$O-C$ diagram. Here we give the result up to the first order in the inner eccentricity,
while a more extended result up to the sixth order in the inner eccentricity can be
found in appendix~\ref{App:longO-C}, where we give also the perturbation equations
directly for the $e_1^m\cos{n\omega_1}$, $e_1^m\sin{n\omega_1}$ expressions.
\begin{eqnarray}
O-C_{P_2}&=&\frac{P_1}{2\pi}A_\mathrm{L}(1-e_1^2)^{1/2}\left\{\left(\frac{4}{5}+\frac{5}{2}e_1^2\mp\frac{6}{5}e_1\sin\omega_1\right)\right. \nonumber \\
&&\times\left[\left(I^2-\frac{1}{3}\right){\cal{M}}+\frac{1}{2}\left(1-I^2\right){\cal{S}}(2v_2+2g_2)\right]\nonumber \\
&&+\left[\frac{51}{20}e_1^2\cos2g_1\mp2e_1\sin(\omega_1-2g_1)\right]\nonumber \\
&&\times\left[\left(1-I^2\right){\cal{M}}+\frac{1}{2}\left(1+I^2\right){\cal{S}}(2v_2+2g_2)\right]\nonumber \\
&&-\left[\frac{51}{20}e_1^2\sin2g_1\mp2e_1\cos(\omega_1-2g_1)\right]I{\cal{C}}(2v_2+2g_2)\nonumber \\
\nonumber \\
&&+\cot i_1\sin i_\mathrm{m}\left\{-\frac{2}{5}\left(1\mp2e_1\sin\omega_1\right)\cos u_\mathrm{m1}I\right. \nonumber \\
&&\times\left[{\cal{M}}-\frac{1}{2}{\cal{S}}(2v_2+2g_2)\right] \nonumber \\
&&\left.\left.+\frac{1}{5}(1\mp2e_1\sin\omega_1)\sin u_\mathrm{m1}{\cal{C}}(2v_2+2g_2)\right\}\right\} \nonumber \\
\nonumber \\
&&-\frac{m_3}{m_{123}}\frac{a_2\sin i_2}{c}\frac{\left(1-e_2^2\right)\sin(v_2+\omega_2)}{1+e_2\cos v_2} \nonumber \\
&&+{\cal{O}}(e_1^2), \label{Eq:O-Clonge1}
\end{eqnarray}
where
\begin{eqnarray}
{\cal{M}}&=&\int 1+e_2\cos v_2\mathrm{d}v_2 \nonumber \\
&=&v_2-l_2+e_2\sin v_2 \label{Eq:kozeppontiegyenlites} \\
&=&3e_2\sin v_2-\frac{3}{4}e_2^2\sin2v_2+\frac{1}{3}e_2^3\sin3v_2+{\cal{O}}(e_2^4) \nonumber \\
&=&3e_2\left(1-\frac{3}{8}e_2^2\right)\sin l_2+\frac{9}{4}e_2^2\sin2l_2+\frac{53}{24}e_2^3\sin3l_2+{\cal{O}}(e_2^4),\nonumber \\
\label{Eq:M-tag}
\end{eqnarray}
and
\begin{eqnarray}
{\cal{S}}(2v_2+x)&\!=\!&\sin(2v_2+x)+e_2\sin(v_2+x)+\frac{1}{3}e_2\sin(3v_2+x) \nonumber \\
&\!=\!&\left(1-4e_2^2\right)\sin(2l_2+x)\nonumber \\
&&+e_2\left[-\left(1-\frac{13}{8}e_2^2\right)\sin(l_2+x)\right. \nonumber \\
&&\left.+\frac{7}{3}\left(1-\frac{207}{56}e_2^2\right)\sin(3l_2+x)\right] \label{Eq:S(2v+x)} \\
&&+e_2^2\left[-\frac{1}{4}\sin{x}+\frac{17}{4}\sin(4l_2+x)\right] \nonumber \\
&&+e_2^3\left[\frac{1}{24}\sin(l_2-x)+\frac{169}{24}\sin(5l_2+x)\right]+{\cal{O}}(e_2^4), \nonumber \\
{\cal{C}}(2v_2+x)&\!=\!&\cos(2v_2+x)+e_2\cos(v_2+x)+\frac{1}{3}e_2\cos(3v_2+x) \nonumber \\
&\!=\!&\left(1-4e_2^2\right)\cos(2l_2+x)\nonumber \\
&&+e_2\left[-\left(1-\frac{13}{8}e_2^2\right)\cos(l_2+x)\right.\nonumber  \\
&&\left.+\frac{7}{3}\left(1-\frac{207}{56}e_2^2\right)\cos(3l_2+x)\right] \label{Eq:C(2v+x)} \\
&&+e_2^2\left[-\frac{1}{4}\cos{x}+\frac{17}{4}\cos(4l_2+x)\right] \nonumber \\
&&+e_2^3\left[-\frac{1}{24}\cos(l_2-x)+\frac{169}{24}\cos(5l_2+x)\right]+{\cal{O}}(e_2^4), \nonumber
\end{eqnarray}
furthermore,
\begin{equation}
u_\mathrm{m1}=\omega_1-g_1,
\label{Eq:um1def}
\end{equation}
i.e. $u_\mathrm{m1}$ is the angular distance of the intersection of the two orbits
from the plane of the sky, or, in other words, the longitude of the (dynamical)
ascending node of the inner orbit along the orbital plane, measured from the sky.

Note, the superscripts refer to the exoplanetary transits (primary minima), and
the subscripts to the secondary occultations (secondary minima). Furthermore, we assumed
the formally second-order $\frac{5}{2}e_1^2$, and $\frac{51}{20}e_1^2$ terms, to be first order, 
as their values exceed $e_1$ for medium eccentricities.
We included also the pure geometrical light-time contribution in the last row. Here $c$ denotes
the speed of light. The minus sign arises because this term reflects the motion of
the inner pair around the common centre of mass, whose true longitude differs by $\pi$
from the one of the third component. In Eq.~(\ref{Eq:kozeppontiegyenlites}) the mean
anomaly of the outer body ($l_2$) appears because of (the constant part of) the difference between the
anomalistic $P$ and sideral (eclipsing, or transiting) $P_\mathrm{s}$ period included
into the first term in the r.h.s. of Eq.~(\ref{eq:apsisexp}).

\section{Discussion of the results}

\begin{figure*}
\centering
\includegraphics[width=7.2cm]{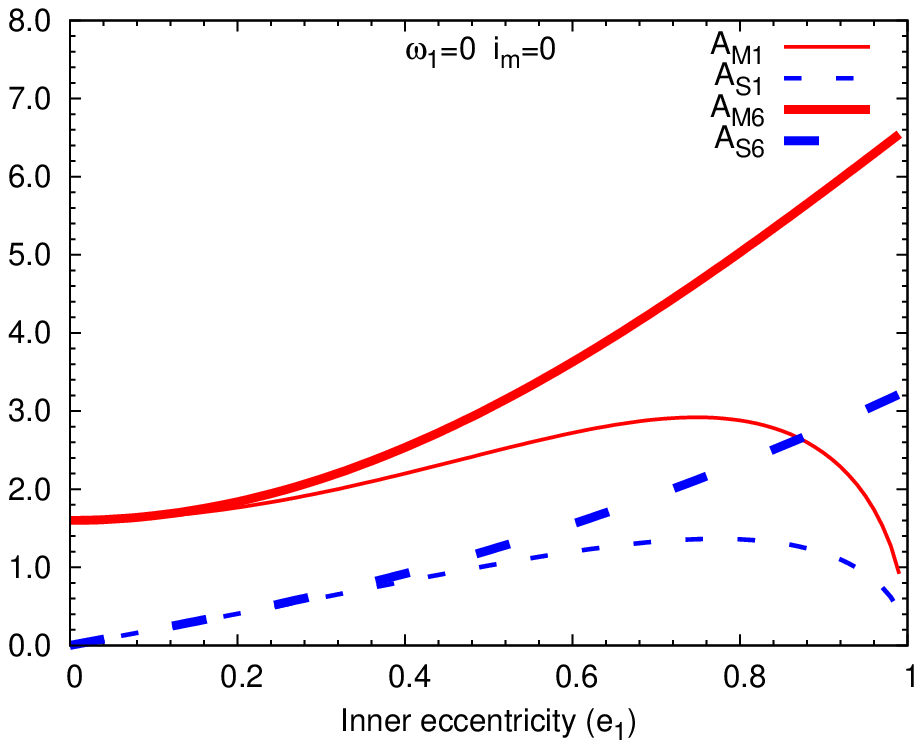}\includegraphics[width=7.2cm]{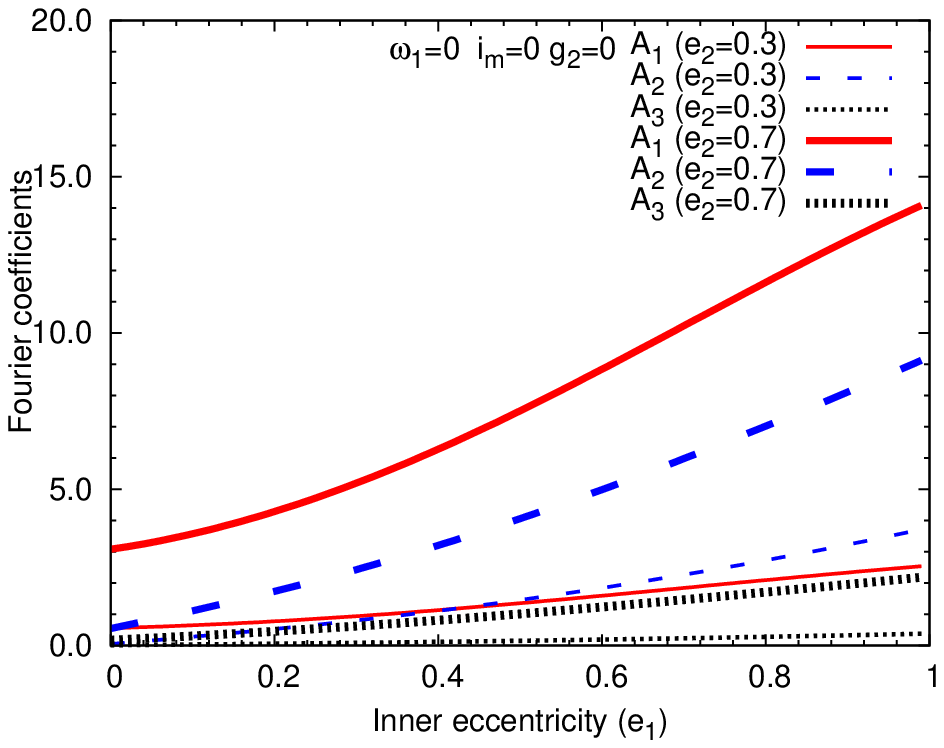}
\includegraphics[width=7.2cm]{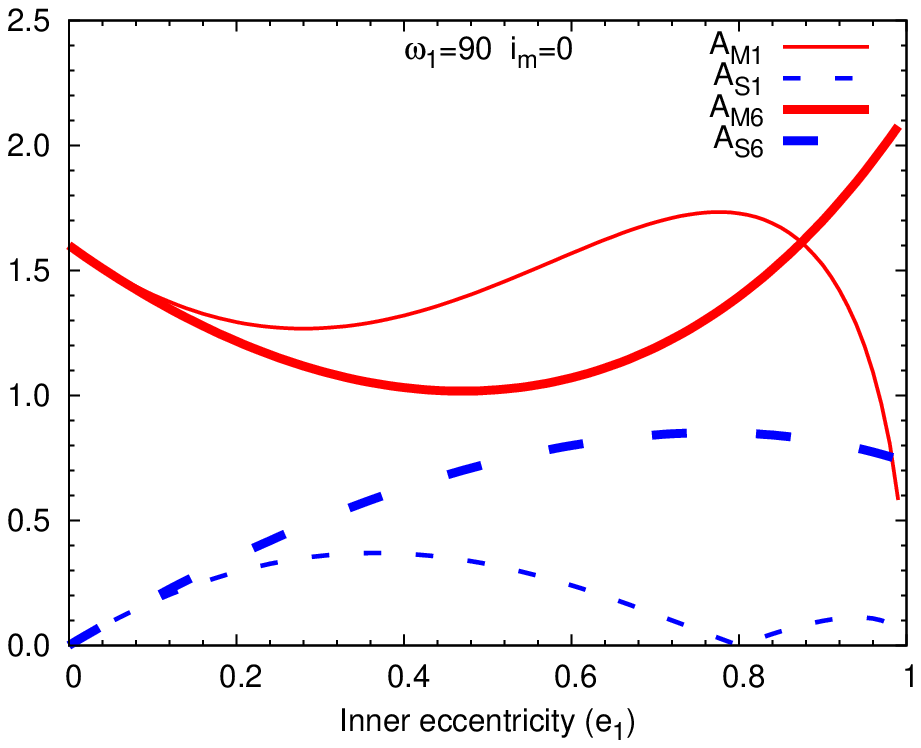}\includegraphics[width=7.2cm]{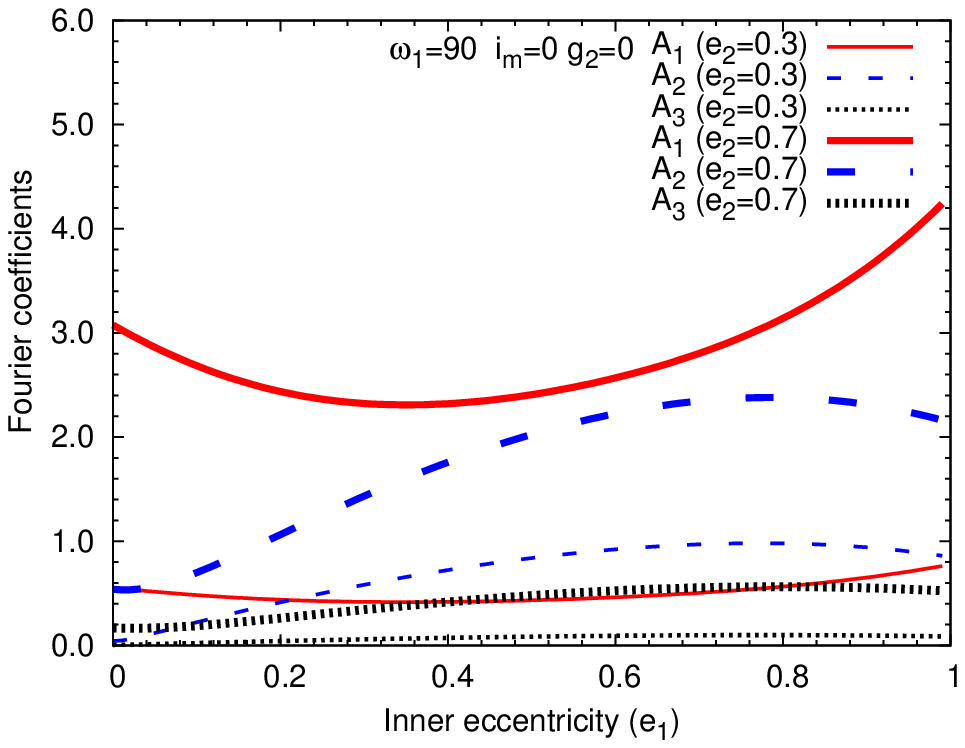}
\includegraphics[width=7.2cm]{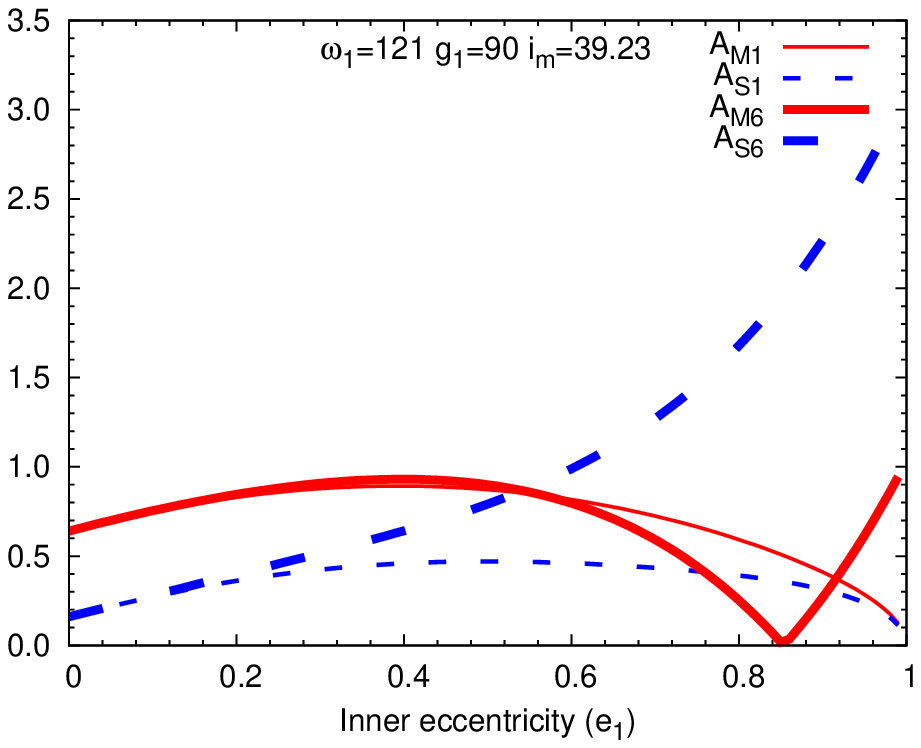}\includegraphics[width=7.2cm]{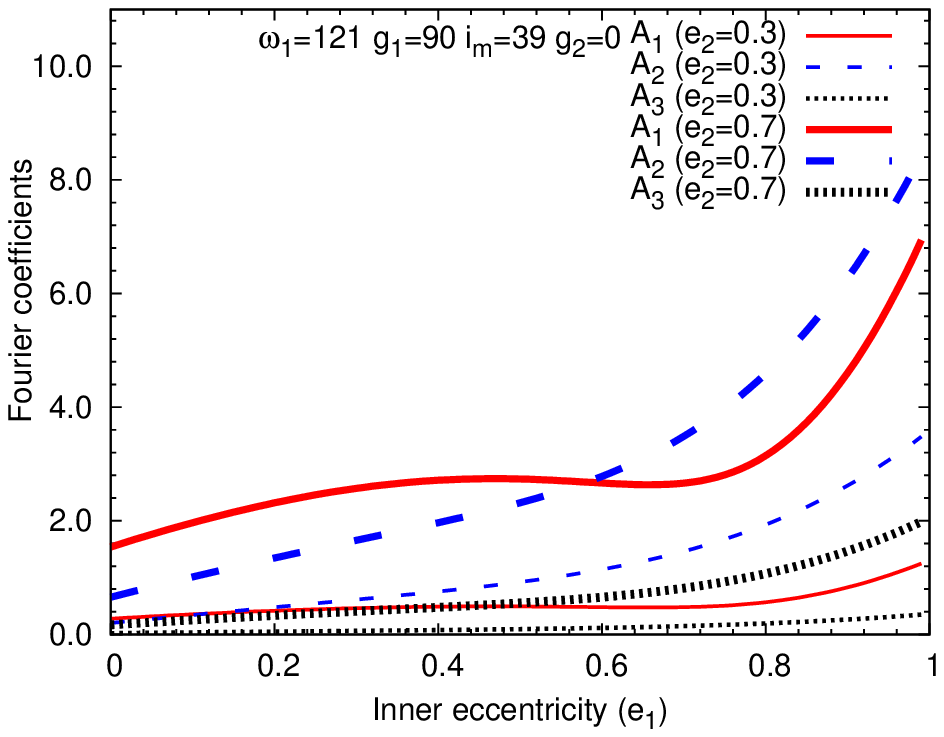}
\includegraphics[width=7.2cm]{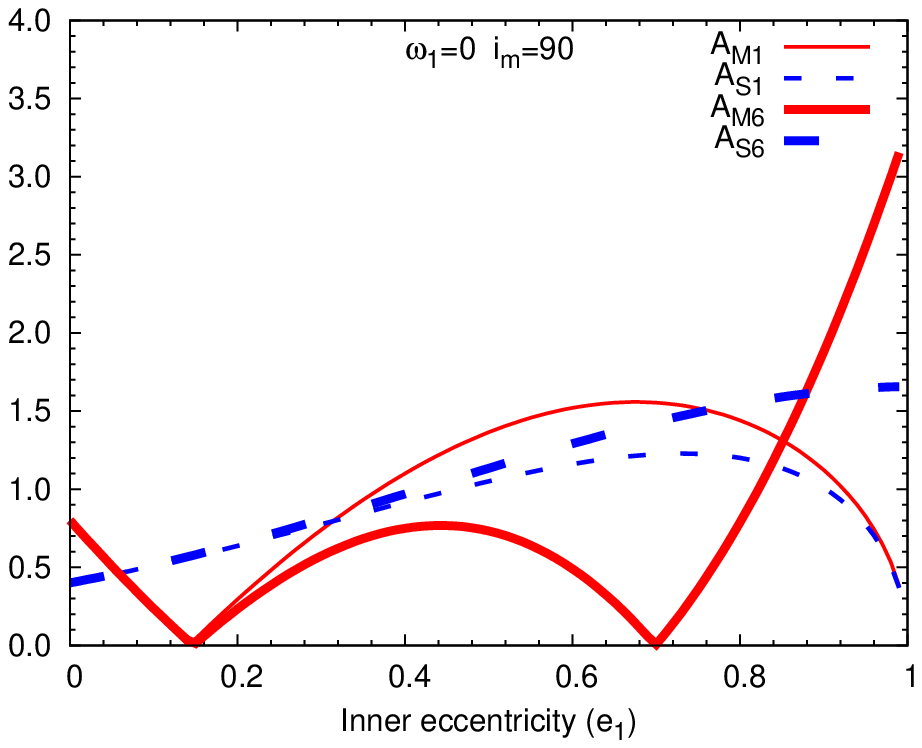}\includegraphics[width=7.2cm]{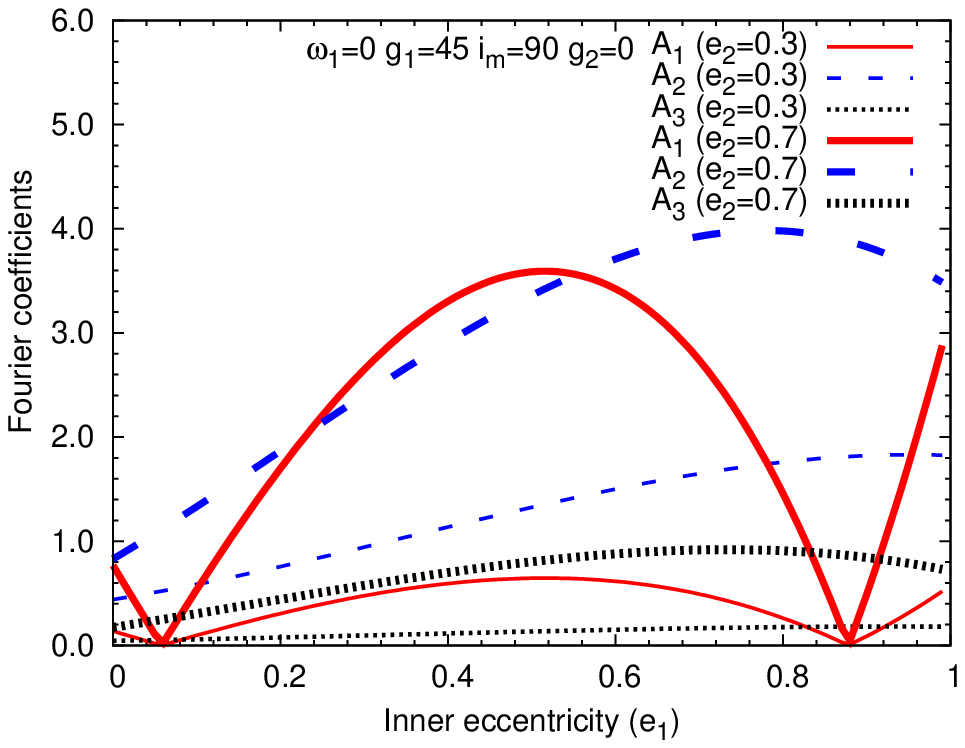}
\caption{{\it Left panels:} The inner eccentricity ($e_1$) dependence of $A_{\cal{M}}$, $A_{\cal{S}}$
amplitudes of ${\cal{M}}$, ${\cal{S}}$ functions of long-period dynamical part of $O-C$ for
specific values of some parameters. 
Note, to compare the ${\cal{M}}$ and ${\cal{S}}$ terms, the former
should be multiplied by $e_2$. The thin lines (indexed by '1') refer to the first order
approximation, while the thick ones (index '6') to the sixth one. 
{\it Right panels:} The corresponding $A_{1,2,3}$
amplitudes of the trigonometric representation (Eq.~\ref{eq:OminCdyntrig}) of the $O-C$ for two different outer
eccentricities ($e_2=0.3$ and $e_2=0.7$) ($g_2$ was set to 0\degr).}
\label{fig:amplitudes-e1}
\end{figure*}

\begin{figure*}
\centering
\includegraphics[width=7cm]{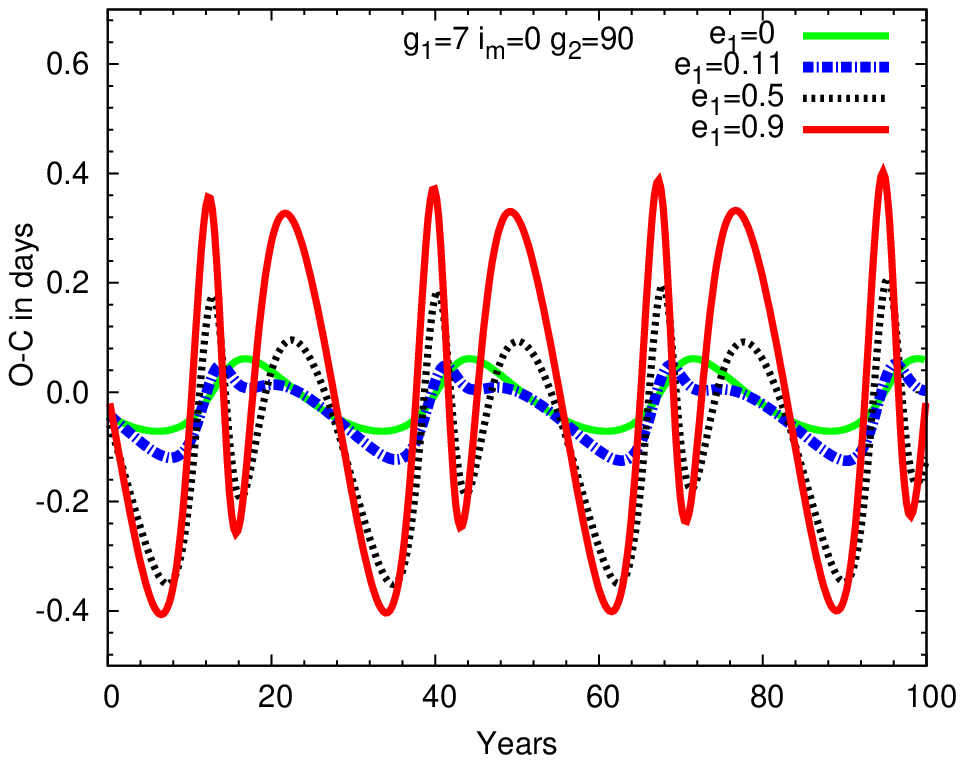}\includegraphics[width=7cm]{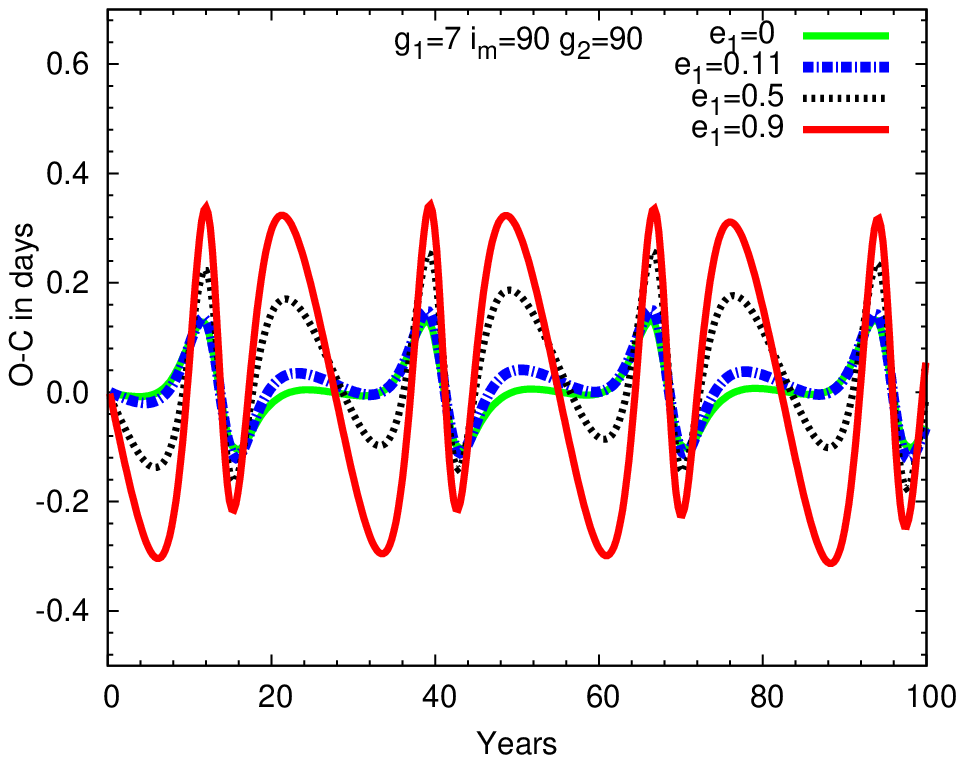}
\includegraphics[width=7cm]{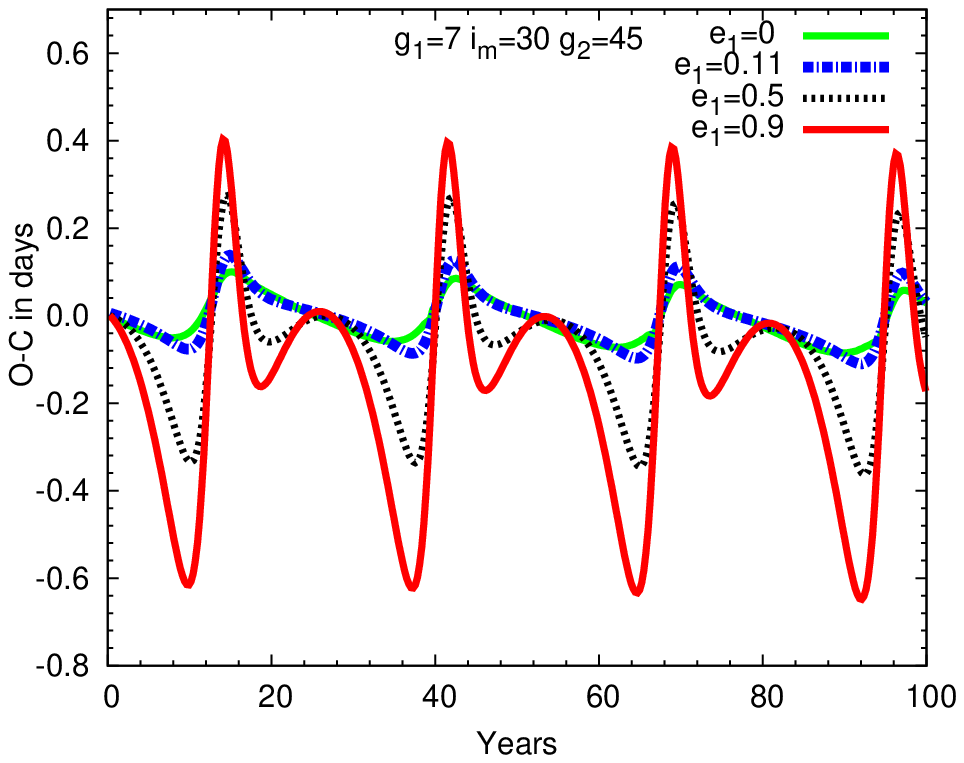}\includegraphics[width=7cm]{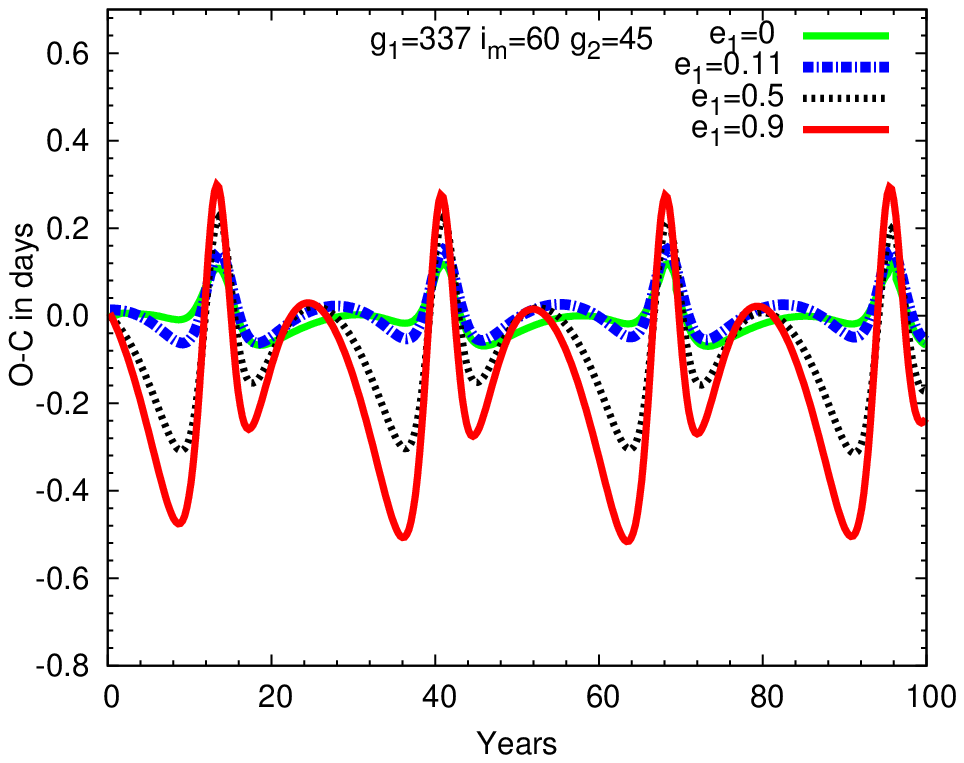}
\includegraphics[width=7cm]{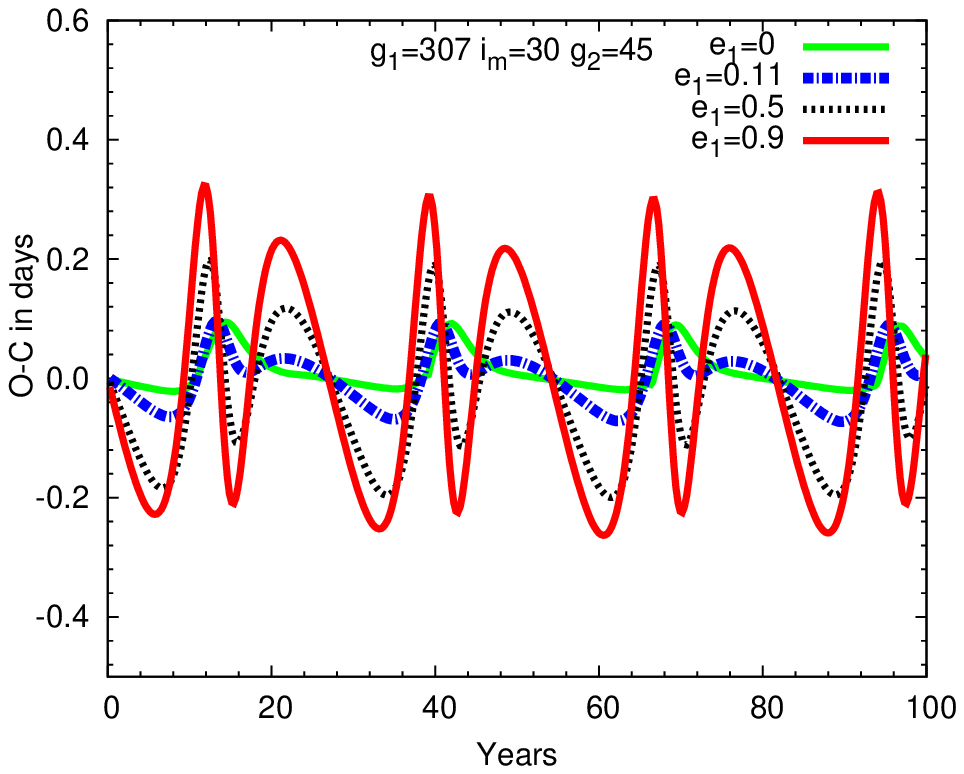}\includegraphics[width=7cm]{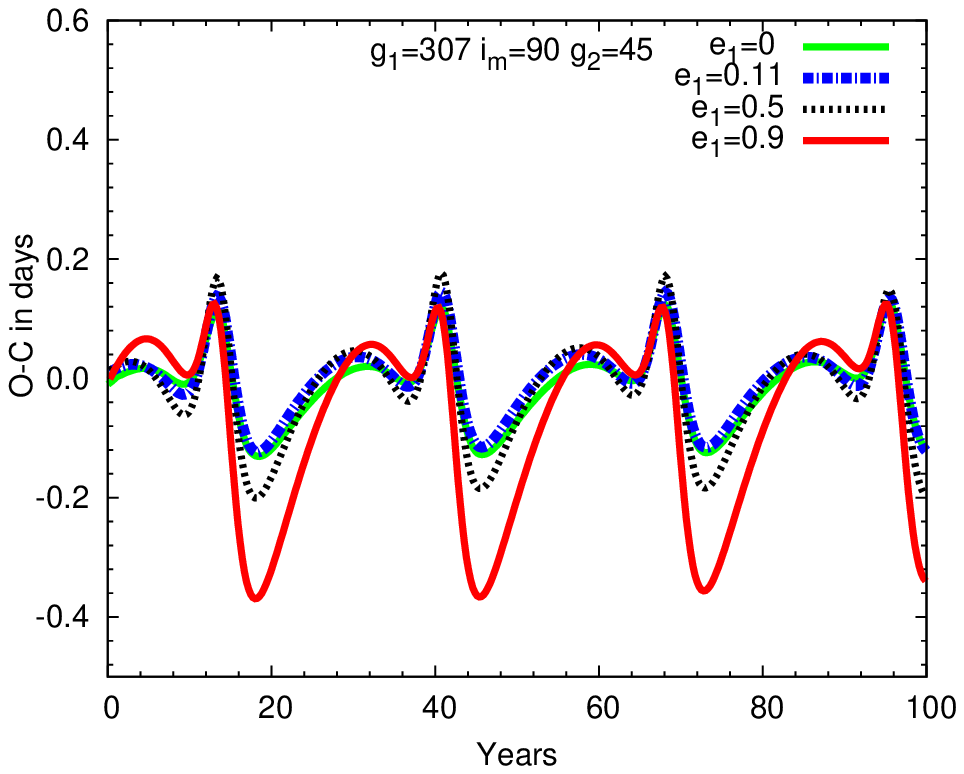}
\includegraphics[width=7cm]{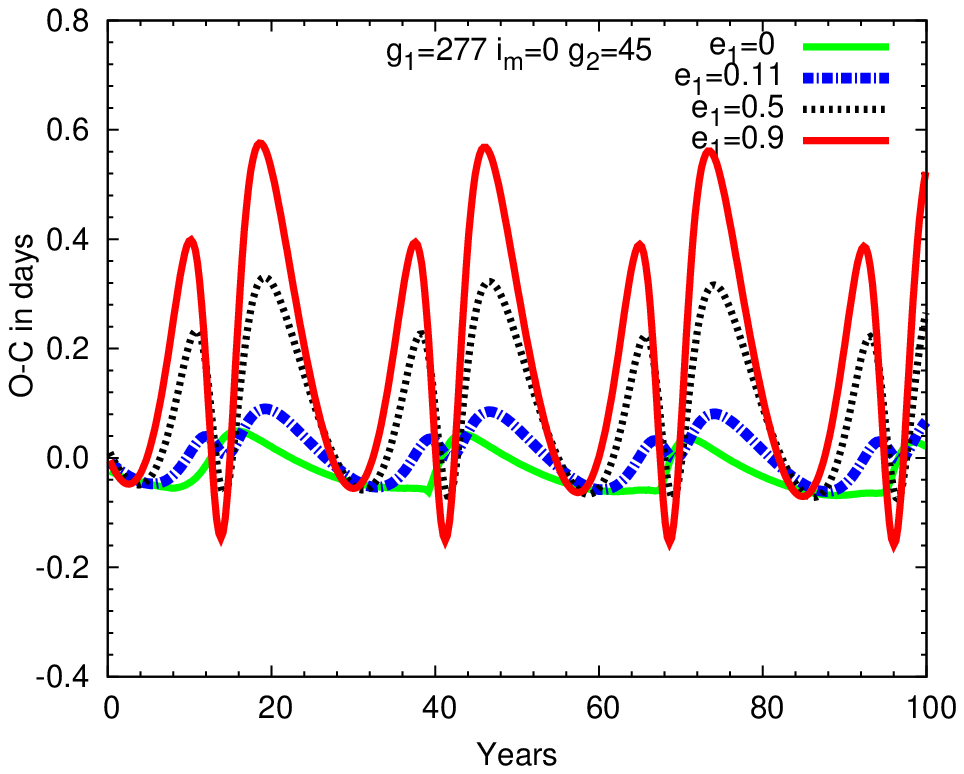}\includegraphics[width=7cm]{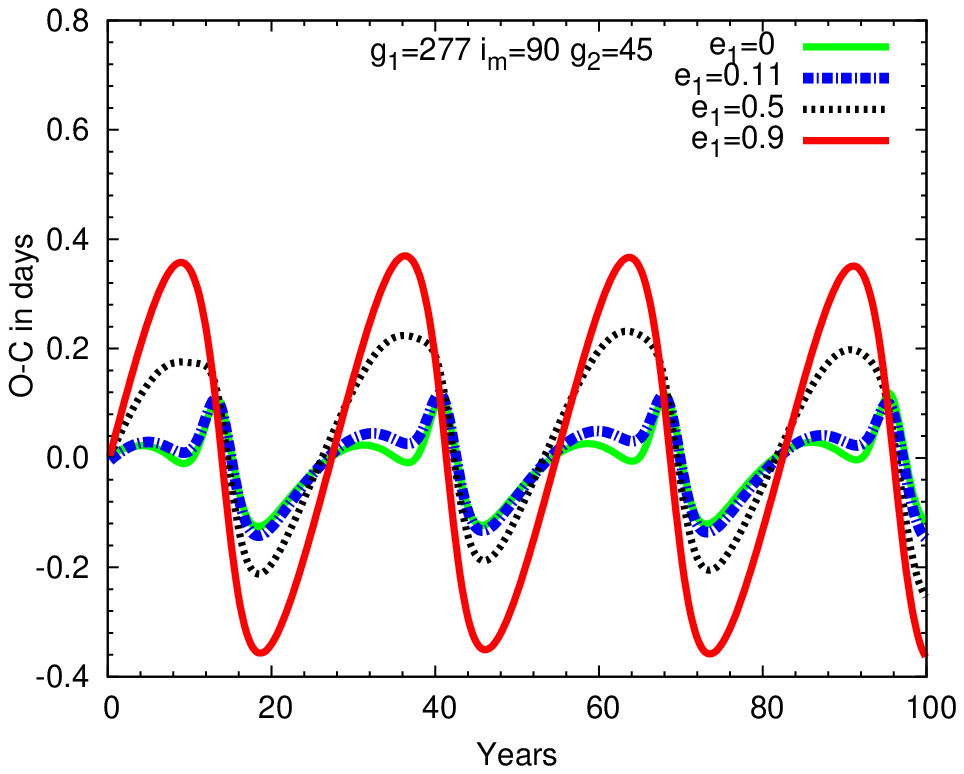}
\caption{Transit timing variatons caused by a hypothetical $P_2=10\,000$ day-period
$m_3=1 M_\mathrm{\sun}$ mass, moderately eccentric $e_2=0.3$ third companion for CoRoT-9b, at different initial orbital elements, 
for four different initial inner eccentricities ($e_1=0$, 0.11, 0.5, 0.9). The various initial elements
of each panel are as follows. Panel $(a)$: $g_1=7\degr$, $g_2=90\degr$, $i_\mathrm{m}=0\degr$; $(b)$: The same,
but for $i_\mathrm{m}=90\degr$; $(c)$: $g_1=7\degr$, $g_2=45\degr$, $i_\mathrm{m}=30\degr$; 
$(d)$: $g_1=337\degr$, $g_2=45\degr$, $i_\mathrm{m}=60\degr$; 
$(e)$: $g_1=307\degr$, $g_2=45\degr$, $i_\mathrm{m}=30\degr$;
$(f)$: same as previous, but for $i_\mathrm{m}=90\degr$.
$(g)$: $g_1=277\degr$, $g_2=45\degr$, $i_\mathrm{m}=0\degr$;
$(h)$: same as previous, but for $i_\mathrm{m}=90\degr$.
(For better comparison the curves are corrected for the different average transit periods, and zero point shifts.)}
\label{fig:C9bsamples}
\end{figure*}

As one can easily see, the result for a circular inner orbit (i.e. $e_1=0$) is identical
with the formula (46) of \citet{borkovitsetal03}. Consequently, the discussion given
in Sect.~4 of that paper is also valid. Nevertheless, as one can see in 
Figs.~\ref{fig:amplitudes-e1}, \ref{fig:C9bsamples}, a significant inner eccentricity produces notably
higher amplitudes, and consequently, is easier to detect. Furthermore, some other
attributes of the transit timing variations also change drastically. For example, in contrast
to the previously studied coplanar ($I=\pm1$), circular inner orbit ($e_1=0$) case, \citep{borkovitsetal03, agoletal05},
as long as the inner orbit is eccentric, the dynamical term does not disappear even if the outer orbit is circular ($e_2=0$).
A further important feature for eccentric
inner orbit, in that the amplitude, the phase and the shape of the $O-C$ variations cease to be
depend simply on the physical (i.e. relative) positions of the celestial bodies, but
also on the orientation of the orbit with respect to the observer.
These latter elements (i. e. $e_1\sin\omega_1$, $e_1\cos\omega_1$ and their combinations)
can also be determined from radial velocity measurements, as well as from the shape
of the transit light-curves, and, in the case of the possible detection of the occultation,
or (in the case of eclipsing binaries), of the secondary minimum, from the time delay between the
two different eclipsing events. So, while the relative, i.e. physical angular parameters
(i.e. periastron distances from the intersection of the two orbits, $g_1$, $g_2$, and
mutual inclination $i_\mathrm{m}$) cannot be aquired from other, generally used methods
(e.g. light-curve, or simple radial velocity curve analysis), these observational geometrical parameters 
could be determined from other sources of information, and then simply can be built
into such a fitting algorithm, which was described in \citet{borkovitsetal03}, or
could be included into such procedures which were presented by \citet{pal10}. 

In the following, as we are mainly interested in transiting systems, where $i_1\approx90\degr$
(which is especially true for relatively longer period, i.e. distant transiting exoplanets),
we omit terms multiplied by $\cot i_1$, i.e. terms arising from nodal motion. 
%Furthermore, 
%for the sake of simplicity, we suppose, that the outer orbit has
%some significant, but not very high eccentricity ($e_2$). In such a case the amplitude
%of the ${\cal{M}}$ and $\frac{1}{2}{\cal{S}}$, $\frac{1}{2}{\cal{C}}$ terms could be approximately
%similar, and, consequently the power spectrum of the dynamical $O-C$ contribution
%is characterised by two fundamental peaks with frequencies at $(1,2)\times P_2^{-1}$, with usually different
%phases. Nevertheless, we shoud emphasize, that due to the significant outer eccentricity,
%the first some harmonics will also occur. (See the last equations of Eqs.~[\ref{Eq:M-tag}]-[\ref{Eq:C(2v+x)}],
%where we gave the direct, linear time dependence of the ${\cal{M}}$, ${\cal{S}}$ and ${\cal{C}}$-terms.) 

In \citet{borkovitsetal03} we considered
only triple stellar systems, where the three masses were usually nearly equal, and
the inner period was of the order of a few days. In such cases light-time term dominates.
The amplitude of the light-time effect is simply 
\begin{eqnarray}
A_\mathrm{LITE}&=&\frac{m_3}{m_{123}}\frac{a_2\sin i_2}{c}\left(1-e_2^2\cos^2\omega_2\right)^{1/2} \nonumber \\
&=&\frac{m_3}{m_{123}}\left(\frac{Gm_{123}}{4\pi^2}\right)^{1/3}\frac{\sin i_2}{c}P_2^{2/3}\left(1-e_2^2\cos^2\omega_2\right)^{1/2} \nonumber \\
&\approx&1.1\times10^{-4}\frac{m_3}{m_{123}^{2/3}}\sin i_2 P_2^{2/3}\left(1-e_2^2\cos^2\omega_2\right)^{1/2},
\label{Eq:AmpliLITE}
\end{eqnarray}
where masses should be expressed in solar units, and $P_2$ in days.

The amplitude of the dynamically forced $O-C$, and consequently, the detectability limit of such perturbations
depends on almost the all dynamical as well as geometrical variables. So, we can give only
some limits on the detectability limit.
Nevertheless, for an easier, and somewhat general study,
we separate the physical and geometrical variables from each other, and furthermore,
we separate also the elements of the inner orbit from those of the outer
perturber (with the exception of the mutual inclination). In order to do this, we introduce the following
quantities, all of which depend on eccentricity ($e_1$), the two types of periastron
arguments ($g_1$, $\omega_1$), and mutual inclination ($i_\mathrm{m}$), via its cosine:
\begin{eqnarray}
\alpha&=&\left\{\frac{2}{5}+\frac{5}{4}e_1^2\left(1+\cos2g_1\right)\mp e_1\left[\frac{3}{5}\sin\omega_1+\sin(\omega_1-2g_1)\right]\right. \nonumber \\
&&-I^2\left\{\frac{2}{5}+\frac{5}{4}e_1^2\left(1-\cos2g_1\right)\right. \nonumber \\
&&\left.\left.\mp e_1\left[\frac{3}{5}\sin\omega_1-\sin(\omega_1-2g_1)\right]\right\}\right\}\left(1-e_1^2\right)^{1/2}, \\
\beta&=&-\left(1-e_1^2\right)^{1/2}2I\left[\frac{5}{4}e_1^2\sin2g_1\mp e_1\cos(\omega_1-2g_1)\right], \\
\gamma&=&\left\{\!-\frac{4}{15}\!+\!\frac{5}{2}e_1^2\left(\cos2g_1-\frac{1}{3}\right)\mp2e_1\left[\sin(\omega_1-2g_1)-\!\frac{1}{5}\sin\omega_1\!\right]\right.\nonumber \\
&&+I^2\left\{\frac{4}{5}+\frac{5}{2}e_1^2\left(1-\cos2g_1\right)\right. \nonumber \\
&&\left.\left.\mp2e_1\left[\frac{3}{5}\sin\omega_1-\sin(\omega_1-2g_1)\right]\right\}\right\}\left(1-e_1^2\right)^{1/2}.
\end{eqnarray}
Then
\begin{equation}
O-C_\mathrm{dyn}\approx\frac{P_1}{2\pi}A_\mathrm{L}\left[\gamma{\cal{M}}+\!\!\sqrt{\alpha^2+\beta^2}{\cal{S}}(2v_2+\phi)\right],
\label{eq:OminCdyntomor}
\end{equation}
where
\begin{equation}
\phi=2g_2+\arctan\frac{\beta}{\alpha}
\end{equation}
or, in trigonometric form
\begin{equation}
O-C_\mathrm{dyn}\approx\frac{P_1}{2\pi}A^*_\mathrm{L}\sum_n A_n\sin(nv_2+\phi_n),
\label{eq:OminCdyntrig}
\end{equation}
where (up to third order in $e_2$)
\begin{eqnarray}
A_1&=&\left(1-e_2^2\right)^{-3/2}e_2\sqrt{A_{\cal{S}}^2+A_{\cal{M}}^2+2A_{\cal{S}}A_{\cal{M}}\cos\phi}, \\
A_2&=&\left(1-e_2^2\right)^{-3/2}\sqrt{A_{\cal{S}}^2+\frac{1}{16}e_2^4A_{\cal{M}}^2+\frac{1}{2}e_2^2A_{\cal{S}}A_{\cal{M}}\cos\phi}, \\
A_3&=&\left(1-e_2^2\right)^{-3/2}\frac{1}{3}e_2\sqrt{A_{\cal{S}}^2+\frac{1}{9}e_2^4A_{\cal{M}}^2+\frac{2}{3}e_2^2A_{\cal{S}}A_{\cal{M}}\cos\phi}, \\
\phi_1&=&\arctan\frac{\sin\phi}{\cos\phi+\frac{A_{\cal{M}}}{A_{\cal{S}}}}, \\
\phi_2&=&\arctan\frac{\sin\phi}{\cos\phi+\frac{1}{4}e_2^2\frac{A_{\cal{M}}}{A_{\cal{S}}}}, \\
\phi_3&=&\arctan\frac{\sin\phi}{\cos\phi+\frac{1}{9}e_2^2\frac{A_{\cal{M}}}{A_{\cal{S}}}},
\end{eqnarray}
and
\begin{eqnarray}
A_{\cal{S}}&=&\sqrt{\alpha^2+\beta^2}, \\
A_{\cal{M}}&=&3\gamma,
\end{eqnarray}
and, furthermore,
\begin{equation}
A^*_\mathrm{L}=A_\mathrm{L}\left(1-e_2^2\right)^{3/2}.
\end{equation}
Note, for a circular outer orbit ($e_2=0$)
\begin{eqnarray}
A_1=A_3&=&0, \\
A_2&=&A_{\cal{S}}.
\end{eqnarray}

As one can see, both sets of amplitudes, i.e. $A_{{\cal{M}},{\cal{S}}}$ and $A_{1,2,3}$
are independent of the real physical parameters of the current exoplanetary system.
Furthermore, the dependence from the elements of the outer orbit (i.e. $e_2$, $g_2$) 
appear only in the $A_{1,2,3}$ Fourier amplitudes. 
The masses (or more accurately mass ratios), and the periods and period ratios (and
indirectly the physical size of the system) occur only as a scaling parameter.
In this way, the following general statements are valid for every hierarchical triple
systems (as far as the initial model assumptions are valid). In order to get the real,
i.e. physical values for the $O-C$ amplitudes in a given system, one must multiply the general system-independent, dimensionless amplitudes with the
system specific number $\frac{15}{16\pi}\frac{P_1^2}{P_2}\frac{m_3}{m_1+m_2+m_3}$.

Considering first the (nearly) coplanar case, i.e. 
when $I\approx\pm1$, then the (half-)amplitude of the two sinusoidals becomes:
\begin{eqnarray}
e_2A_{{\cal{M}}0}&\approx&\frac{8}{5}\left(1-e_1^2\right)^{1/2}\left(1+\frac{25}{8}e_1^2\mp\frac{3}{2}e_1\sin\omega_1\right)e_2, \\
A_{{\cal{S}}0}&\approx&2\left(1-e_1^2\right)^{1/2}e_1\sqrt{1\mp\frac{5}{2}e_1\sin\omega_1+\frac{25}{16}e_1^2}.
\label{Eq:AmpliP2dyncoplanar}
\end{eqnarray}
This illustrates the above-mentioned fundamental differences to the previously-studied coplanar, $e_1=0$ case \citep{borkovitsetal03, agoletal05},
as according to our new result for eccentric inner orbits, the dynamical term does not
disappear even if the outer orbit is circular ($e_2=0$). Furthermore, the amplitudes
strongly depend on the orientation of the orbital axis with respect to the observer.
When the apsidal line coincides (more or less) with the line of sight (i.e.
$\omega_1=\pm90\degr$) there can be very significant differences both in the shape
and amplitude of the primary (transit) and secondary (occultation) $O-C$ curves. However,
when the apsidal line lies nearly in the sky, then these differences disappear.
A further interesting feature of the $\omega_1=\pm90\degr$ configuration, is that for eclipse events which occur around apastron there is a full square
under the square-root sign in the ${\cal{S}}$-term, with the root of $e_1=0.8$, which
means that in this situation this term would disappear. Nevertheless, for such a
high eccentricity the first order approximation is far from being valid, as is
illustrated in Fig.~\ref{fig:amplitudes-e1}.

For highly inclined ($I\approx0$) orbits the (half-)amplitudes are as follows:
\begin{eqnarray}
e_2A_{{\cal{M}}90}&\approx&\frac{4}{5}\left(1-e_1^2\right)^{1/2}\left\{-1-\frac{25}{8}e_1^2\left(1-3\cos2g_1\right)\right. \nonumber \\
&&\left.\pm\frac{3}{2}e_1\left[\sin\omega_1-5\sin(\omega_1-2g_1)\right]\right\}e_2, \\
A_{{\cal{S}}90}&\approx&\frac{2}{5}\left(1-e_1^2\right)^{1/2}\left\{1+\frac{25}{8}e_1^2\left(1+\cos2g_1\right)\right. \nonumber \\
&&\left.\mp\frac{3}{2}e_1\left[\sin\omega_1+\frac{5}{3}\sin(\omega_1-2g_1)\right]\right\},
\label{Eq:AmpliP2dynperpendic}
\end{eqnarray}
furthermore, the phase of the ${\cal{S}}$-term is simply
\begin{equation}
\phi=2g_2.
\end{equation}
In this case a further parameter, namely, the periastron distance of the inner planet 
from the intersection of the two orbital planes ($g_1$) also plays an important role.

Finally we also mention a very specific case, namely for the maximum eccentricity phase
of the Kozai mechanism driven $e$-cycles. During this phase, $\cos2g_1$ takes
one definite value, namely $\cos2g_1=-1$. Furthermore,
the mutual inclination of the two orbits here reaches its minimum. The actual value
depends on both the maximum mutual inclination $i_\mathrm{m}$, or more strictly, on $j_1$,
and the minimal inner eccentricity $e_1$. Nevertheless, in the case of an initially almost circular
inner orbit, the minimum mutual inclination is almost independent of
its maximum value, and takes $j_1(\approx i_\mathrm{m})\approx39\fdg23$ (or its retrograde
counterpart, $j_1(\approx i_\mathrm{m})\approx140\fdg77$), i.e. $I^2=3/5$. 
For this scenario:
%\begin{itemize}
%\item[(a)]{$g_1=90\degr+k\cdot180\degr$}
\begin{eqnarray}
e_2A_{{\cal{M}}\mathrm{Kozai},90}&\approx&\frac{16}{25}\left(1-e_1^2\right)^{1/2}\left(1-\frac{25}{16}e_1^2\pm\frac{9}{4}e_1\sin\omega_1\right)e_2, \\
%A_{{\cal{S}}\mathrm{Kozai},90}&\approx&\frac{3}{4\pi}\frac{m_3}{m_{123}}\frac{P_1^2}{P_2}\frac{\left(1-e_1^2\right)^{1/2}}{\left(1-e_2^2\right)^{3/2}}\nonumber \\
%&&\times\sqrt{\frac{1}{25}+\frac{257}{100}e_1^2+\frac{225}{64}e_1^4\pm\frac{17}{25}e_1\left(1-\frac{75}{8}e_1^2\right)\sin\omega_1+\frac{43}{100}e_1^2\cos2\omega_1}.
\alpha_{\mathrm{Kozai},90}&=&\left(1-e_1^2\right)^{1/2}\left(\frac{4}{25}-\frac{5}{3}e_1^2\pm\frac{34}{25}e_1\sin\omega_1\right), \nonumber \\
\beta_{\mathrm{Kozai},90}&=&\mp\left(1-e_1^2\right)^{1/2}\frac{2}{5}\sqrt{15}e_1\cos\omega_1.
\label{Eq:AmpliP2dynKozai-g190}
\end{eqnarray}

In order to get a better overview of the parameter dependence of the formulae above, 
we investigate the $A_{{\cal{M}},{\cal{S}}}$, as well as the $A_{1,2,3}$ amplitudes graphically.
Due to the complex dependence of these amplitudes on many parameters, it is difficult
to give general statements. Therefore we investigate only the dependence of
the amplitudes on the inner eccentricity, while the effects of other parameters
will be considered in Sect.~\ref{Sect: case studies} for specific systems, where
some of the parameters can be fixed.

Fig.~\ref{fig:amplitudes-e1} shows the inner eccentricity ($e_1$) dependence of
the amplitudes for some specific values of the other parameters. In general, one sees that the $e_1$ increase with amplitude, 
but this growth usually remains within one order of magnitude.
So, although the shapes of the individual $O-C$ curves may differ significantly, 
the net amplitudes vary over a narrow range. The graphs also suggest another,
perhaps suprising fact, that the amplitude of $O-C_\mathrm{P2dyn}$
depends only weakly on the mutual inclination ($i_\mathrm{m}$). This is especially valid
for medium inner eccentricities, since in this case, at least for the cases shown in Fig.~\ref{fig:amplitudes-e1}, all $A_{{\cal{M}},{\cal{S}}}$ amplitudes have similar values.
Moreover, the numerically generated sample $O-C$ curves of Fig.~\ref{fig:C9bsamples},
also suggest such a conclusion.
Nevertheless, there are several exceptions. 
Some particular configurations some of the amplitudes, or even both of them
(and consequently, the corresponding terms) can disappear. Such a situation will be
discussed in Sect.~\ref{Sect: C9b}.
We will return to this question in detail in Sect.~\ref{Sect: case studies}, where
the dependence of the amplitudes on other parameters will be also studied for the
systems of \object{CoRoT-9b} and \object{HD 80606b}.

Returning now to the LITE amplitude, 
and comparing it with the dynamical case, an increment of $P_2$ (keeping $P_1$ constant)
results in an increment of $A_\mathrm{LITE}/A_\mathrm{P2-dyn}$ by $P_2^{5/3}$. 
Consequently, for more distant systems
the pure geometrical effect tends to exceed the dynamical one, as is the case in
all but one (\object{$\lambda$ Tau}, \citealp[see e. g.][]{soderhjelm75}) of the known hierarchical eclipsing triple stellar systems.
Nevertheless, as we will illustrate in this paper (see Fig.~\ref{fig:C9bg169g290OminC}), we have a good chance of finding the opposite
situation in some of the recently discovered transiting exoplanetary systems. We can make the
following crude estimation. Consider a system with a solar-like host star, and
two approximately Jupiter-mass companions) ($m_{123}\approx m_1=1\mathrm{M}_\odot$, 
$m_2=m_3=10^{-3}\mathrm{M}_\odot$) choosing $A=10^{-3}$ day for the case of a certain detection,
then the LITE term for the most ideal case gives $P_2\ge10^6\mathrm{d}\approx2\,700\mathrm{y}$. Alternatively, setting
$m_3=10^{-2}\mathrm{M}_\odot$, and allowing $A=10^{-4}$ for detection limit,
the result is $P_2\ge10^3\mathrm{d}\approx2.7\mathrm{y}$. Similarly, for two Jupiter-mass
planet, in the coplanar case, for small $e_1$, the $A=10^{-3}$ day limit gives
\begin{equation}
P_2\le2\times10^3m_3e_2\left(1-e_2^2\right)^{-3/2}P_1^2
\label{eq:P_2coplimit}  
\end{equation}
condition for the detectability of a third companion by its long-term dynamical
perturbations. For the perpendicular case the same limit is
\begin{equation}
P_2\le\frac{1}{2}\times10^3m_3\left(1-e_2^2\right)^{-3/2}\left(1+\frac{2}{3}e_2\right)P_1^2.
\label{eq:P_2perlimit}  
\end{equation}
We note that for $m_3<<m_1$ the above equations are linear for $m_3$, 
so it is very easy to give the limiting period $P_2$ in the function of $m_3$. 
Nevertheless, we emphasize again that
there are so many terms with different periodicity and phase, that these equations give
only a very crude, first estimation. For $P_1=$1, 10, 100 days,
(at zero outer eccentricity) gives $P_2\le$0.5, 50, $5\,000$ days, respectively,
 
These results refer to the total amplitude of the $O-C$ curve, i.e. the variation of
the transit times during a complete revolution of the distant companion, which can
take as long as several years or decades. Naturally, the perturbations in the transit times
could be observed within a much shorter period, from the variation of the interval
between consecutive minima. 
This estimation can be calculated e.g from Eq.~(\ref{eq:OminCdyntomor}), or even directly from Eq.~(\ref{Eq:dlambda1dv2}). 
According to the meaning of the $O-C$
curve, the transiting or eclipsing period ($\overline{P}$) between two consecutive (let us assume the $n$-th and $n+1$-th) minima is
\begin{eqnarray}
\overline{P}_n=t_{n+1}-t_{n}&=&\left(O-C\right)(t_{n+1})-\left(O-C\right)(t_n) \nonumber \\
&\approx&P_\mathrm{s}+\frac{P_1}{2\pi}A_\mathrm{L}\left(1-e_1^2\right)^{1/2}\Delta v_{2n}\left(1+e_2\cos v_{2n}\right) \nonumber \\
&&\times\left[\gamma+2\sqrt{\alpha^2+\beta^2}\cos(2v_{2n}+\phi)\right],
\label{Eq:onetransit}
\end{eqnarray}
where up to third order in $e_2$
\begin{eqnarray}
\Delta v_2&\approx&\Delta l_2\left(1+2e_2^2+2e_2\cos v_2+\frac{1}{2}e_2^2\cos2v_2+3e_2^3\cos 3v_2\right) \nonumber \\
&\approx&2\pi\frac{\overline{P}}{P_2}(1+\delta v_2) \nonumber \\
&\approx&2\pi\frac{P_1}{P_2}(1+\delta v_2),
\end{eqnarray}
i.~e. we approximated the variation of the mean anomaly of the outer companion ($l_2$)
by the constant value of
\begin{equation}
\Delta l_2\approx2\pi\frac{P_1}{P_2}.
\end{equation}
(Here, and in the following text we neglect the pure geometrical LITE contribution.)
Then the variation of the length of the consecutive transiting periods becomes
\begin{eqnarray}
\Delta\overline{P}&\approx&-\frac{15}{4}\pi\frac{m_3}{m_{123}}\frac{P_1^4}{P_2^3}\frac{\left(1-e_1^2\right)^{1/2}}{\left(1-e_2^2\right)^{3/2}}\nonumber \\
&&\times\left\{\gamma\left[3e_2\left(1+\frac{9}{2}e_2^2\right)\sin v_2-6e_2^2\sin2v_2-\frac{9}{2}e_2^3\sin3v_2\right]\right. \nonumber \\
&&+2\sqrt{\alpha^2+\beta^2}\left[2\left(1+8e_2^2\right)\sin(2v_2+\phi)\right. \nonumber \\
&&+\frac{7}{2}e_2\left(1+\frac{9}{2}e_2^2\right)\sin(v_2+\phi)\nonumber \\
&&+\frac{13}{2}e_2\left(1+\frac{9}{2}e_2^2\right)\sin(3v_2+\phi)\nonumber \\
&&\left.\left.+\frac{1}{4}e_2^3\sin(-v+\phi)+\frac{19}{4}e_2^3\sin(5v_2+\phi)\right]\right\}.
\label{Eq:deltaPv2}
\end{eqnarray}
Comparing (the amplitude of) this result with Eqs.~(1,2) of \citet{holmanmurray05}, one can see that the power
of the period ratio differs. (Furthermore, in the original paper $\pi$ stands
at the same place, i.e. in the numerator (as above) but later, in \citealp{holman10} this was declared as an error,
and it was put into the denominator.) There is a principal difference in the background.
\citet{holmanmurray05} state that they ``estimate the variation in transit intervals
between successive transits'', but what they really calculate is the departure of a transiting
period (i.e. the interval between two successive transits) from a constant
mean value. This quantity was estimated in our Eq.~(\ref{Eq:onetransit}). In other words,
the Lagrangian perturbation equations gives the instantaneous period of the perturbed system.
So, its integration gives the time left between two consecutive minima. The first derivative
of the perturbation equation gives the instantaneous period variation, and so the
variation of the length of two consecutive transiting period can be deduced by integrating the
latter. As a consequence, the best possibility for the detection of the transit timing variation,
and so, for the presence of some perturber occurs when the absolute value of the second derivative of the
$O-C$ diagram, or practically Eq.~(\ref{Eq:deltaPv2}) (the period variation during a revolution)
is maximum. We will illustrate this statement in the next Section for specific systems.

\section{Case studies \label{Sect: case studies}}

\subsection{\object{CoRoT-9b} \label{Sect: C9b}}

\begin{figure*}
\centering
\includegraphics[width=6cm]{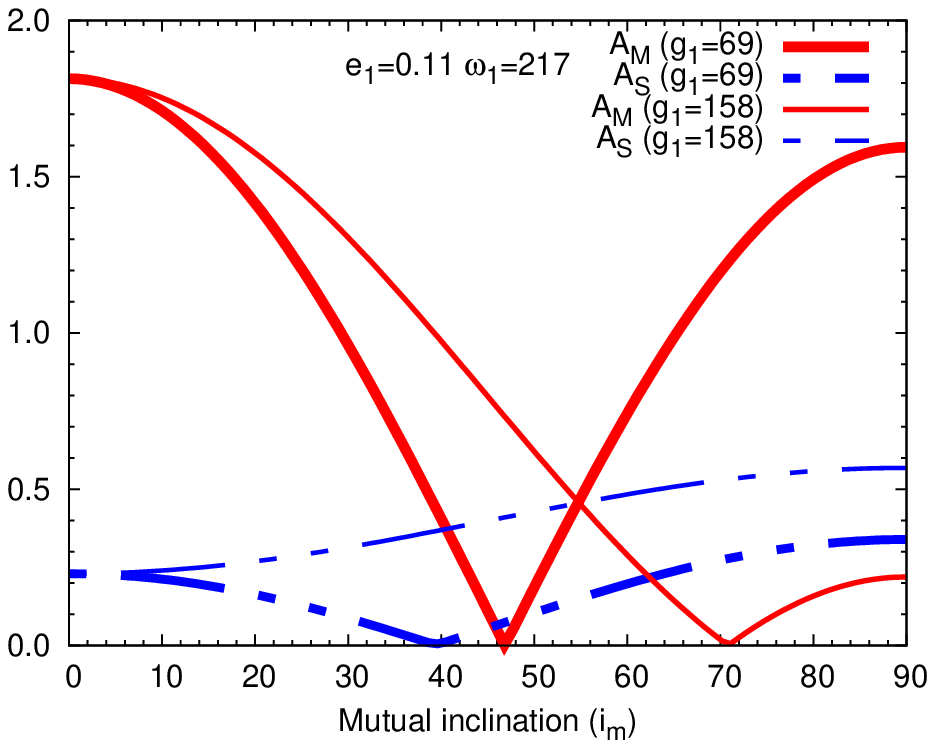}\includegraphics[width=6cm]{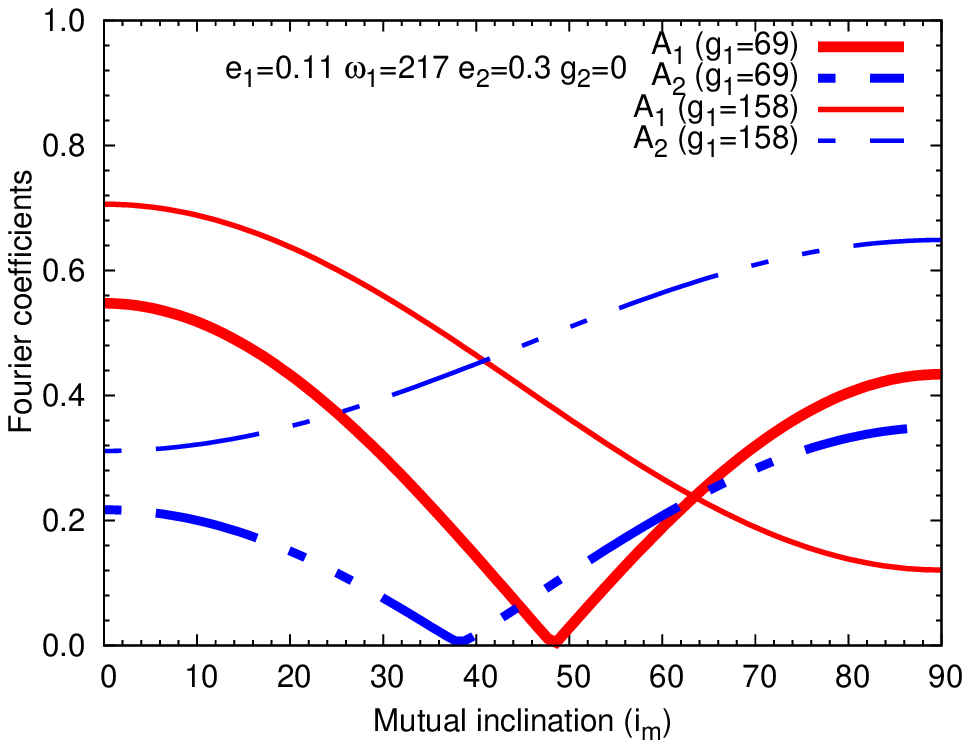}\includegraphics[width=6cm]{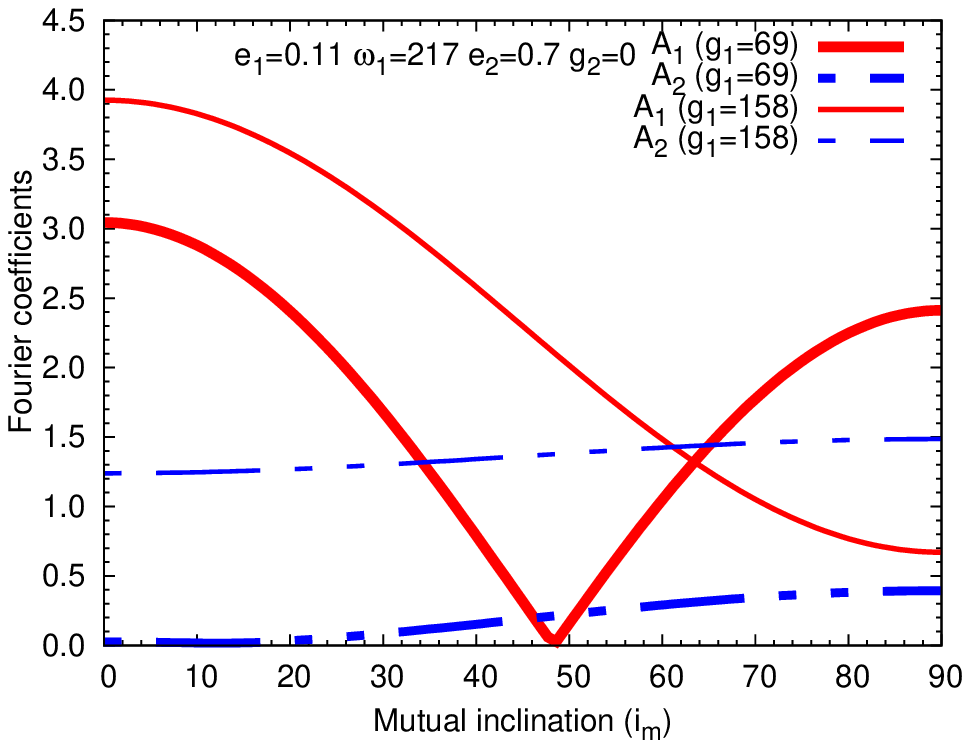}
\includegraphics[width=6cm]{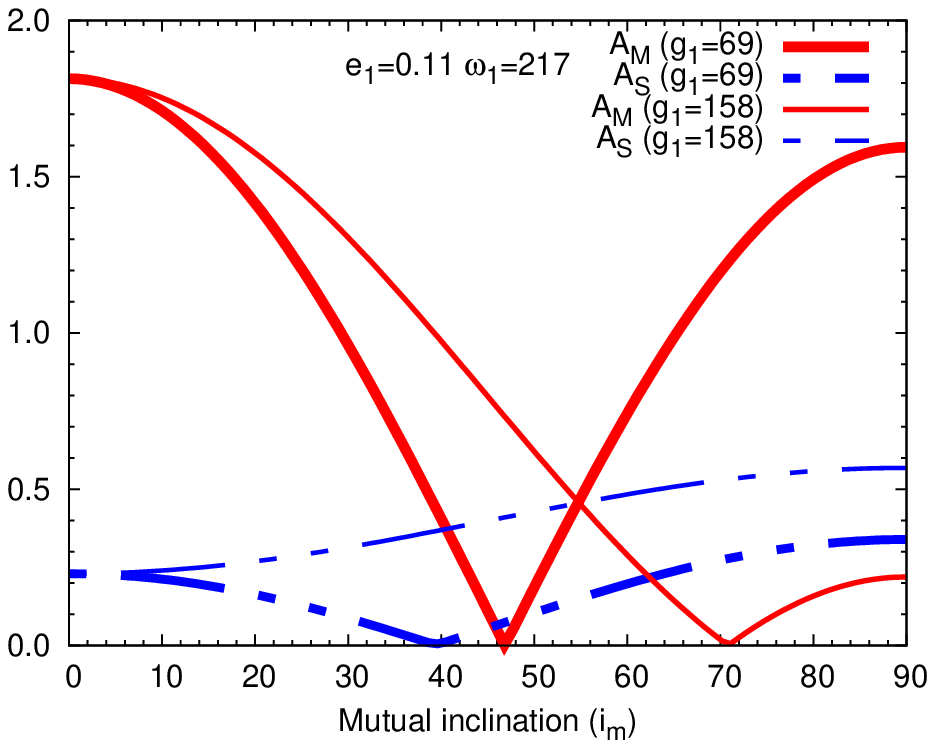}\includegraphics[width=6cm]{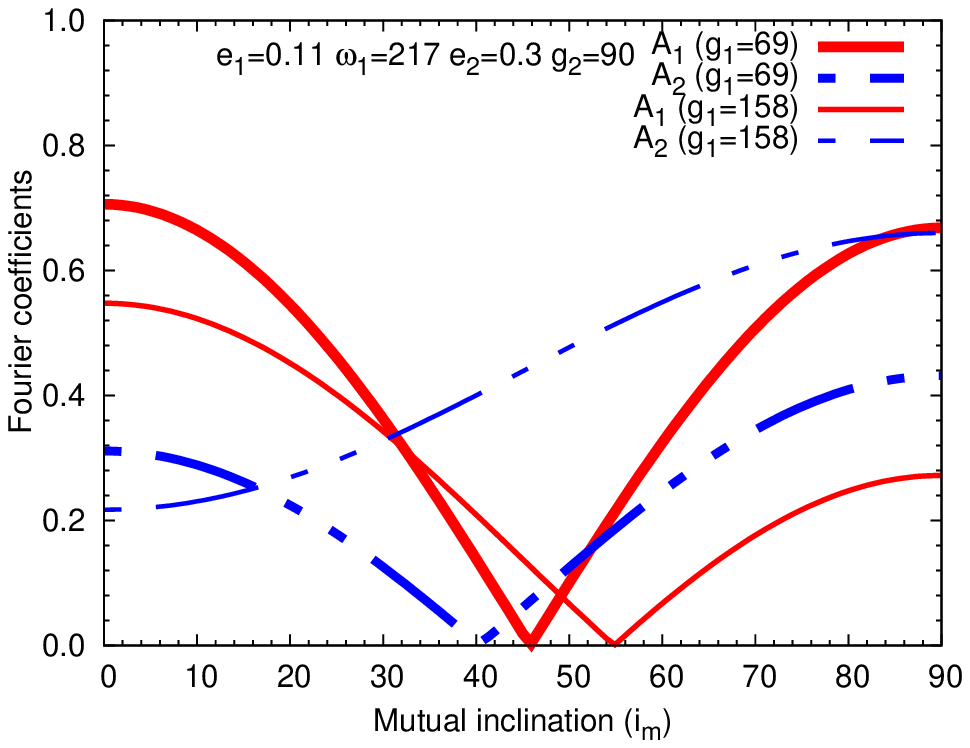}\includegraphics[width=6cm]{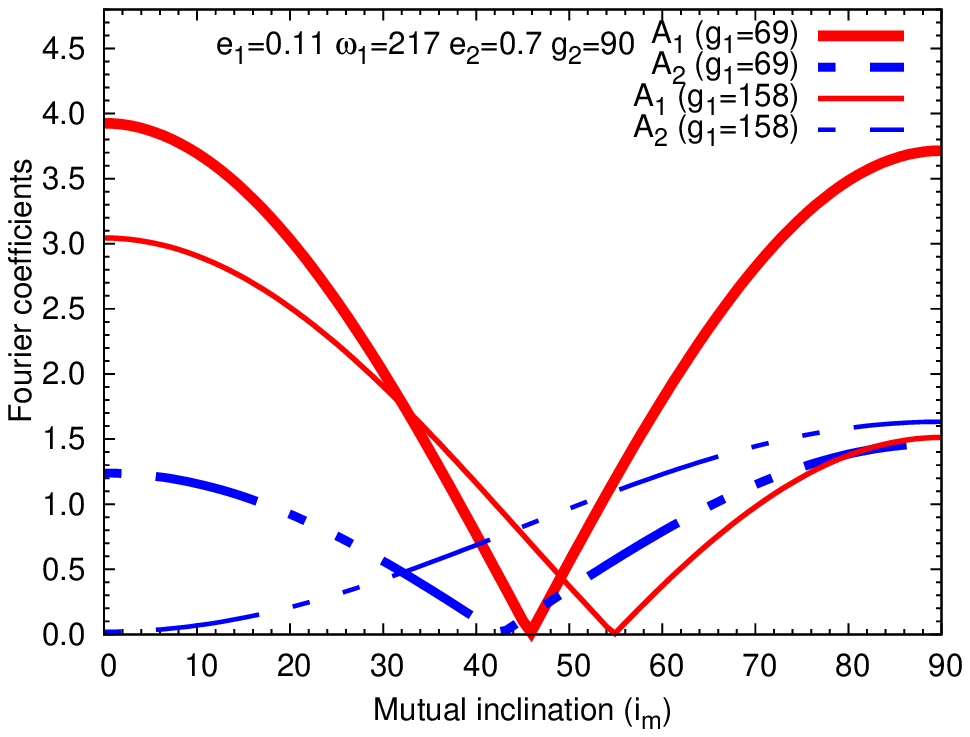}
\caption{{\it Left panels:} The mutual inclination ($i_\mathrm{m}$) dependence of $A_{\cal{M}}$, $A_{\cal{S}}$
amplitudes of ${\cal{M}}$, ${\cal{S}}$ functions of the long-period dynamical component of $O-C$ for
CoRoT-9b, for $g_1=69$\degr and $g_1=158$\degr. For other $g_1$ values the corresponding curves run within
the area limited by these lines. (The two left panels are identical.)
{\it Middle and right panels:} The corresponding $A_{1,2}$
amplitudes of the trigonometric representation (Eq.~\ref{eq:OminCdyntrig}) of the $O-C$ for two different outer
eccentricities $e_2=0.3$ (middle) and $e_2=0.7$ (right); and outer dynamical (relative) argument
of pericentrum arguments $g_2=0$\degr (up) and $g_2=90$\degr (down). For the sake of clarity, we did not
plot the small $A_3$ coefficients. (Note, for $e_2=0.0$ the $A_{\cal{S}}$ amplitudes of the 
two identical left panels are equal to $A_2$, while $A_1=A_3=0$.)}
\label{fig:C9bparameterspace}
\end{figure*}

\object{CoRoT-9b} is a transiting giant planet which revolves around its host star approximately at the distance
of Mercury \citep{deegetal10}. Consequently, tidal forces (including the possible rotational
oblateness) can safely be neglected in this system. Furthermore, due to the relatively large
absolute separation of the planet from its star, we can expect a large amplitude signal from the perturbations
of a hypothetical, distant (but not too distant) further companion. To illustrate this possibility,
and, furthermore, to check our formulae, we both calculated and plotted the amplitudes
with the measured parameters of this specific system, and carried out short-term numeric integrations
for comparison.
The physical, and some of the orbital elements of CoRoT-9b were taken from \citet{deegetal10}.
These data are listed in Table~\ref{tab:binaryfixparams}. 
(Note, we added $180$\degr to the $\omega_1$ published in that paper, as the spectroscopic $\omega_1$
refers to the orbit of the host star around the common centre of mass of the star-planet
double system, and consequently, differ by $180$\degr from the observational argument
of periastron of the relative orbit of the transiting planet around its host star.)
In the case of the numerical integrations, the inner planets were started from periastron,
while the outer one from its apastron.

Fixing the observationally aquired data, $A_{{\cal{M}},{\cal{S}}}$
depend only on two parameters, namely $g_1$ and $i_\mathrm{m}$. 
In the left panels of
Fig.~\ref{fig:C9bparameterspace} we plotted the $i_\mathrm{m}$ versus $A_{{\cal{M}},{\cal{S}}}$ graphs
for $g_1=69$\degr and $g_1=158$\degr. We found that the amplitudes reach their extrema
around these periastron arguments, i.e. for other $g_1$ values results occur within the areas
limited by these lines.\footnote{\bf Strictly speaking this latter argument is true only if
negative values are also allowed for the coefficients.} In the middle and right panels the effect of the two
further free parameters, i.e. outer eccentricity
($e_2$), and dynamical (relative) argument of periastron ($g_2$) were also considered. 
The middle panels show $A_{1,2}$ for $e_2=0.3$, the
right panels for $e_2=0.7$, for both $g_2=0$\degr (upper panels), and  $g_2=90$\degr (bottom panels). 
(Note, that $A_{\cal{S}}$, plotted in the left panel
is identical to $A_2$ for $e_2=0$, in which case $A_{1,3}=0$.)

These figures clearly show again that the amplitudes, and consequently, the actual
full amplitude of the dynamical $O-C$ curve remains within one order of magnitude
over a wide range of orbital parameters. This once more verifies the very crude
estimations given by Eqs.~(\ref{eq:P_2coplimit},\ref{eq:P_2perlimit}). 
As a consequence, for a given system we
can estimate very easily the expected amplitude of the $O-C$ variations induced
by a further companion. For a planet-mass third body, the 
\begin{equation}
A_\mathrm{dyn}\approx\frac{1}{2\pi}\frac{m_3}{P_2}\frac{P_1^2}{m_\mathrm{host-star}}
\end{equation}
gives a likely estimation at least in magnitude.  
For example, for \object{CoRoT-9b}
\begin{eqnarray}
{\cal{A}}=\frac{1}{2\pi}\frac{P_1^2}{m_\mathrm{host-star}}&\approx&1459\mathrm{d}^2\mathrm{M}_\mathrm{\sun}^{-1} \nonumber \\
&\approx&1.39\mathrm{d}^2\mathrm{M}_\mathrm{J}^{-1},
\end{eqnarray}
i.e. a Jupiter-mass additional planet could produce $0\fd001$
half-amplitude already from the distance of the Mars. (Of course, if $e_1$
and $\omega_1$ are known, a more precise estimation can be provided easily using the formulae of the present paper.)

\begin{figure*}
\centering
\includegraphics[width=6cm]{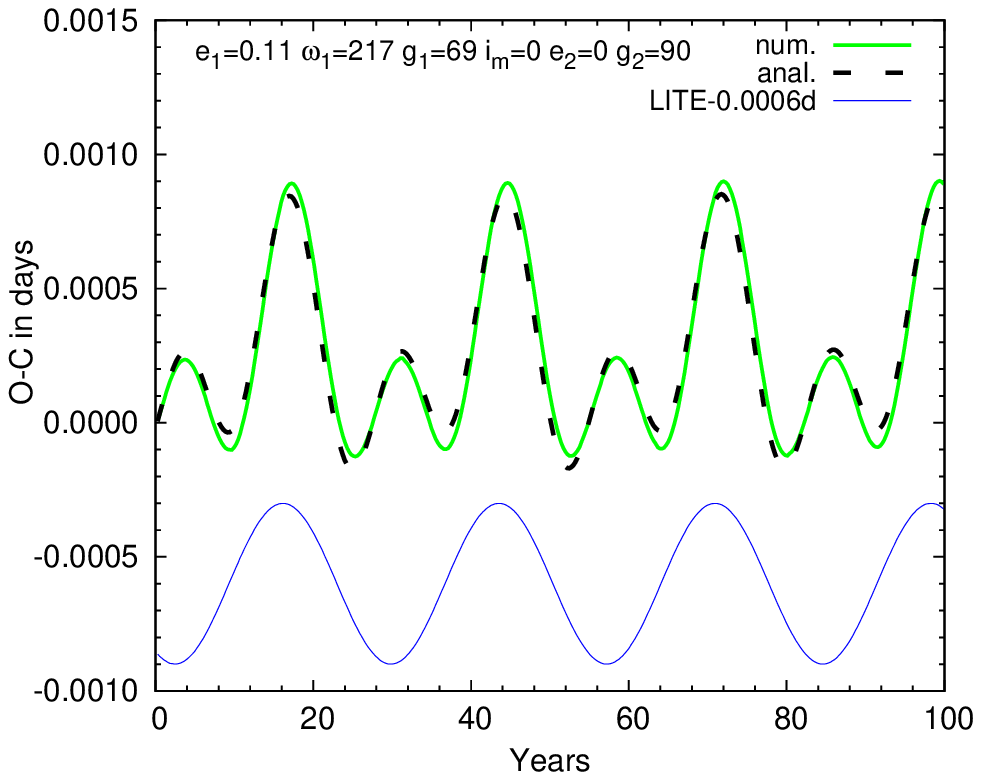}\includegraphics[width=6cm]{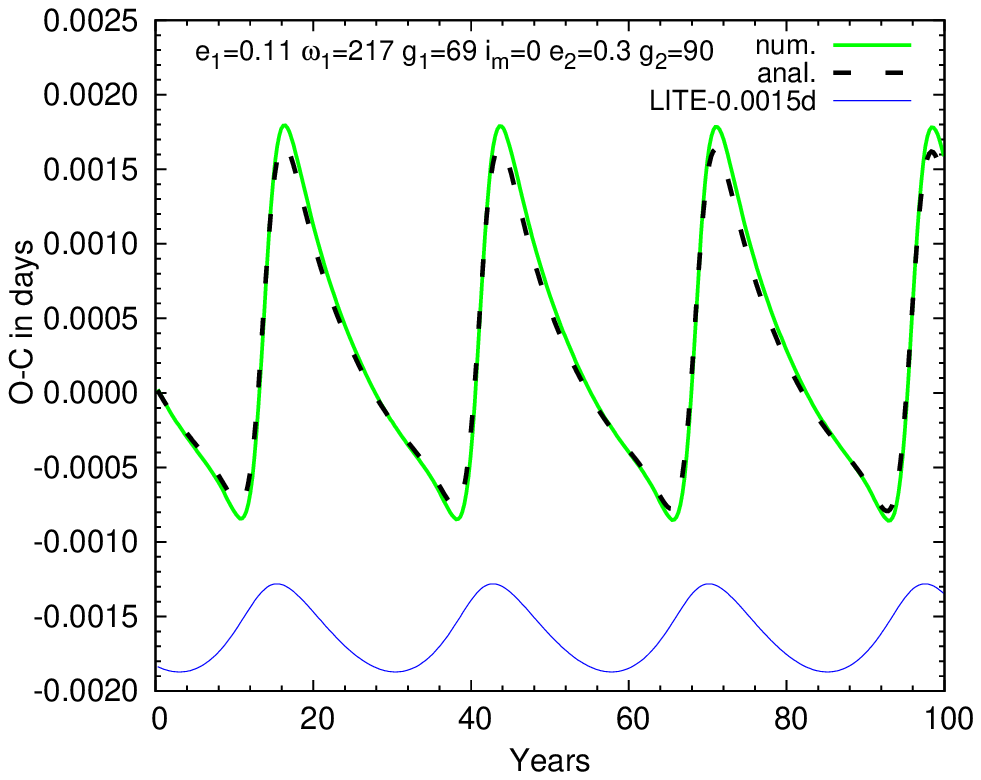}\includegraphics[width=6cm]{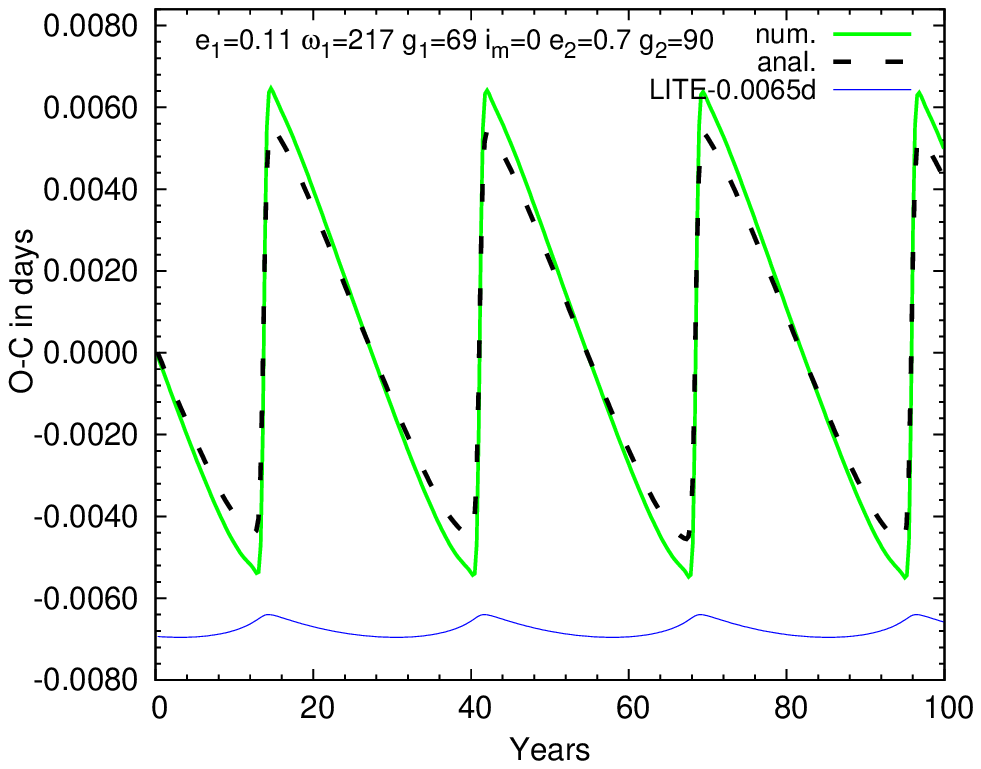}
\includegraphics[width=6cm]{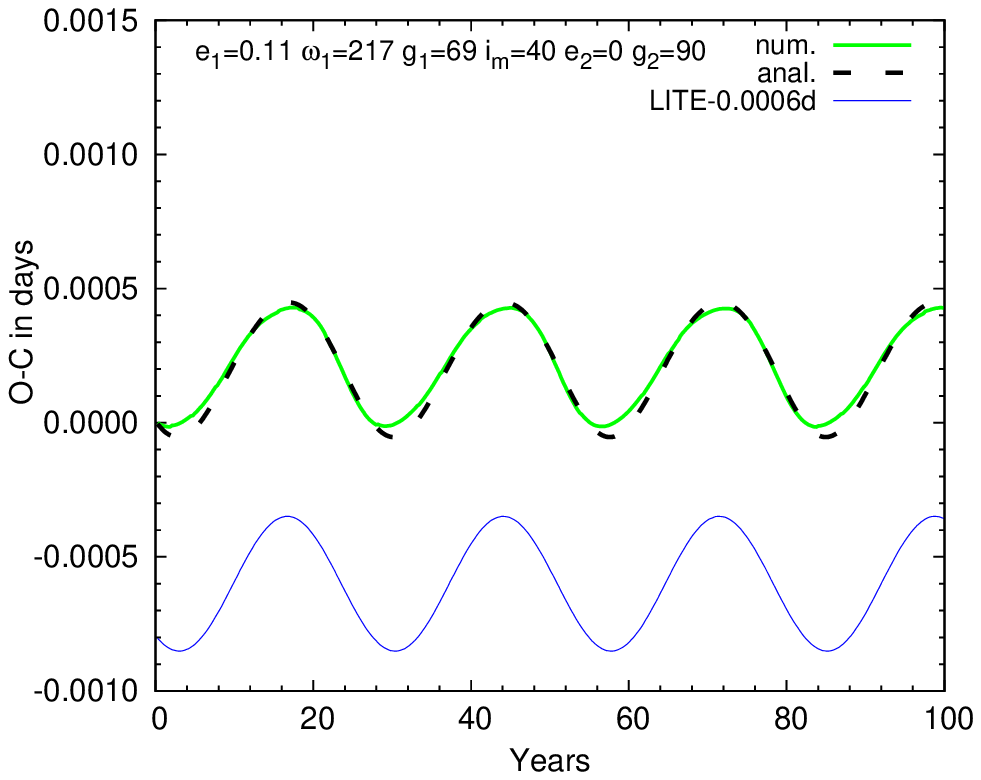}\includegraphics[width=6cm]{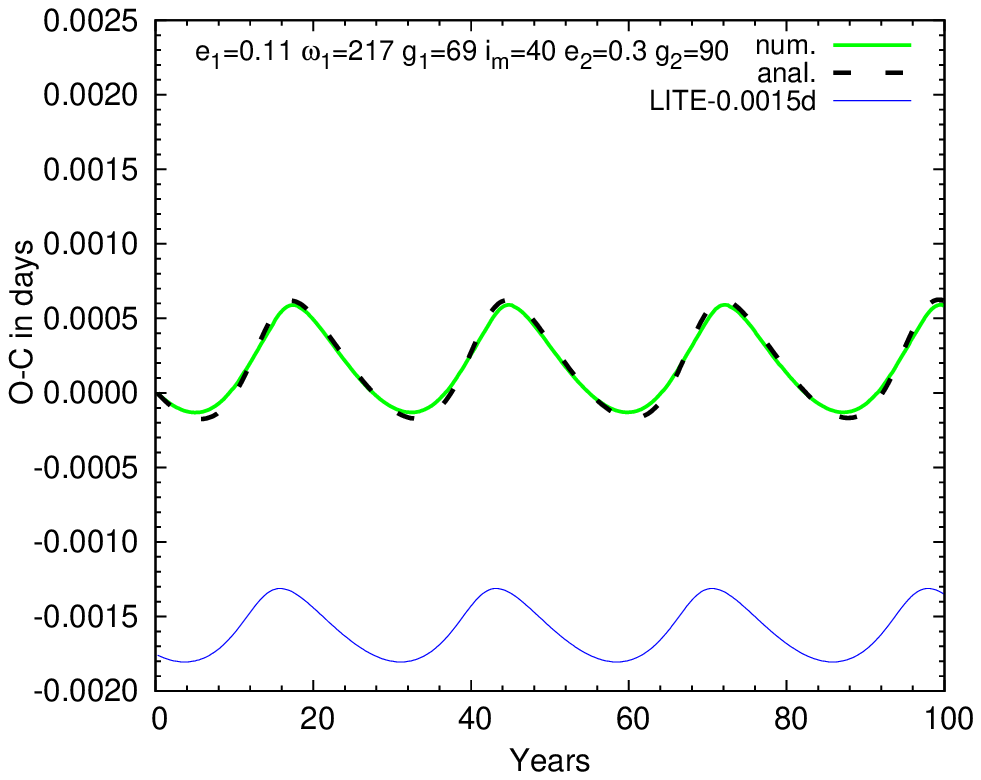}\includegraphics[width=6cm]{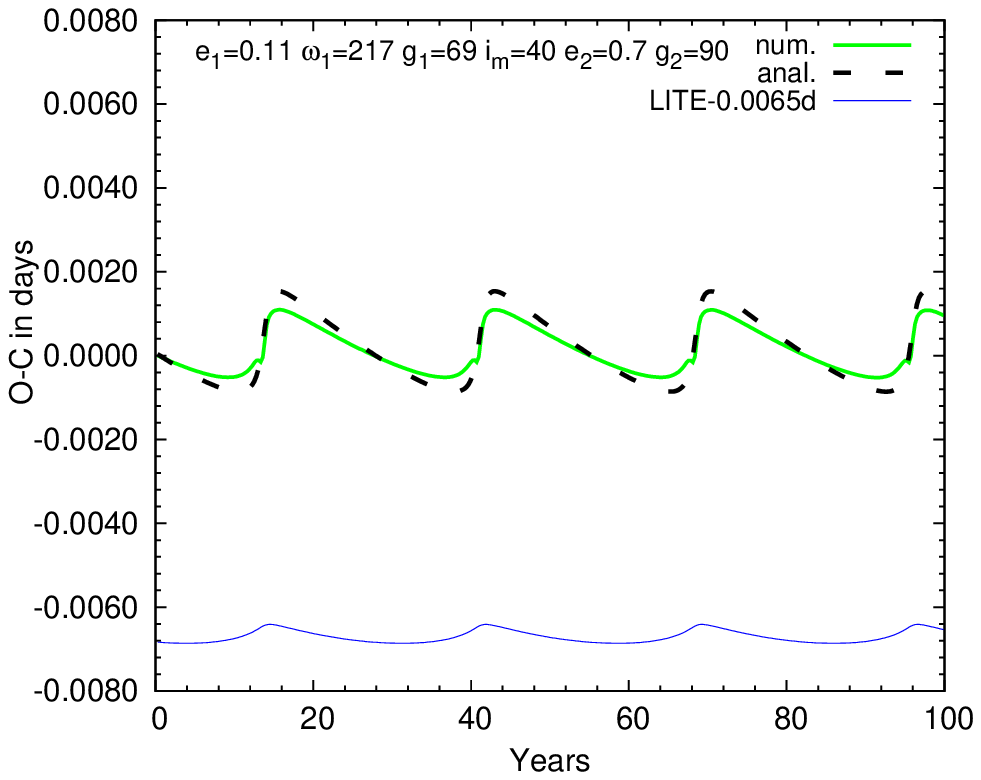}
\includegraphics[width=6cm]{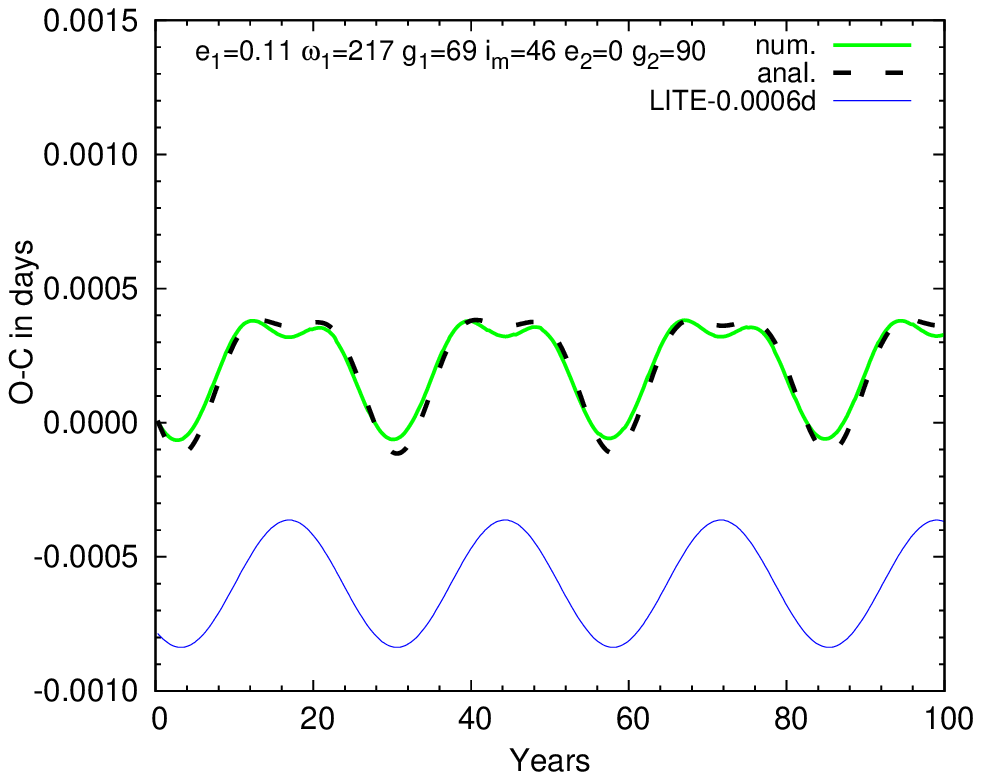}\includegraphics[width=6cm]{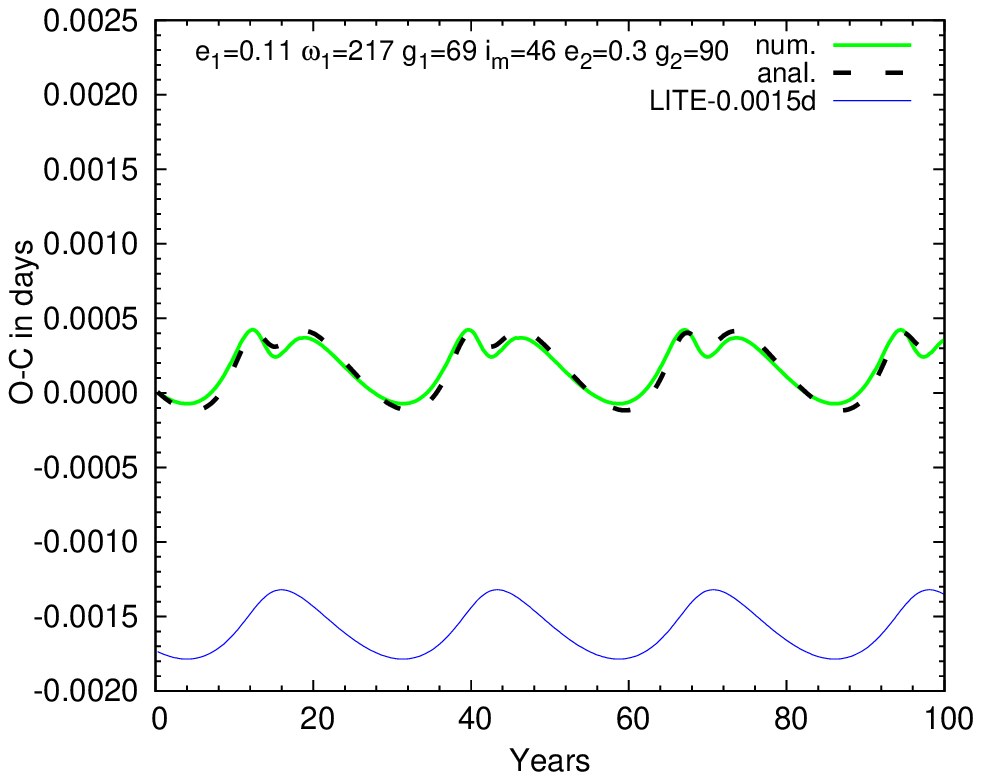}\includegraphics[width=6cm]{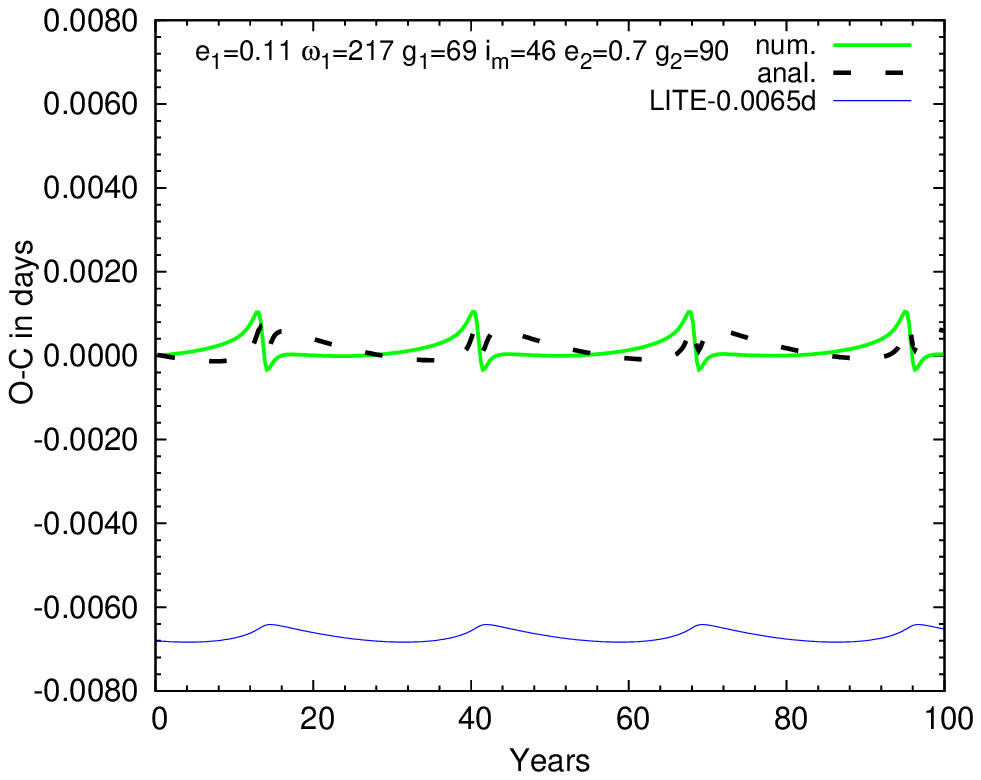}
\includegraphics[width=6cm]{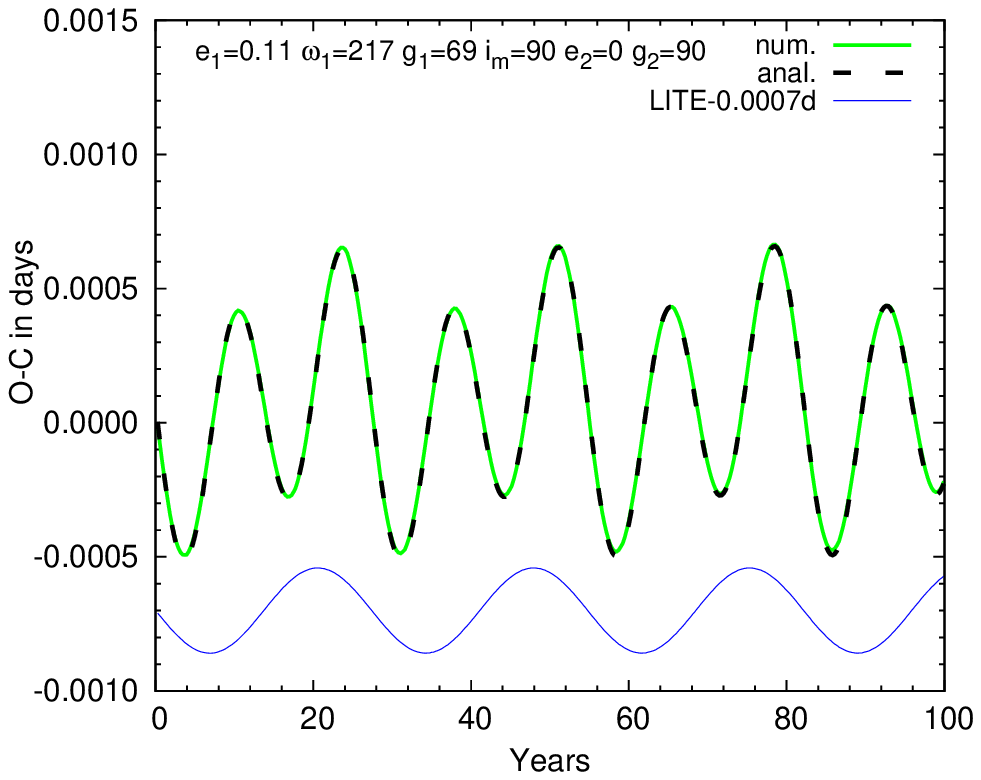}\includegraphics[width=6cm]{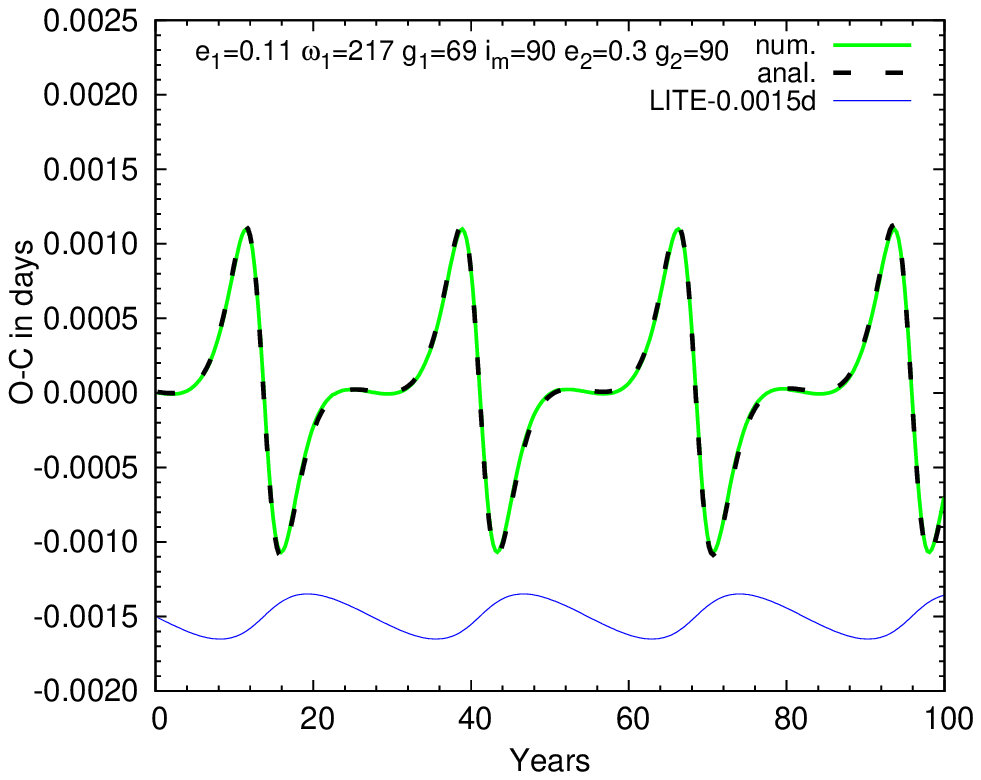}\includegraphics[width=6cm]{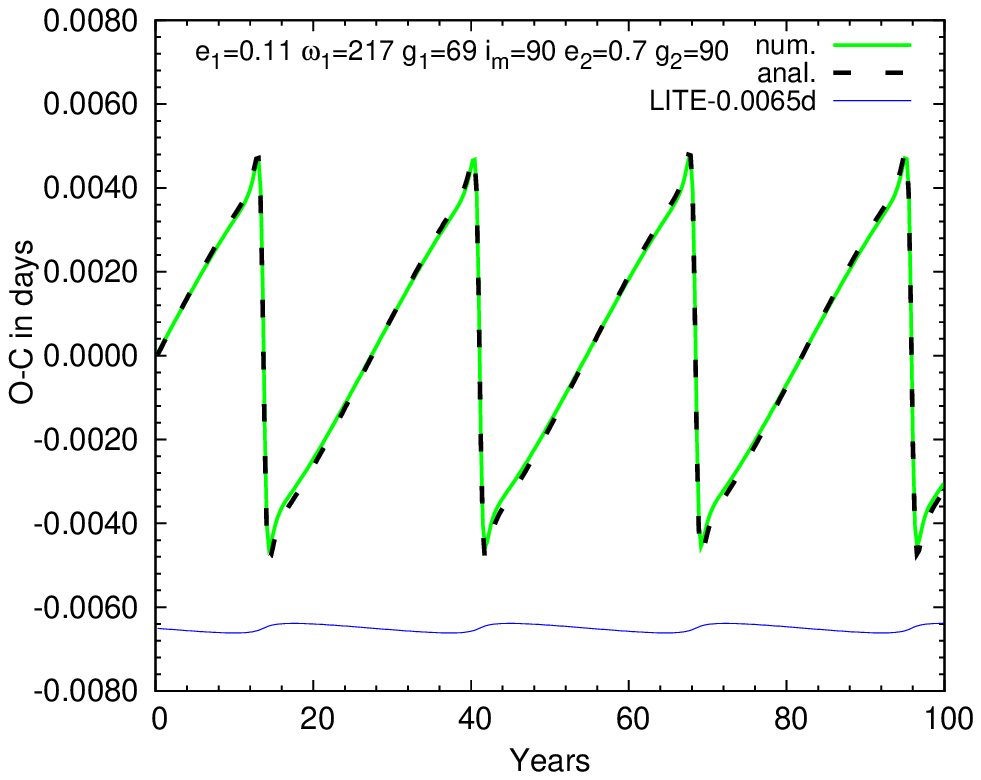}
\caption{Sample of transit timing variations caused by a hypothetical $P_2=10\,000$ day-period
$m_3=0.005 M_\mathrm{\sun}(\approx 5 M_\mathrm{J})$ mass third companion for \object{CoRoT-9b} at
four different initial mutual inclinations ($i_\mathrm{m}=0$\degr, 40\degr, 46\degr, 90\degr, from up to down),
and three different initial outer eccentricities ($e_2=0$, 0.3, 0.7, from left to right).
The dynamical (relative) arguments of periastrons are set to $g_1=69$\degr, $g_2=90$\degr.
The curves show the sum of the geometrical LITE, and the dynamical terms obtained both from
numeric integrations, and analytic calculations up to sixth order in $e_1$. The pure
LITE contributions are also plotted separately. Note that the vertical scale
of the individual columns (i.e. different $e_2$-s) are different.}
\label{fig:C9bg169g290OminC}
\end{figure*}

Nevertheless, while the net amplitudes usually vary in a narrow range, the dominances
of the two main terms (with period $P_2$ and $\frac{1}{2}P_2$) can alternate and, although it is not shown, their relative phase can also vary so, the shape of the actual curves may show great diversity, as is illustrated
e.g. in Figs.~\ref{fig:C9bsamples}, \ref{fig:C9bg169g290OminC}. Furthermore, for specific values of the parameters,
one or the other amplitudes might disappear. At \object{CoRoT-9b} particularly interesting 
is the $g_1=69$\degr $i_\mathrm{m}\sim45$\degr configuration since in this case both $A_{\cal{M}}$
and $A_{\cal{S}}$ disappear very close to each other. This means that for specific $e_2$, 
$g_2$ values the dynamical $O-C$ almost disappears. This possibility warns us of the
fact, that from the absence of $TTV$ in a given system one cannot automatically exclude
the presence of a further planet (which should be observed according to its parameters).

This can be seen clearly in Fig.~\ref{fig:C9bg169g290OminC}, where we plotted the corresponding
$O-C$ curves for coplanar, perpendicular, and the above mentioned interesting $i_\mathrm{m}=40$\degr and 46\degr
configurations. In this latter figure we plotted $O-C$ curves obtained from both numerical integrations
of the three-body motions, and analytical calculations with our sixth order formula.
(Note, according to Fig.~\ref{fig:amplitudes-e1}, for the small inner eccentricity of
\object{CoRoT-9b} ($e_1=0.11$), the first order approximation would have given practically the same results.) 
Comparing the analytical and the numerical curves, the best similarity can be seen in
the perpendicular case (last row). In the coplanar case some minor discrepancies
can be observed both in the shape and amplitude, while the discrepancies are
more expressed in the particular $i_\mathrm{m}=46$\degr case, where for large
outer eccentricity (right panel in the third row) our solution fails. (Nevertheless,
the total amplitude of the {\bf analytical curve is similar to the numerical one} in this case, too.) Although
a thorough analysis of the sources of the discrepancies is beyond the scope
of this paper, we suppose that in these situations the discrepancies come from
the higher order perturbative terms. As was shown e.g. by \citet{soderhjelm84,fordetal00}
the higher order contributions are the most significant for $e_1\sim 0$, $i_\mathrm{m}=(0,1)\times180$\degr.
Although the above authors considered only the secular, or apse-node term perturbations,
the same might be the case for the long-period ones. This might explain
the better correspondance in the perpendicular configuration than in the coplanar one.
We suppose some similar reason in the $i_\mathrm{m}=46$\degr case. As in this situation
the first order contributions almost disappear, whereas the small higher order
terms can also be more significant. Furthermore, numeric integrations in this latter case
show a disappearence of the identity between the $g_1$, $g_1+180$\degr initial conditions,
which also suggests a significance of the higher order terms as in this case
we can expect the appearance of trigonometric functions with $g_1$ and $3g_1$ in
their arguments.

\begin{figure*}
\centering
\includegraphics[width=8cm]{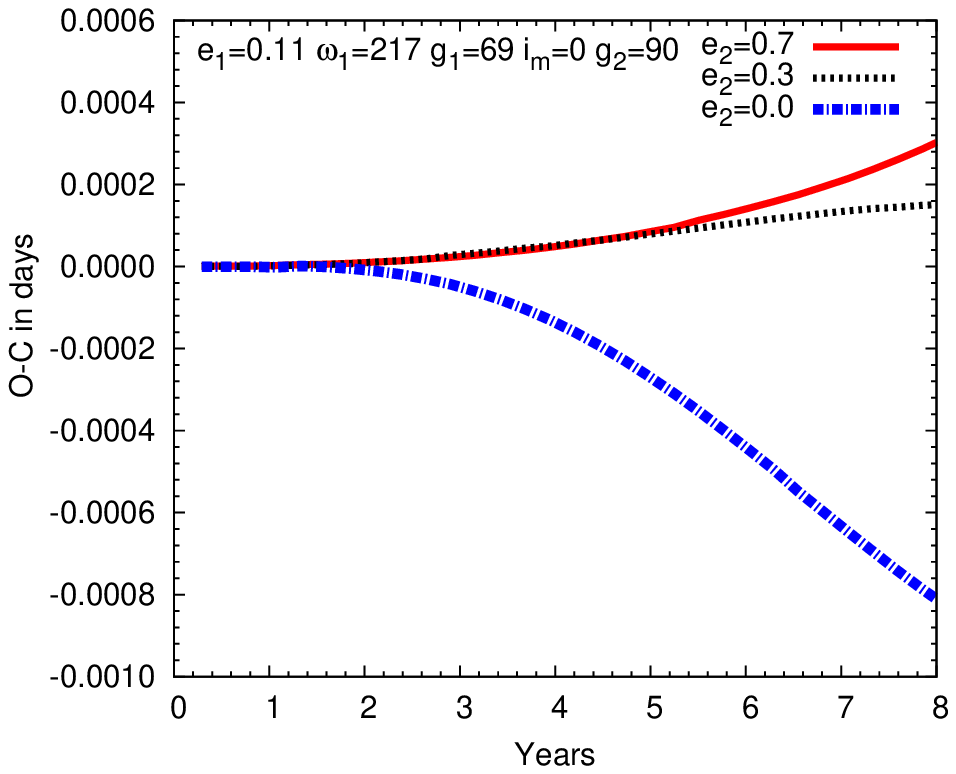}\includegraphics[width=8cm]{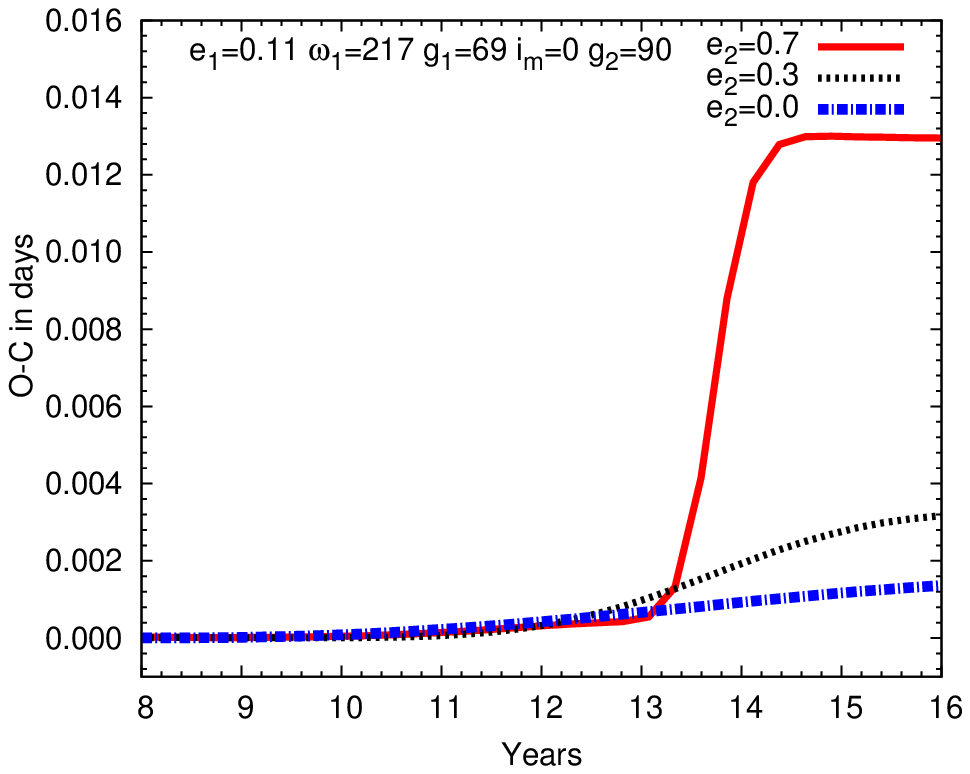}
\includegraphics[width=8cm]{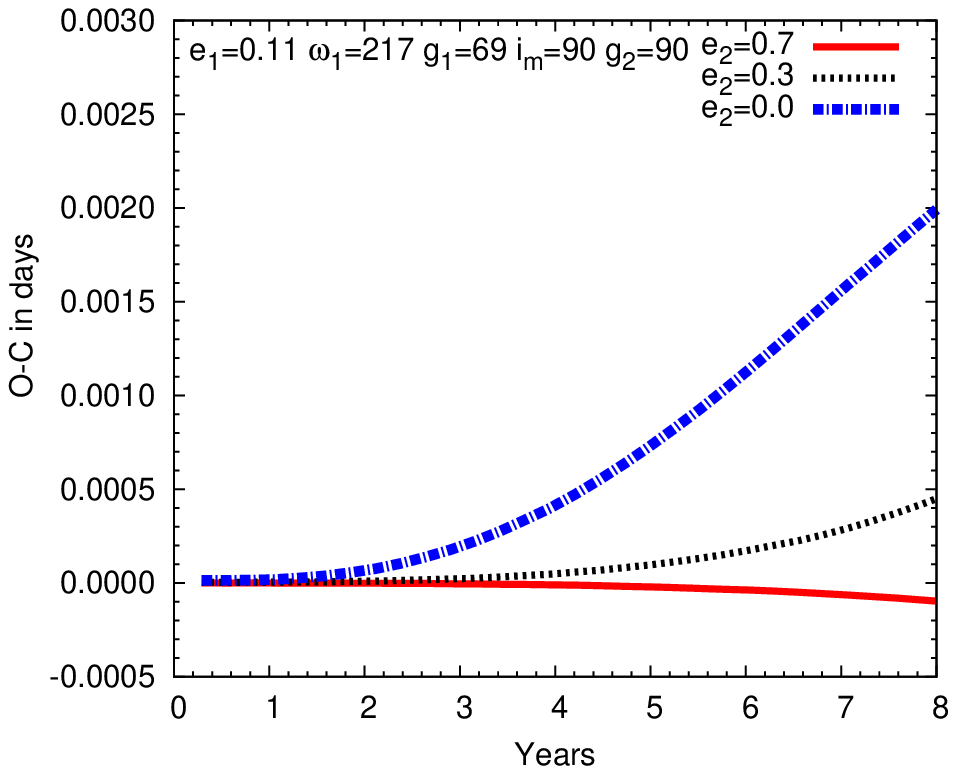}\includegraphics[width=8cm]{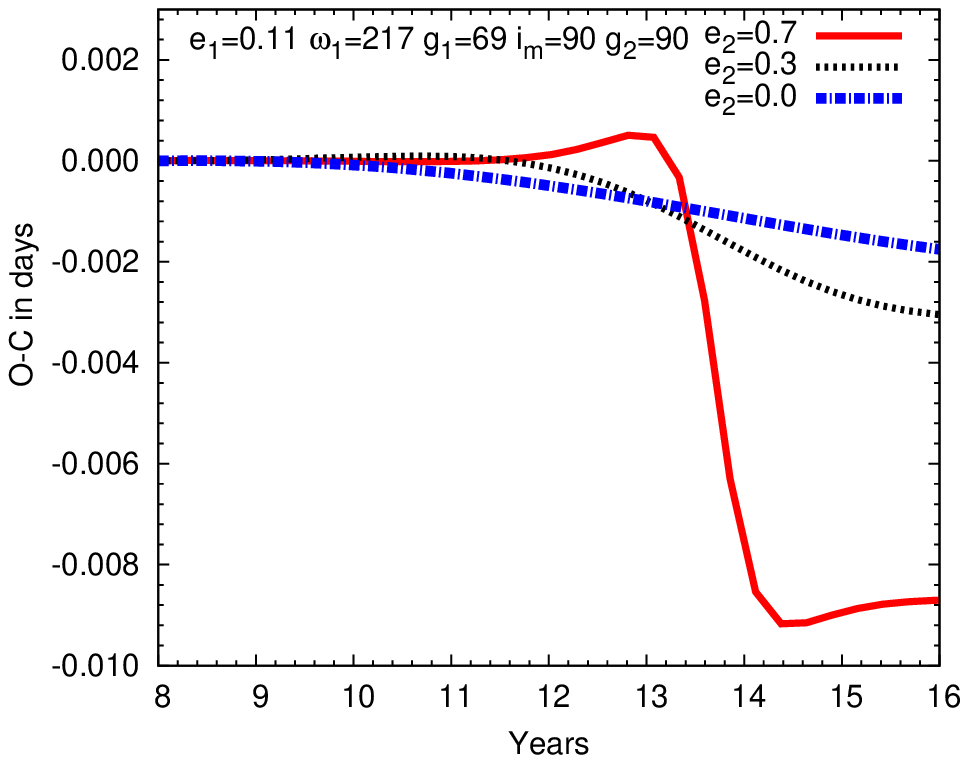}
\caption{The first and the second 8 years of $O-C$-s plotted in the first and the last rows of Fig.~\ref{fig:C9bg169g290OminC}. 
The periods of the individual curves were set equal to the respective initial transiting periods.}
\label{fig:C9b816year}
\end{figure*}

As
expected, the highly eccentric-distant-companion scenario produces the largest amplitude
TTV, at least when a complete revolution is considered. Nevertheless, on a shorter
time-scale, the length of the observing window necessary for the detection
depends highly on the phase of the curve. To illustrate this, in Fig.~\ref{fig:C9b816year}
we plotted the first, and second 8 years of the three primary transit $O-C$ curves
for both the coplanar, and the perpendicular cases, shown in the first and last
rows of Fig.~\ref{fig:C9bg169g290OminC}.
The transiting periods for each curve were calculated in the usual, observational manner,
i.e. the time interval between the first (some) transits were used. According to Fig.~\ref{fig:C9b816year},
in the present situation, in the first 8 years (i.e. which begins at the apastron of the
outer planet) it would be unlikely to detect the
perturbations in the largest (total) amplitude $e_2=0.7$ cases, as well as in the coplanar
$e_2=0.3$ case. The most certain detection would be possible in the two smallest
amplitude circular $e_2=0$ configurations. Nevertheless, if the observations starts
at those phases plotted on the right panels (8-16 years), the pictures completely
differ. In this latter interval, the circular cases produce the smallest curvature $O-C$-s,
and the discrepancy from the linear trend reaches $0\fd001$ days (which can be considered
as a limit for certain detection) occuring towards the end of the interval. On the other hand, in the case
of the highly eccentric configurations, the 'moment of truth' comes after some years.
Nevertheless, we have to stress, that although we plotted the $O-C$ curves with
continuous lines, in reality they would contain only 3-4 points during the phase of
the seemingly abrupt jump, is a further complicating factor with regards to 'certain' detection.

\begin{figure*}
\centering
\includegraphics[width=7cm]{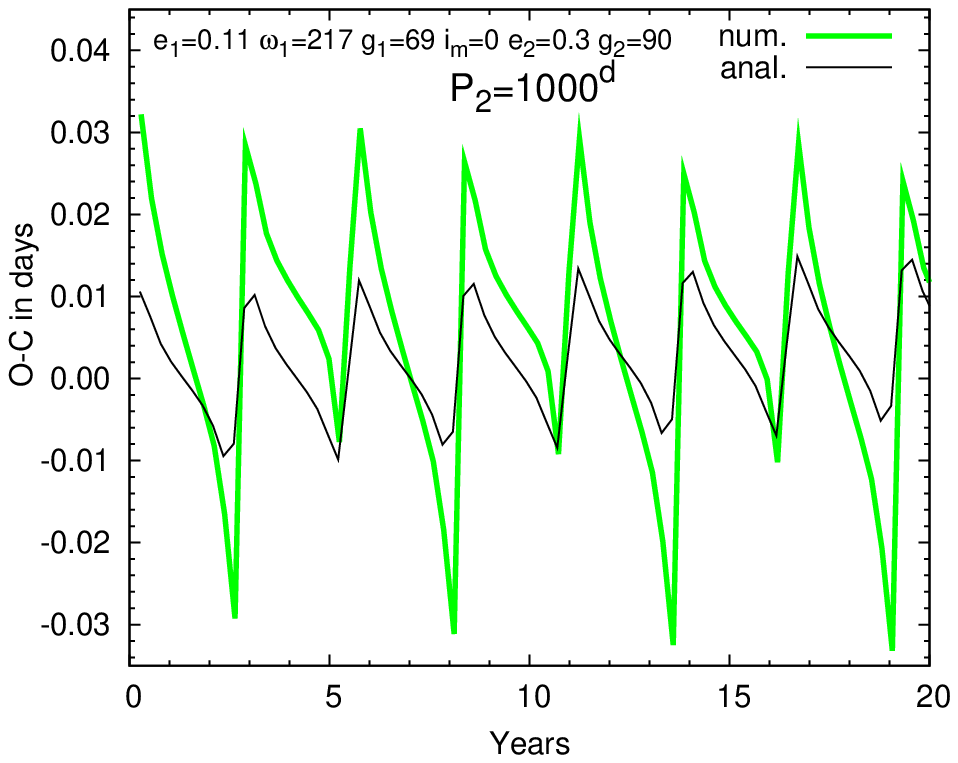}\includegraphics[width=7cm]{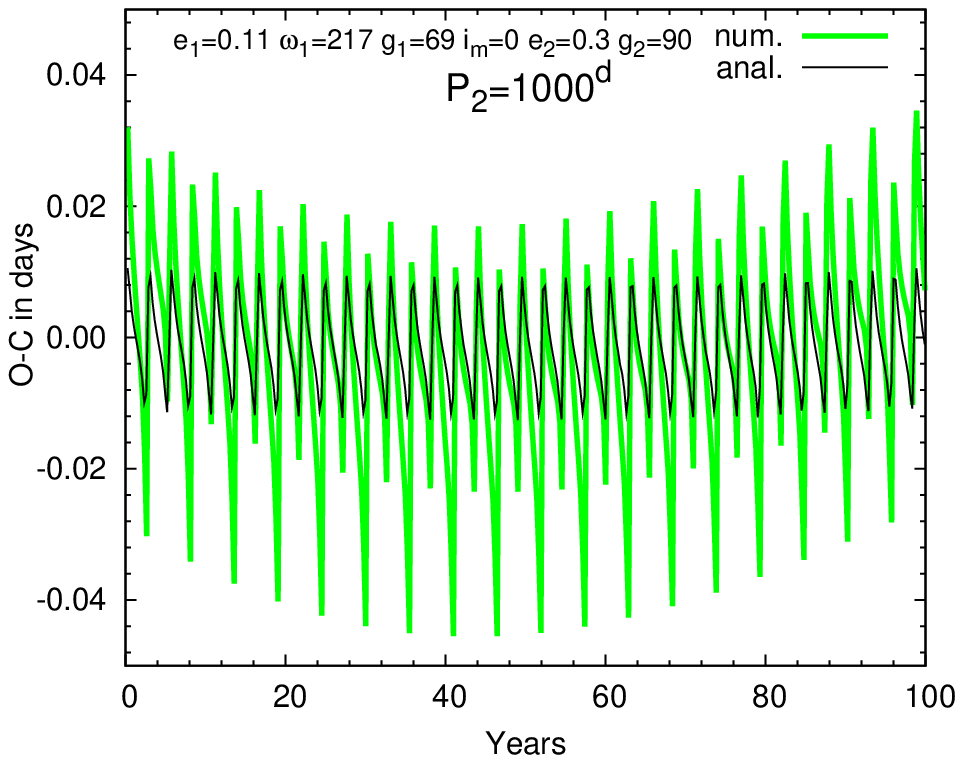}
\includegraphics[width=7cm]{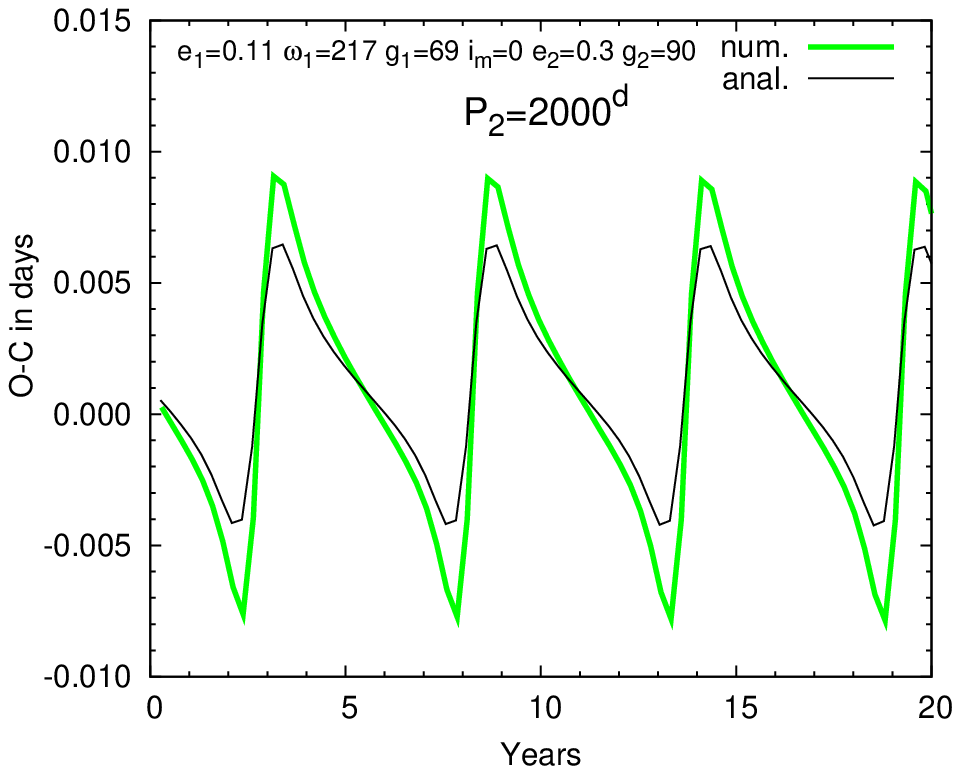}\includegraphics[width=7cm]{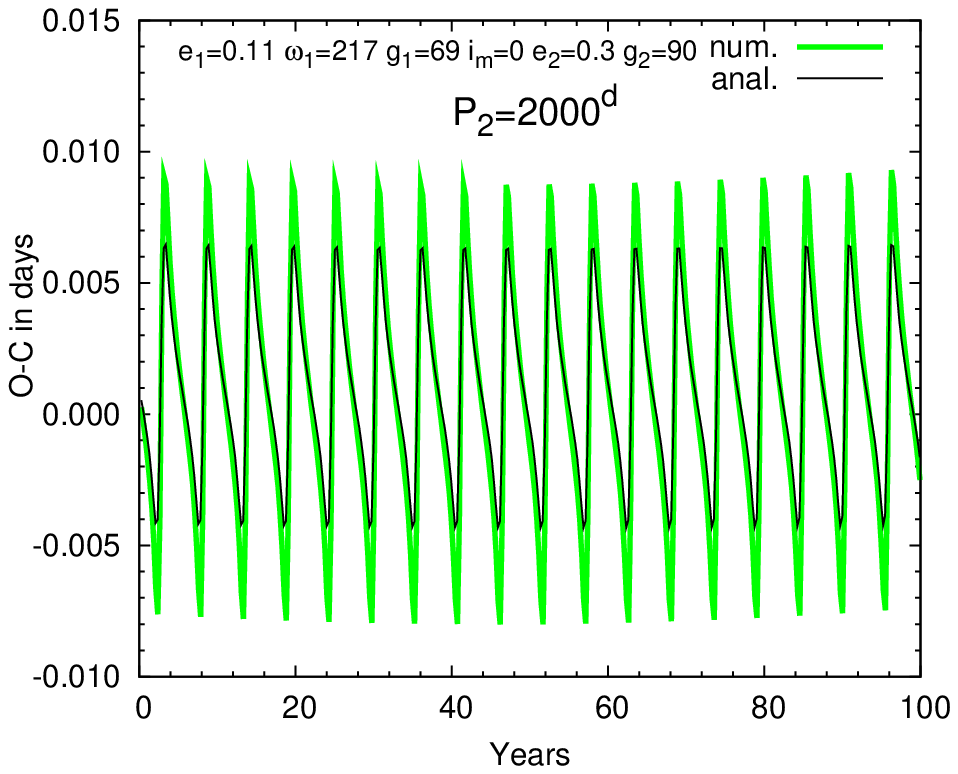}
\includegraphics[width=7cm]{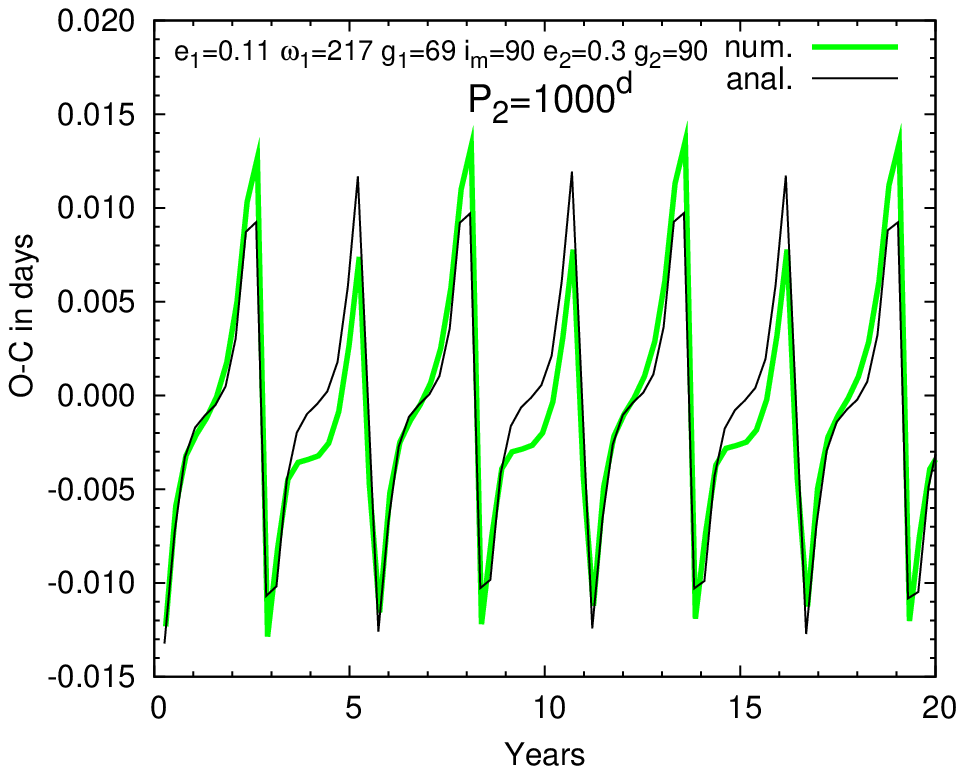}\includegraphics[width=7cm]{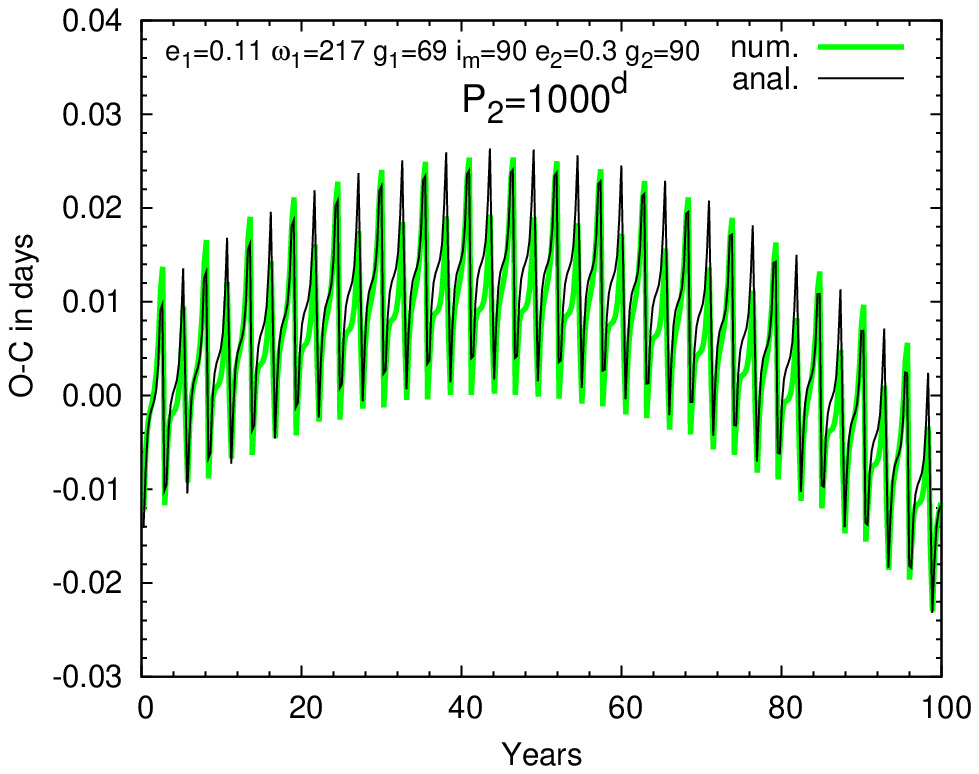}
\includegraphics[width=7cm]{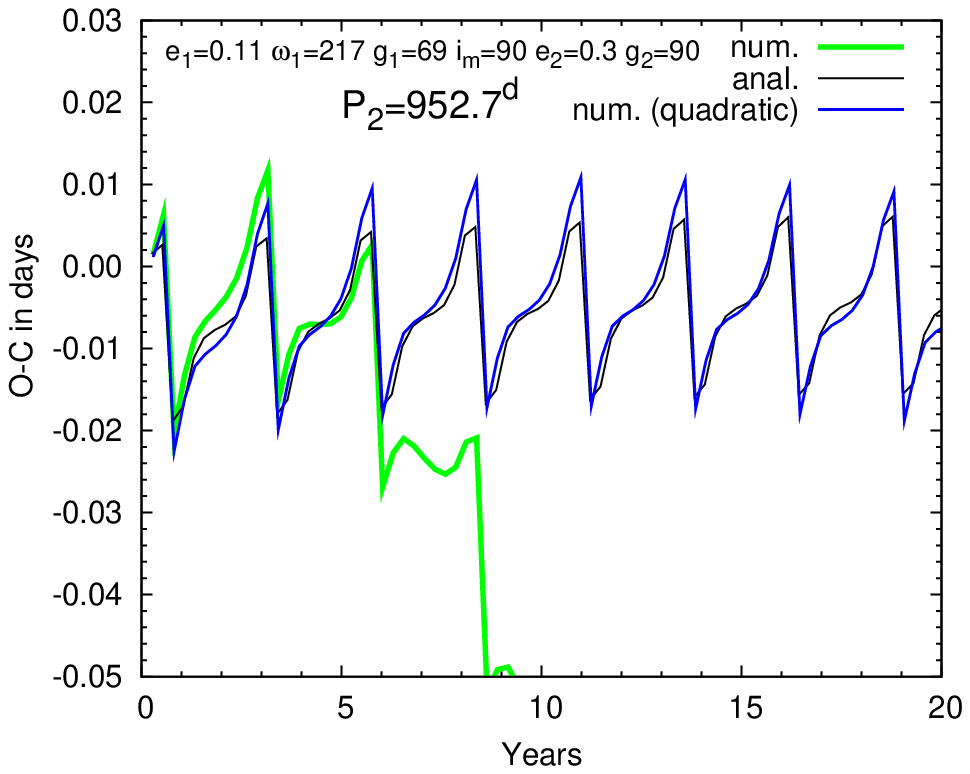}\includegraphics[width=7cm]{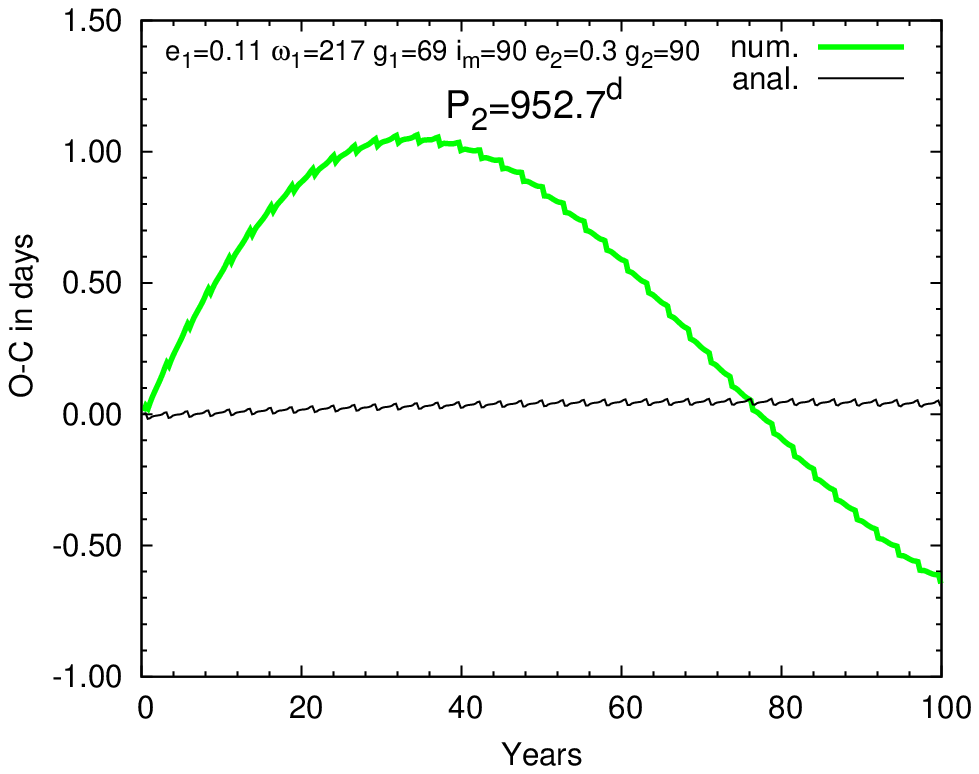}
\caption{\bf Checking the validity of hierarchical approximation for closer systems. In the first two rows
the initial conditions were set to be the same as at the uppermost middle panel of Fig.~\ref{fig:C9bg169g290OminC}, with
the exception of $P_2=1000$ {\it (first row)} and $2000$ days {\it (second row)}, i.e. $a_2/a_1\approx4.8$ and $\approx7.6 $, respectively,
while the third and last rows have initial conditions similar to the middle panel of Fig.~\ref{fig:C9bg169g290OminC}, with
the exception of $P_2=1000$ {\it (third row)} and $952.7$ days {\it (last row)}. 
This latter illustrates the case of a $1:10$ mean-motion resonance. The left panels
represent a 20-year-long time-scale, while the right ones show the TTV behaviour during a century.
In the left panel of the last row the blue (quadratic) curve shows the $O-C$ curve
calculated by including a quadratic term. See text for details.}
\label{fig:C9bszoros}
\end{figure*}

{\bf In the sample runs above a moderately hierarchic scenario $a_2\approx22a_1$ was studied.
In order to get some picture about the lower limit of the validity of our low-order, hierarchical approximation,
we carried out further integrations for less hierarchic configurations. In Fig.~\ref{fig:C9bszoros}
we show the results of some of these runs, which were carried out with the same initial conditions what
were used in the middle panel of the first and last rows (coplanar and perpendicular cases, respectively) 
of Fig.~\ref{fig:C9bg169g290OminC}, but for $P_2=1000$ days (i.e. $a_2\approx4.8a_1$), $P_2=2000$ days ($a_2\approx7.6a_1$),
and $P_2\approx952.7$ days ($a_2\approx4.6a_1$), in which latter case the two planets orbit in $1:10$ mean motion resonance.
In the left panels of Fig.~\ref{fig:C9bszoros} we plot 20-year-long intervals, while in the right ones century-long
time-scales are shown. As one can see, on this latter time-scale, apse-node 
effects already reach or exceed the magnitude of the long period
ones. This naturally arises from the fact that the typical time-scales of these latter high-amplitude
perturbations are proportional to $P_2^2/P_1$, i.e. in the present cases these are $\sim100\times$
faster than in the previously investigated case. Note, that for the sake of a better comparison, 
the analytical curves in the right panels were calculated with the inclusion of these apse-node terms, 
although the latter will be presented only in a forthcoming paper. 
Turning back to the 20-year-long integrations, one can see that
the limit of the validity of the present approximation does strongly depend on the mutual
inclination. While for the $i_\mathrm{m}=90\degr$ configurations the long period analytical results
are in remarkably good agreement with the numerical curves even for $a_2<5a_1$ (left panels of
third and forth rows), for the coplanar ($i_\mathrm{m}=0\degr$) case our approximation
is clearly insufficient for such small $a_2/a_1$ ratios, and even for the doubled outer period case
(i.e. $a_2\approx8a_1$) the amplitude of the analytical curve is highly underestimated.

We also investigated the case of the $1:10$ mean-motion resonance. Our results for the
perpendicular case are plotted in the last row. In this case the numerical integration 
shows very high amplitude apse-node scale variations that does not occur in the analytical curve.
In order to get a better comparison between the analytical and numerical long-term variations in
this case, we removed the apse-node effect from the numerical curve, by the use of a
quadratic term, i.e. the (blue) $O-C$ curve was calculated in the form of 
\begin{equation}
O-C=c_0+c_1E+c_2E^2,
\end{equation}
where $E$ is the cycle number. As one can see, this quadratic (blue) curve shows
similar agreement with the analytical curve, which was found in the similar
$a_2/a_1$ ratio non-resonant case. Consequently, we can state that our long term
formulae are capable to produce the same accuracy even around mean-motion resonances.

Nevertheless, from these few arbitrary trial runs we cannot give general statements
about the limits of our approximations. A detailed discussionof this point is postponed to
a forthcoming paper, when we include the apse-node time scale terms.}

While \object{CoRoT-9b} served as an illustration for the Transit Timing Variations
in the low inner eccentricity case, our next sample exoplanet \object{HD 80606b}
represents the extremely eccentric case.

\begin{table}
\caption[]{The initial parameters of the transiting planetary subsytems. 
(The masses are given in solar mass, periods in days, and angular elements in degrees.)
The parameters are taken from \citet{deegetal10} for \object{CoRoT-9b}, and from \citet{pontetal09} for \object{HD~80606b}.
(Note that the argument of periastron ($\omega_1$) for the relative orbit
of the planet around its host star differ by 180\degr from the value deduced from
radial velocity data.)}
\label{tab:binaryfixparams}
$$
\begin{array}{lllllll}
\hline
\hline
\noalign{\smallskip}
\mathrm{System} & m_1 & m_2 & P_1 & e_1 & \omega_1 & i_1\\
\hline
\noalign{\smallskip}
\mathrm{\object{CoRoT-9b}}& 0.99 & 0.0008 & 95.2738 & 0.11 & 217 & 89.99 \\
\mathrm{\object{HD~80606b}}&0.97 & 0.0038 &111.4357 & 0.93 & 121 & 89.32 \\
\hline
\noalign{\smallskip}
\end{array}
$$
\end{table}

\subsection{\object{HD 80606b}}

The high-mass gas giant exoplanet \object{HD 80606b} features an almost 4-month long period and an
extremely eccentric orbit around its solar-type host-star. It was discovered spectroscopically by \citet{naefetal01}.
Recently both secondary occultation (with the Spitzer space telescope, \citealp{laughlinetal09}), and
primary transit (\citealp{moutouetal09,fosseyetal09}) have been detected. A thorough analysis of the
collected data around the February 2009 primary transit led to the conclusion that there
is a significant spin-orbit misalignment in the system, i.e. the orbital plane of
\object{HD 80606b} fails to coincide with the equatorial plane of its host star \citep{pontetal09}.
These facts suggest that this planet might be seen at an instant close to the maximum
eccentricity phase of a Kozai cycle induced by a distant, inclined third companion
\citep[cf.][]{wumurray03,fabryckytremaine07}. Note that although \object{HD 80606} itself forms
a binary with \object{HD 80607}, due to the large separation, one may expect a further,
not so distant companion for an effective Kozai mechanism. Such a very distant companion may nevertheless play an indirect role in the initial triggering of the Kozai mechanism, by
the effect described in \citet{takedaetal09}. The most important parameters of the
system are summarized in Table~\ref{tab:binaryfixparams}.

Due to its very high eccentricity, the periastron distance of this planet is $q_1=a_1(1-e_1)\approx0.03\mathrm{AU}$
at which distance the tidal forces and especially tidal dissipation should be effective. 
(As a comparison we note, that this separation corresponds to the semi-major axis of an approx. 2 day period orbit.) 
Furthermore, we can expect also a significant relativistic contribution to the apsidal motion.
Although such circumstances do not invalidate the following conclusions (as the time-scale
of these effects are significantly longer than the ones investigated) we nevertheless provide a quantitive
estimation of their effects. Furthermore, the effect of dissipation will also be considered
numerically, at the end of this section. 

As is well known, the apsidal advance speed, averaged for one
orbital revolution, can be written in the following form:
\begin{equation}
\dot g_1=A+B\cos2g_1,
\label{eq:gpont}
\end{equation}
where the non-zero contributions of the third-body, tidal and relativistic terms are as follows:
\begin{eqnarray}
A_{3^\mathrm{rd}}&=&A_\mathrm{L}\frac{P_1}{P_2}\left(1-e_1^2\right)^{-1/2}\left[I^2-\frac{1}{5}\left(1-e_1^2\right)+\frac{2}{5}\left(1+\frac{3}{2}e_1^2\right)\frac{C_1}{C_2}I\right], \nonumber \\
\\
B_{3^\mathrm{rd}}&=&A_\mathrm{L}\frac{P_1}{P_2}\left(1-e_1^2\right)^{-1/2}\left[\left(1-e_1^2\right)-I^2-e_1^2\frac{C_1}{C_2}I\right], \\
A_\mathrm{tidal}&=&\frac{5{\cal{T}}}{2a_1^5}\frac{1+\frac{3}{2}e_1^2+\frac{1}{8}e_1^4}{\left(1-e_1^2\right)^5}+\frac{\cal{R}}{a_1^2\left(1-e_1^2\right)^2}, \\
A_\mathrm{rel}&=&6\pi\frac{Gm_1}{c^2a_1(1-e_1^2)},
\end{eqnarray}
where
\begin{eqnarray}
C_1&=&\frac{m_1m_2}{m_{12}}\sqrt{Gm_{12}a_1\left(1-e_1^2\right)}, \nonumber \\
&=&L_1\sqrt{1-e_1^2}, \\
C_2&=&\frac{m_{12}m_3}{m_{123}}\sqrt{Gm_{123}a_2\left(1-e_2^2\right)}, \nonumber \\
&=&L_2\sqrt{1-e_2^2},
\end{eqnarray}
moreover,
\begin{eqnarray}
{\cal{T}}&=&6\left(\frac{m_2}{m_1}k_2^{(1)}R_1^5+\frac{m_1}{m_2}k_2^{(2)}R_2^5\right), \nonumber\\
&=&6\frac{m_2}{m_1}R_1^5\left[k_2^{(1)}+k_2^{(2)}\left(\frac{\overline{\rho}_1}{\overline{\rho}_2}\right)^2\frac{R_1}{R_2}\right], \\
{\cal{R}}&=&\frac{k_2^{(1)}R_1^5\omega^2_{z'_1}}{Gm_1}+\frac{k_2^{(2)}R_2^5\omega^2_{z'_2}}{Gm_2} \nonumber \\
&=&\frac{3k_2^{(1)}}{4\pi G\overline{\rho}_1}v^2_{\mathrm{e}_1}+\frac{3k_2^{(2)}}{4\pi G\overline{\rho}_2}v^2_{\mathrm{e}_2}.
\end{eqnarray}
In these equations $m_1$, $R_1$, $\overline{\rho}_1$, $k_2^{(1)}$, $\omega_{z'_1}$, $v_{\mathrm{e}_1}$ refer to the mass, 
radius, average density, first apsidal motion constant, uni-axial rotational angular velocity, and
equatorial rotational velocity of the host star respectively, while subscript $_2$
denotes the same quantities for the inner planet. Note also that the rotational term
is valid only for non-aligned rotation, and consequently, it provides only a crude
estimation for \object{HD 80606b}. Nevertheless, for the present purpose it seems satisfactory.
First consider the third-body term. Substituting $\cos 2g=-1$, $I^2=3/5$,
\begin{eqnarray}
\dot g_{3^\mathrm{rd}}&=&\frac{3\pi}{2}\frac{m_3}{m_{123}}\left(\frac{P_1}{P_2}\right)^2\left(1-e_2^2\right)^{-3/2}\left(1-e_1^2\right)^{-1/2} \nonumber \\
&&\times\left[3e_1^2\pm\sqrt{\frac{3}{5}}\left(1+4e_1^2\right)\frac{L_1}{L_2}\sqrt{\frac{1-e_1^2}{1-e_2^2}}\right],
\label{eq:g1p_3rd}
\end{eqnarray}
from which the last term usually omittable in hierarchical systems, since $C_1<<C_2$,
which is more expressively true for high $e_1$. For example, in the present situation
\begin{equation}
\frac{C_1}{C_2}\approx0.06\left(1-e_2^2\right)^{-1/2}.
\end{equation}
So, for \object{HD 80606b} we obtain that:
\begin{equation}
\Delta g_{3^\mathrm{rd}}\approx0\fdg43\times\left(1-e_2^2\right)^{-3/2}\mathrm{century}^{-1}.
\end{equation}
(This result is in excellent correspondance with Fig.~\ref{fig:HD80606dynevol}.)

There are several uncertainties in the calculation of the tidal contribution. While the $k_2$
constant is relatively well-known for ordinary stars, it has a great ambiguity for
exoplanets. Furthermore, we do know nothing about the rotational velocity of \object{HD 80606b}.
So, according to the tables of \citet{claretgimenez92} we set $k_2^{(1)}=0.02$ for the host star, 
and assumed $k_2^{(2)}=0.2$ for its planet, which is the same order of magnitude as for \object{Jupiter} and of \object{WASP-12b}
\citep{campoetal10}. The stellar rotation was set to $V_\mathrm{rot}=1.8\mathrm{~km}\,\mathrm{s}^{-1}$ (i.e. $P_\mathrm{rot}=27\fd5$;
$\omega_{z'_1}=0.228\mathrm{~day}^{-1}$) \citep{fischervalenti05}, while for the planet we supposed (arbitrarily) a one-day
rotation period, i.e. $\omega_{z'_1}=6.283\mathrm{~day}^{-1}$. By the use of these values,
the classical tidal contribution to the apsidal motion becomes
\begin{equation}
\Delta g_\mathrm{tidal}\approx0\fdg007\mathrm{~century}^{-1},
\end{equation}
which can be neglected.

Finally, the relativistic contribution is estimated to be:
\begin{equation}
\Delta g_\mathrm{rel}\approx0\fdg06\mathrm{~century}^{-1},
\end{equation}
i.e. it is smaller by one magnitude than the third-body term, and consequently, does also not play any important role.

\begin{figure*}
\centering
\includegraphics[width=8cm]{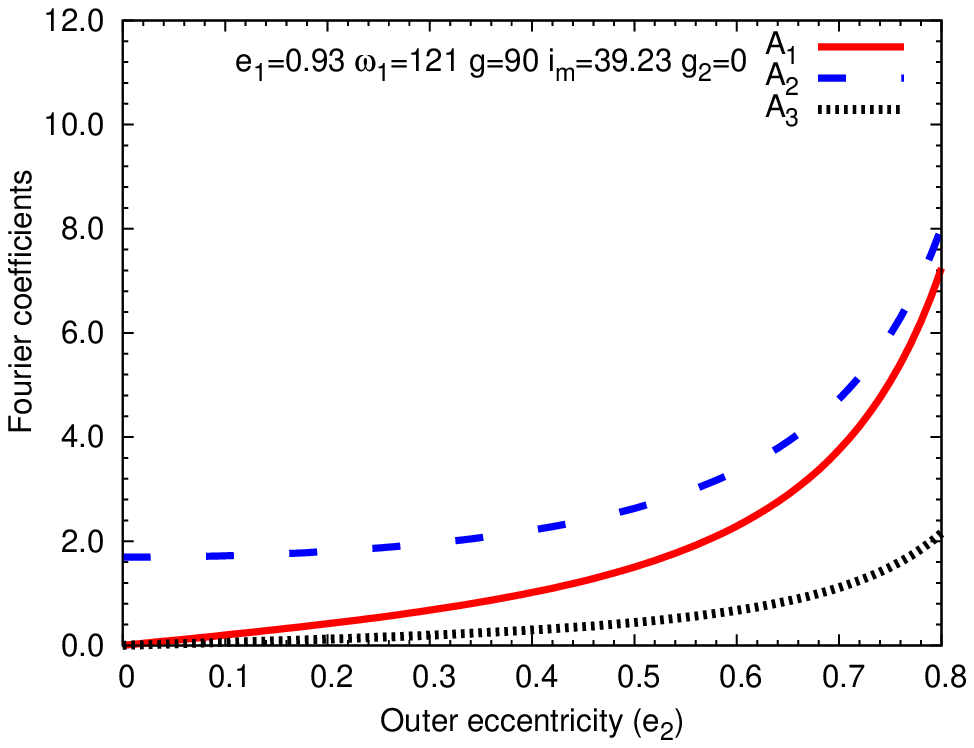}\includegraphics[width=8cm]{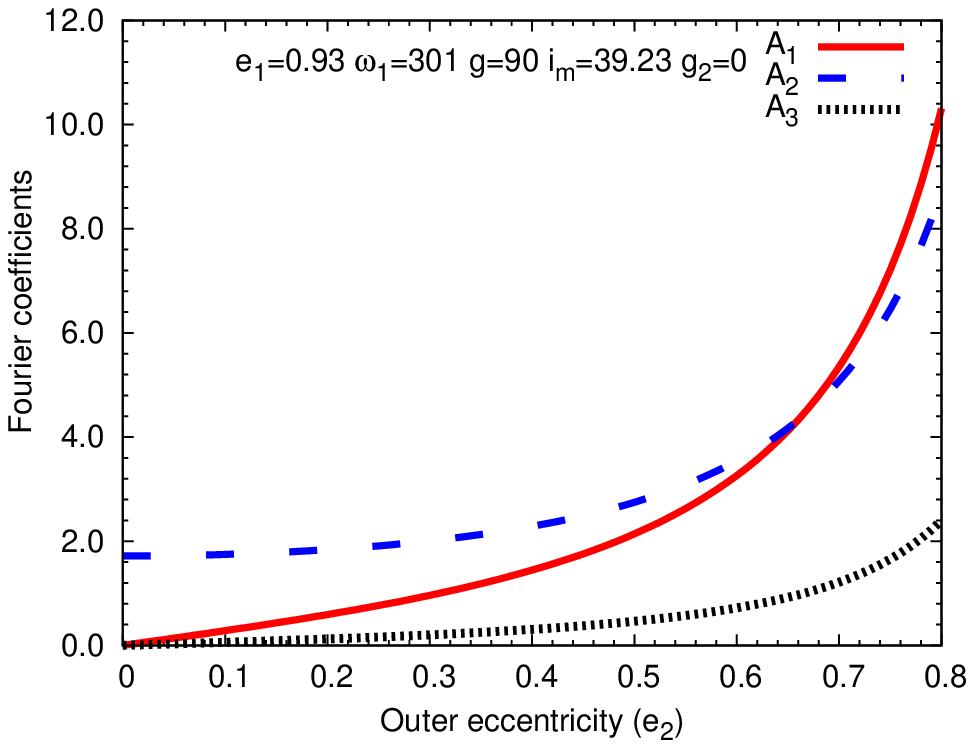}
\caption{The $A_{1,2,3}$ amplitudes for \object{HD 80606b} for primary transits (left), and secondary occultations (right), 
supposing that the planet is seen at the instant of the maximum phase of a Kozai-cycle.}
\label{fig:HD80606bphasespace}
\end{figure*}

\begin{figure*}
\centering
\includegraphics[width=5.6cm]{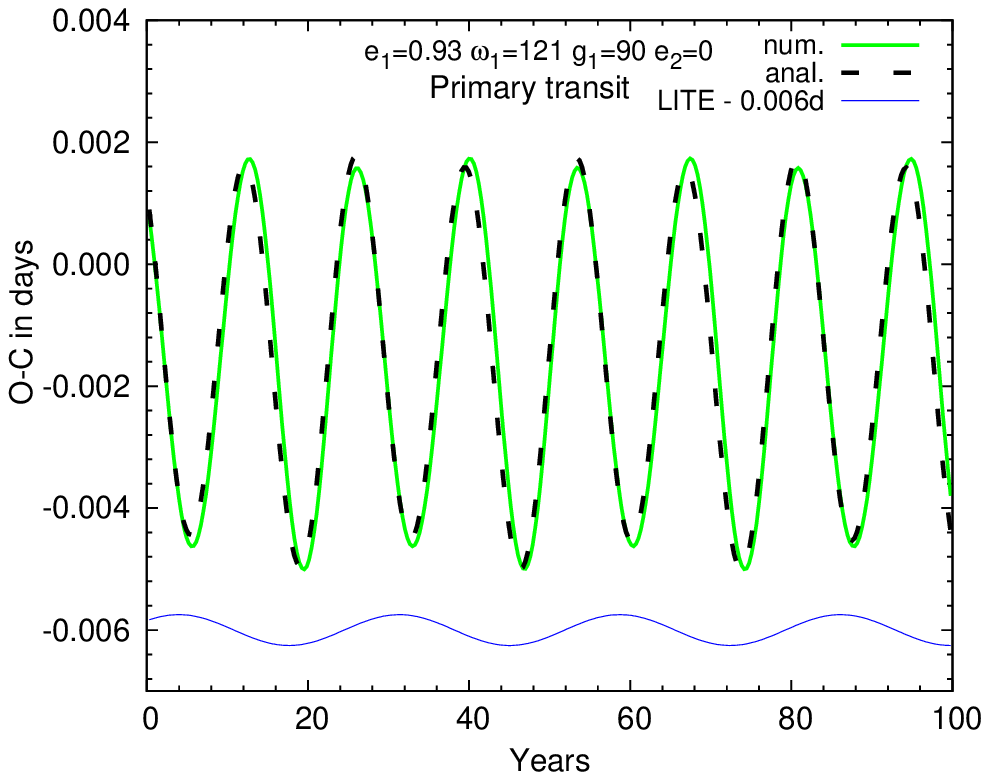}\includegraphics[width=5.6cm]{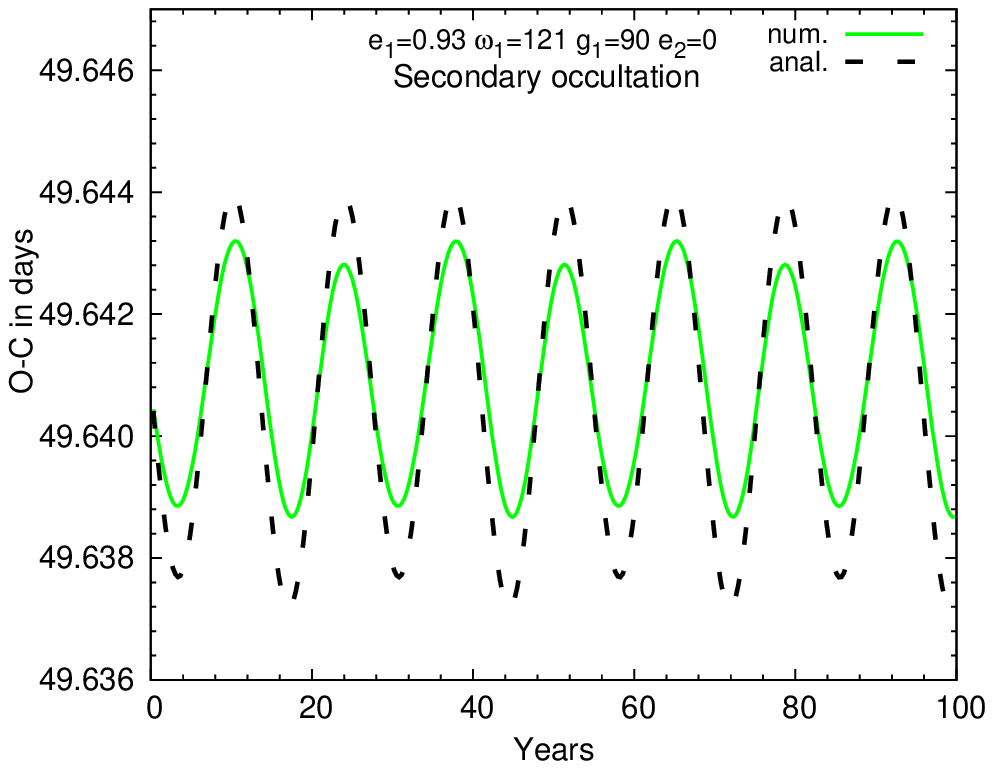}\includegraphics[width=5.6cm]{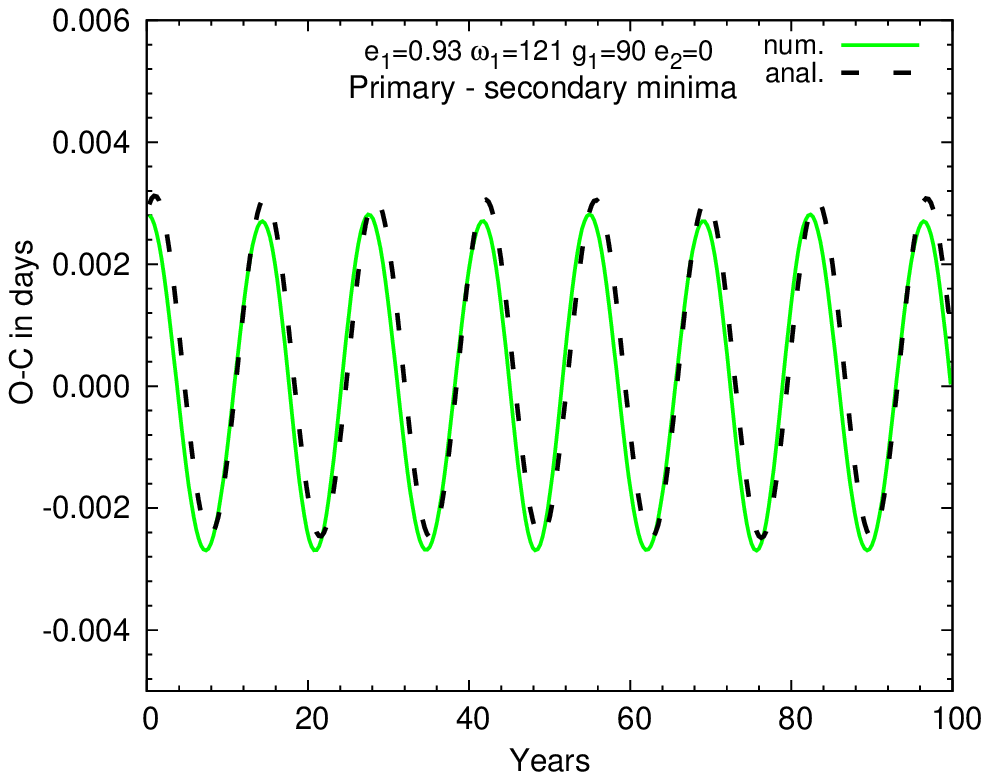}
\includegraphics[width=5.6cm]{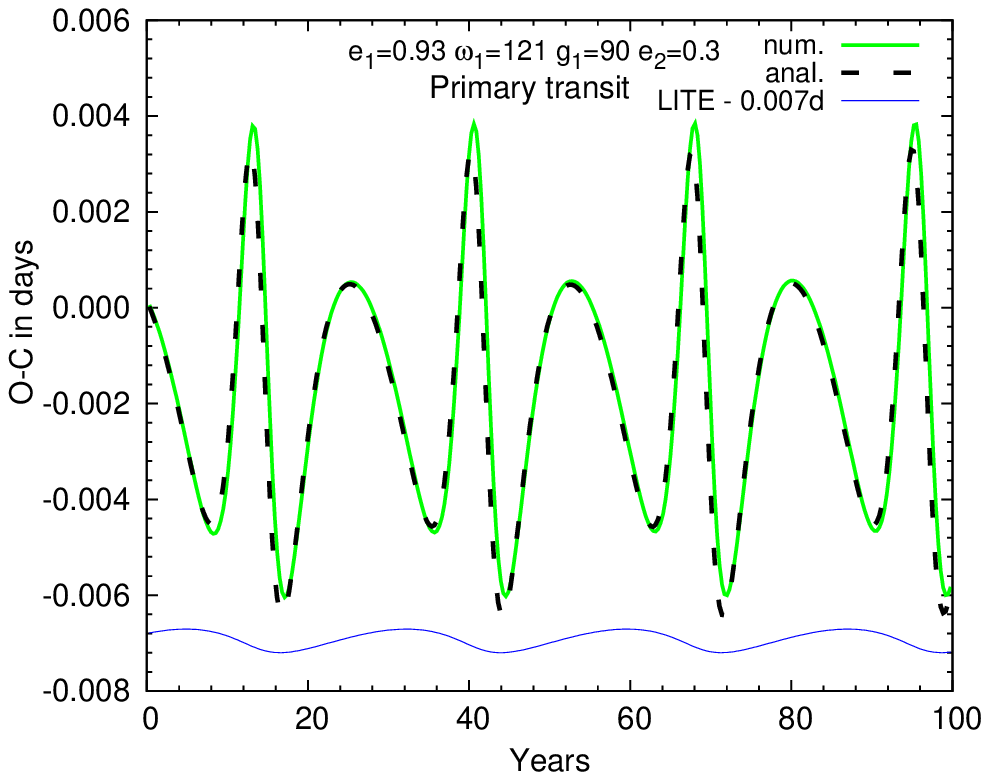}\includegraphics[width=5.6cm]{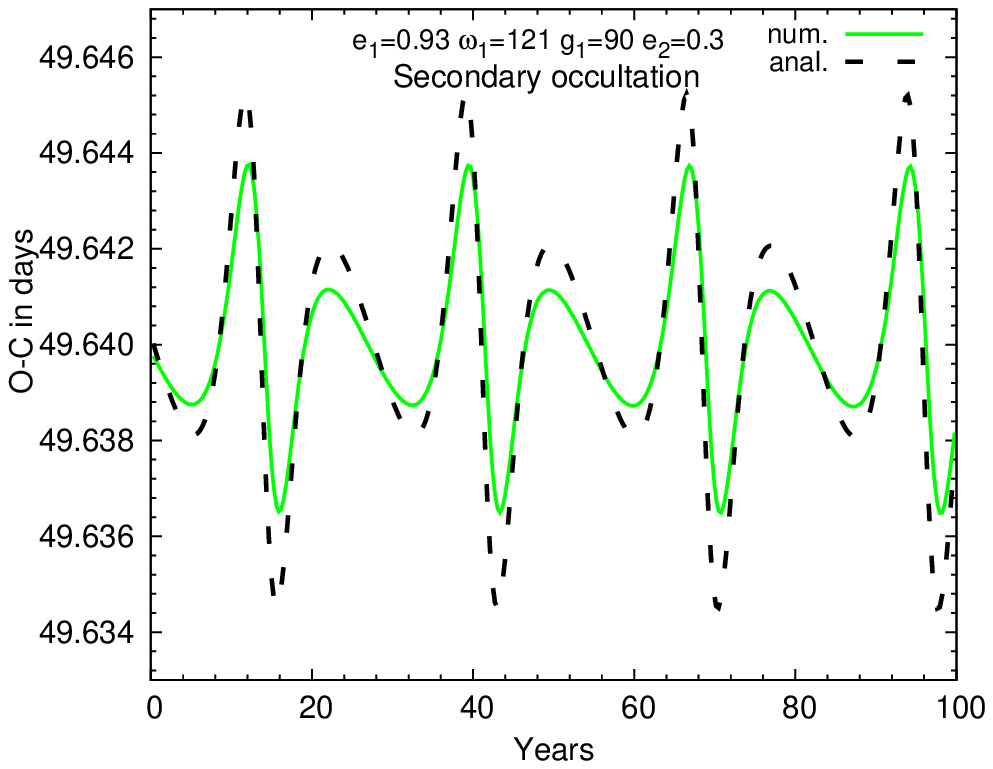}\includegraphics[width=5.6cm]{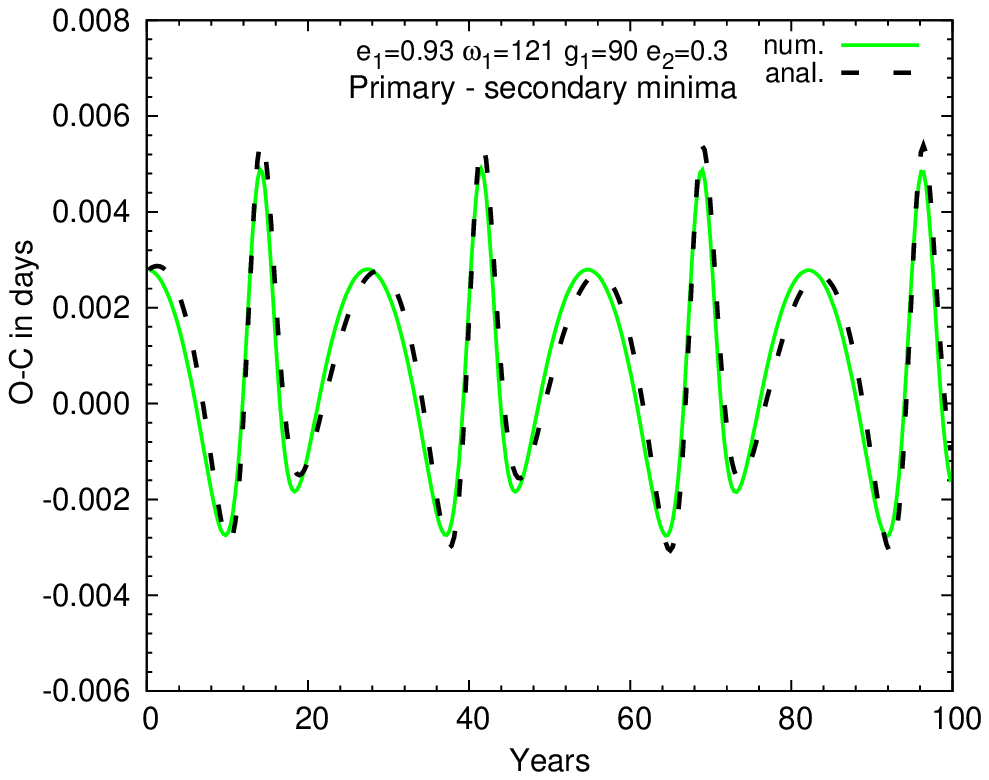}
\includegraphics[width=5.6cm]{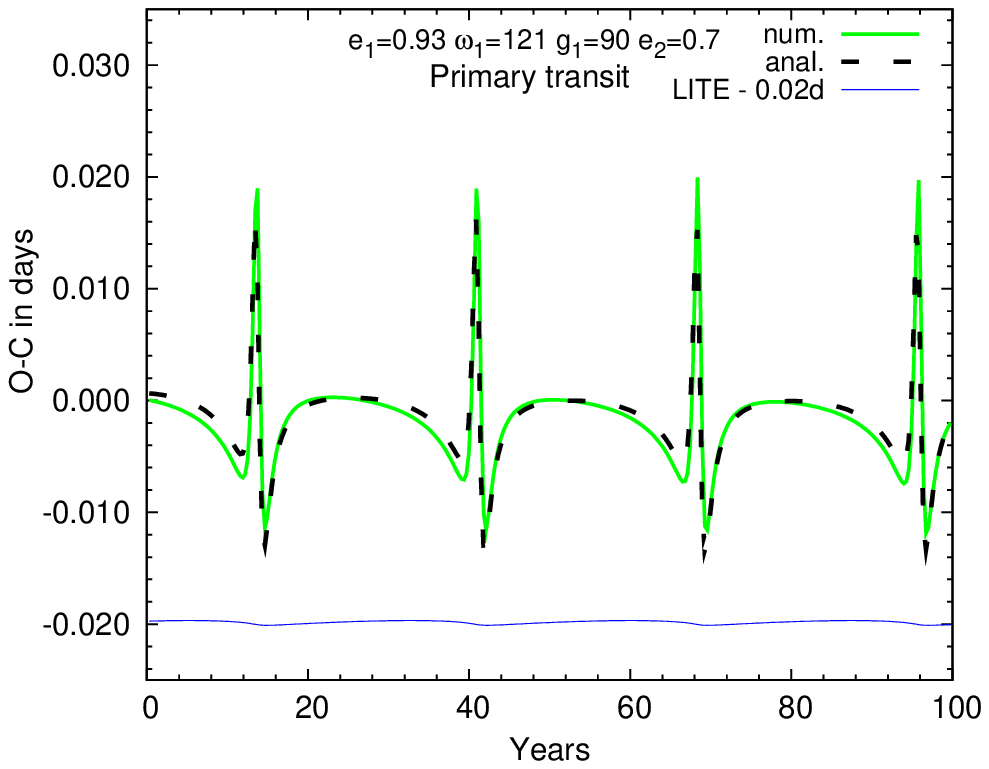}\includegraphics[width=5.6cm]{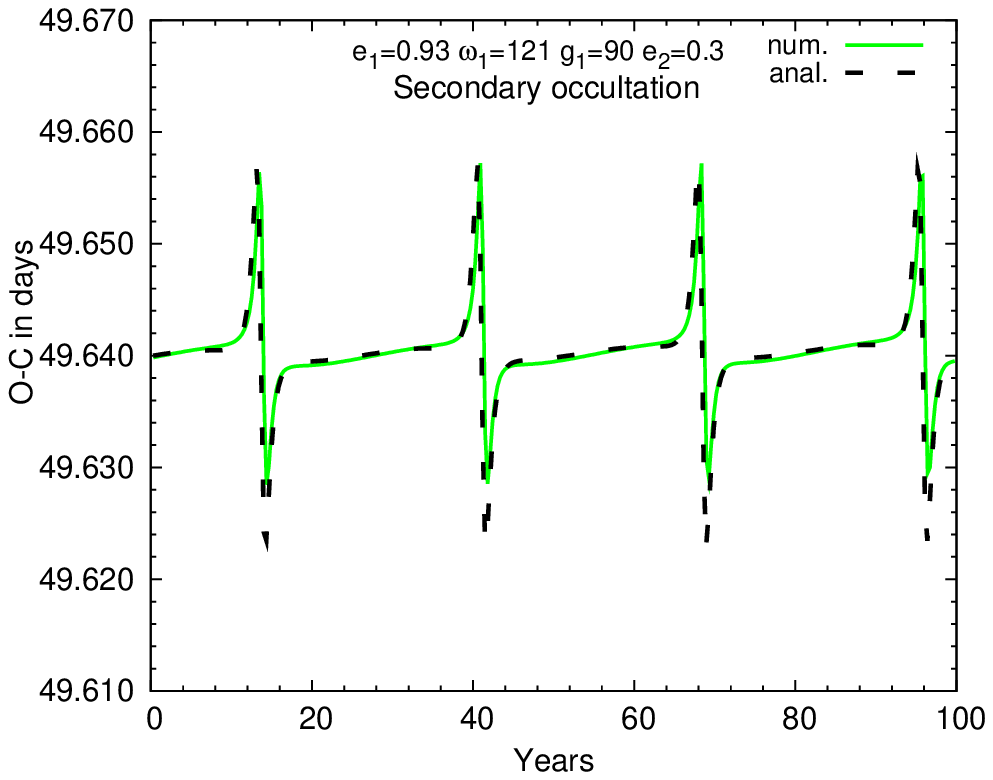}\includegraphics[width=5.6cm]{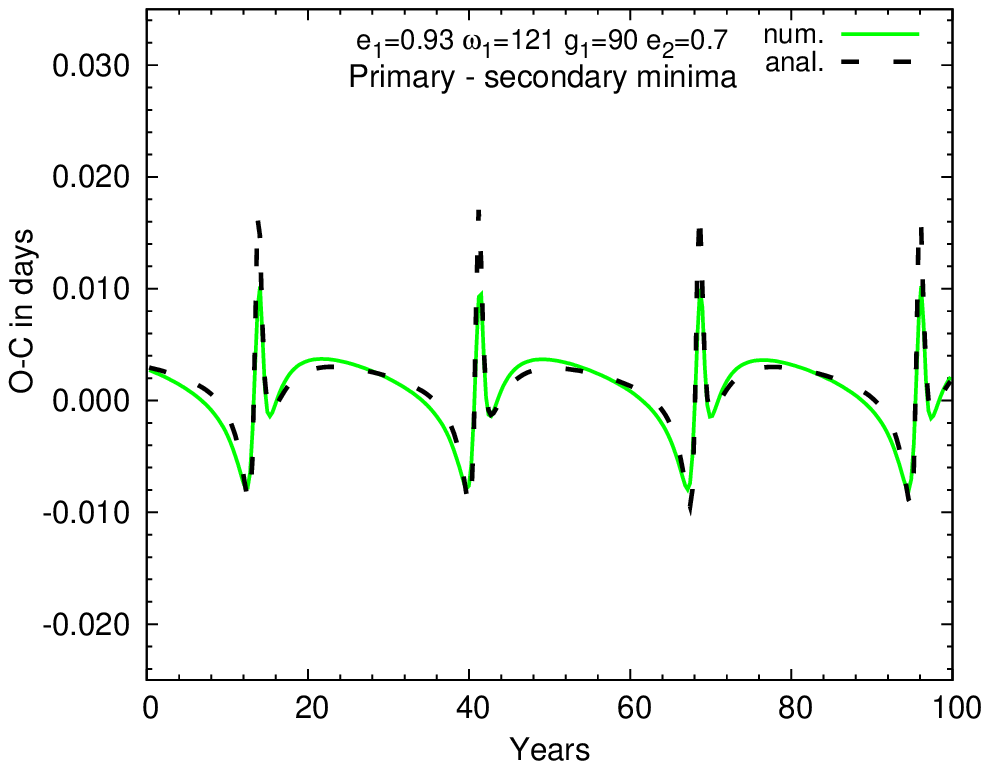}
\caption{Transit timing variatons caused by a hypothetical $P_2=10\,000$ day-period
$m_3=0.005 M_\mathrm{\sun}(\approx 5 M_\mathrm{J})$ mass third companion for \object{HD 80606b}
at the maximum eccentricity phase of the induced Kozai-cycles. Top: $e_2=0$; middle: $e_2=0.3$; bottom: $e_2=0.7$
($g_2=0\degr$ in all cases);
{\it left panels:} primary transits; {\it middle panels:} secondary occultations; {\it right panels:} primary minus secondary difference. 
See text for further details.
(For better comparison the curves are corrected for the different average transit periods, and zero point shifts.)}
\label{fig:HD80606bO-C}
\end{figure*}

\begin{figure*}
\centering
\includegraphics[width=8cm]{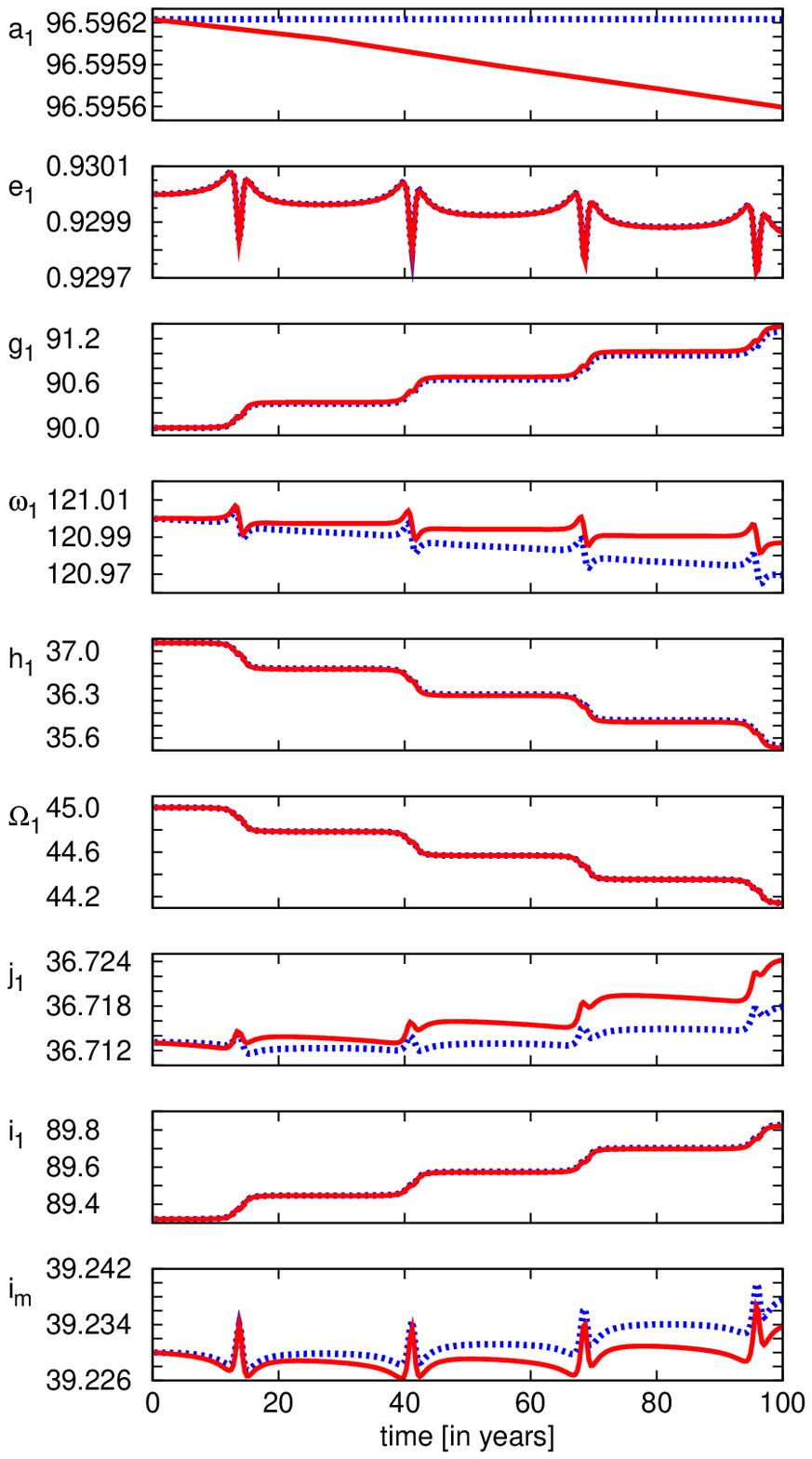}\includegraphics[width=8cm]{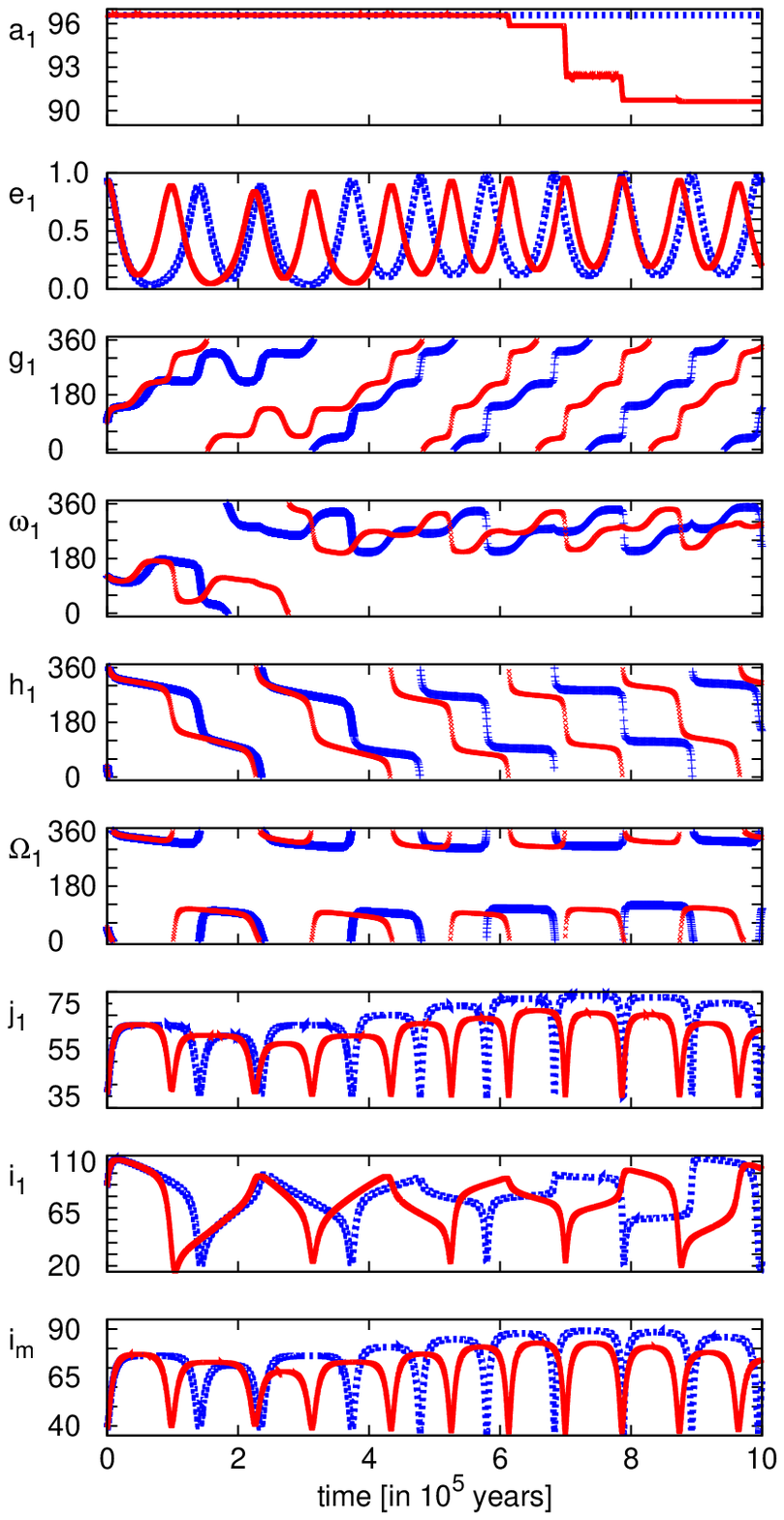}
\caption{Dynamical evolution of the orbital elements of \object{HD 80606b} in the presence of
a hypothetical $P_2=10\,000$ day-period $m_3=0.005 M_\mathrm{\sun}(\approx 5 M_\mathrm{J})$ mass third companion.
The initial orbital elements correspond to the last row of Fig.~\ref{fig:HD80606bO-C}.
{\it Red curves}: tidal effects and dissipation are considered; {\it blue curves}:
three point-mass model. (Relativistic contributions are omitted.) Note, the semi-major axis
($a_1$) is given in R$_\mathrm\sun$.}
\label{fig:HD80606dynevol}
\end{figure*}

Returning to the $P_2$ time-scale variations of the TTV-s, we carried out our calculations and integration runs with the supposition that a second, similar mass
giant planet is the source of this comet-like orbit, which is seen in the instant of
the maximum eccentricity phase of the Kozai-cycle. Consequently, we set $g_1=90$\degr,
and $i_\mathrm{m}=39\fdg23$. With these values from the sixth order formula we get
$A_{\cal{M}}=-0.38$, $A_{\cal{S}}=1.69$ for primary transits, and $A_{\cal{M}}=-1.06$, 
$A_{\cal{S}}=1.72$ for secondary occultations.
Consequently, for primary transits the $O-C$ is evidently dominated
by the ${\cal{S}}$-term. (The negative $A_{\cal{M}}$ indicates a simple 180\degr phase-shift.)
In Fig.~\ref{fig:HD80606bphasespace} the $A_{1,2,3}$ amplitudes are plotted as a
function of the outer eccentricity ($e_2$). Due to the more than four and a half-times larger
primary transit ${\cal{S}}$ amplitude, the $g_2$ dependence of the amplitudes here are weak, 
and therefore we show only $A_{1,2,3}$-s for $g_2=0$\degr.
In Fig.~\ref{fig:HD80606bO-C} we present both the numerically
generated short-term $O-C$ curves, as well as the analytically calculated cases up to the sixth
order in eccentricity (see Appendix) for three different eccentricities ($e_2=0$, 0.3, 0.7) of 
the outer perturber's orbit. We plotted the $O-C$-s for both primary transits and secondary
occultations. Also shown are the corresponding primary minus secondary curves.

As one can see, the sixth order formulae gave satisfactory results even for such
high eccentricities, although the analytical amplitudes are somewhat overestimated. 
Nevertheless, a more detailed analysis shows that the accuracy of our formulae for such high inner eccentricities strongly depends on the other orbital parameters.
This is illustrated even in the present situation, where, for the secondary-occultation
curves, (which correspond to $\omega_1=301$\degr), the discrepancies are clearly
larger. From this point of view, the $e_2=0$ case (first row) is the more interesting,
as in this situation, due to the non-zero, very similar $A_\mathrm{\cal{S}}=A_2$
amplitudes, we would expect almost identical $O-C$ curves (since the dashed analytical
ones are very similar), but this, in fact, is not the case.

\begin{figure*}
\centering
\includegraphics[width=8cm]{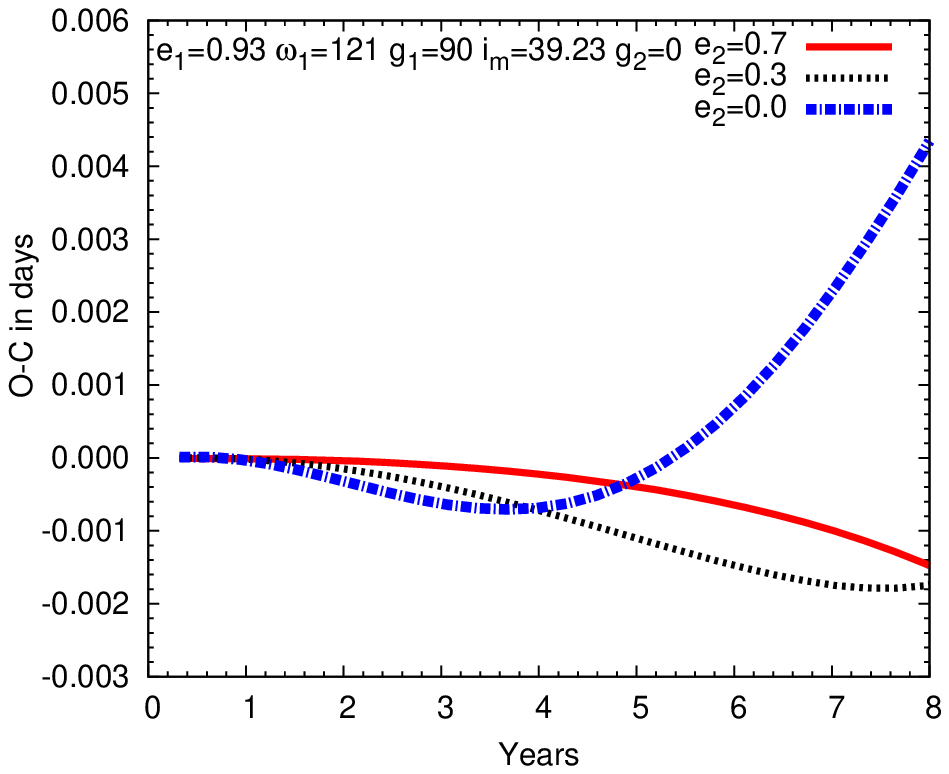}\includegraphics[width=8cm]{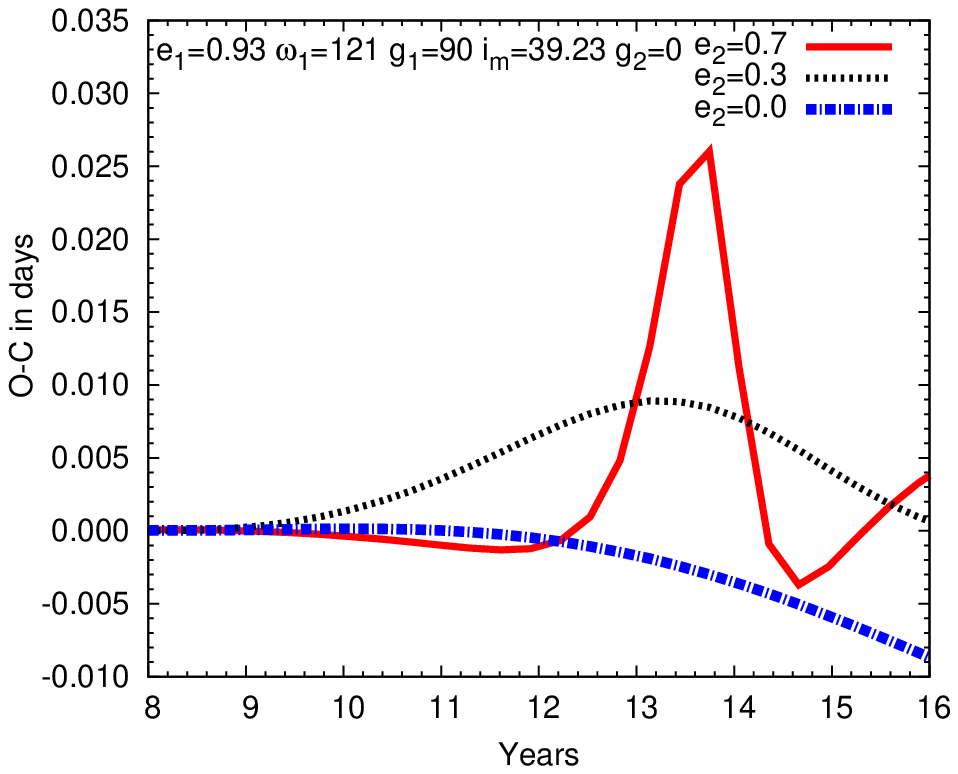}
\caption{The first and second 8 years of the left panels of Fig.~\ref{fig:HD80606bO-C}. The periods of
the individual curves was set equal to the each initial transiting periods.}
\label{fig:HD80606b8year}
\end{figure*}

Considering the case of fastest possible detection of the amplitudes, in Fig.~\ref{fig:HD80606b8year}
we plotted also the first and second 8 years of the three primary transit $O-C$ curves, shown in Fig.~\ref{fig:HD80606bO-C}.
The transiting periods for each curves were calculated on the usual, observational manner,
i.e. the time interval between the first (some) transits were used. According to Fig.~\ref{fig:HD80606b8year},
in the first 8 year, i.e. after the apastron of the outer body, the fastest detection would be possible in the smallest (total) amplitude
circular third body-case, while the curvature of the highly eccentric
curve is so small, that it needs almost 7 years to exceed the $0\fd001$ difference
which could promise certain detection. (Of course, this is a non-realistic ideal case, when
all the transits are measured, without any observational error.) Around periastron (right panel),
the situation is completely different, similarly to the case of \object{CoRoT-9b}.

We also carried out numerical integrations to investigate the possible orbital evolution
of \object{HD 80606b} in the presence of such a perturber.
We have shown that both relativistic and tidal effects are
omittable in the present configuration of the system. However, this assumption is
only correct in those cases where tidally forced dissipation is not considered.
Consequently, we integrated the dynamical evolution of the system including both tidal effects (both dissipative
and conservative tidal terms), and without them (i.e. in the frame of pure three, point-mass gravitational
interactions). The equations of the motion (including tidal and dissipative terms,
and stellar rotation) are given in \citet{borkovitsetal04}, where the description of our integrator
can also be found. In the dissipative case our dissipation constant was selected in such a
way that it produced $\Delta t_1\approx2\fd16\times10^{-7}$ tidal lag-time for the host-star,
and $\Delta t_1\approx5\fd21\times10^{-4}$ for the planet, which are equivalent to
$Q_1\approx4.1\times10^{7}$, $Q_2\approx1.7\times10^4$ dissipation parameters, respectively.
Fig.~\ref{fig:HD80606dynevol}
shows the variation of most of the orbital elements (both dynamical and observational)
for 100 and 1 million years in the dynamically most-excited $e_2=0.7$ case.
Thick lines represent the dissipative case, while thin curves show the point-mass result.
The century-long left panel demonstrates, that apart from the shrinking semi-major
axis, there is no detectable variation in the orbital elements during such a short
time-interval. And, of course, if this is true for the extremum of the Kozai cycle, it is more
expressly valid for other situations, as this phase produces the fastest orbital element variations.
Furthermore, on such a short time-scale the orbital variations in a point-mass or non-Keplerian,
tidal framework are indistinguishable. (Again, we do not take into account
the semi-major axis.) This provides further verification of the effects previously-discussed, namely that we neglected
the tidal and relativistic effects completely, and considered all the orbital elements
as constant.

Now, we consider orbital shrinking due to dissipation. As one can see, in the
present situation the decrease in $a_1$ during the first 100 years is $\Delta a_1\approx6\times10^{-4}\mathrm{~R}_\mathrm{\sun}$.
Converting this into a period variation suggests $\Delta P_1\approx10^{-3}\mathrm{~d}$. This
gives for one transiting period a $\dot P_1\approx 3\times10^{-6}\mathrm{~day\,cycle}^{-1}$ rate.
From the point of view of an eclipsing binary observer, this is an incredibly large
value. For comparison, a typical secular period variation rate measured for many of
the eclipsing binaries is about $10^{-9}-10^{-11}\mathrm{~day\,cycle}^{-1}$.
Such a high rate would produce $0\fd001$ departure in transit time during $\sim26$
cycles, i.e. during approximately 8 years. Note, this period variation ratio
is close to that ratio produced by typical long-term perturbations within
a few years. This illustrates that if the data-length is significantly shorter than
the period of the long-term periodic perturbations, then the effect arose from such
perturbations, and other effects, coming from i.e. orbital shrinking can overlap each-other,
and as a result they can be misinterpreted easily. (The question of such kinds
of misinterpretations or false identifications were considered in general
in Sect.~2 of \citealp{borkovitsetal05}).

Finally, we consider the right panel of Fig.~\ref{fig:HD80606dynevol}, which  
illustrates the frequently mentioned Kozai-mechanism (with and without tidal friction) in operation. 
Note, that in the present situation due to the high outer eccentricity, and as well as the relatively 
weak hierarchicity of the system (i.e. $a_1/a_2\approx0.05$) the higher order terms
of the perturbation function are also significant, which results in very different
consecutive cycles even in the point-mass case, too. This manifests not only as different 
periods and maximum eccentricities,
but e.g. in the fact that the argument of (dynamical) periastron ($g_1$) shows both
circulation, and libration, alternately \citep[cf.][]{fordetal00}. Nevertheless, the
detailed investigation of the apse-node timescale behaviour of the TTV, as well as other orbital
elements and observables will be the subject of a succeeding paper.

\section{Conclusions}

We have studied the long-term $P_2$ time-scale transit timing variations in transiting
exoplanetary systems which feature a further, more distant ($a_2>>a_1$)
either planetary, or stellar companion. We gave the analytical form of the $O-C$
diagram which describes such TTV-s. Our result is an extention of our previous work, namely \citet{borkovitsetal03}
for arbitrary orbital elements of both the inner transiting planet and the outer
companion. We showed that the dependence of the $O-C$ on the orbital and physical
parameters can be separated into three parts. Two of these are independent of the
real physical parameters (i.e. masses, separations, periods) of a concrete system,
and depend only on dimensionless orbital elements, and so, can be analysed in general.
For the two other kinds of parameters, which are amplitudes and phases of trigonometric
functions, we separated the orbital elements of the inner from the outer planet. 
The practical importance of such a separation is that in the case of any actual transiting
exoplanets, if eccentricity ($e_1$) and the observable argument of periastron ($\omega_1$)
are known e.g. from spectroscopy, then the main characteristics of any, caused by a possible third-body, transit timing variations can be mapped simply by the variation
of two free parameters (dynamical, relative argument of periastron, $g_1$, and
mutual inclination, $i_\mathrm{m}$), which then can be refined by the use of
the other, derived parameters, including two additional parameters fot the
possible third body (eccentricity, $e_2$, and dynamical, relative argument of periastron, $g_2$).
Moreover, as the physical attributes of a given system occur only as scaling
parameters, the real amplitude of the $O-C$ can also be estimated for a given
system, simply as a function of the $m_3/P_2$ ratio.

{\bf At this point it would be no without benefit to compare our results with the
conclusions of \citet{nesvornymorbidelli08} and \citet{nesvorny09}. These authors investigated the
same problem, i.e. the fast detectability of outer perturbing planets, and determination of
their orbital and physical parameters from their perturbations on the transit timing
of the inner planet, by the help of the analytical description of the perturbed
transiting $O-C$ curve. For the mathematical description they used the explicite perturbation theory
of \citet{hori66} and \citet{deprit69} based on canonical transformations and on the use of Lie-series.
This theory does not require the hierarchical assumption, i.e. the $\alpha=a_1/a_2$ parameter,
although less than unity, is not required to be small. This suggests a somewhat greater generality of the results,
i.e. it is well applicable systems similar to our solar system. Nevertheless, perhaps
a small disadvantage of this method with respect to the hierarchical approximation is,
that for large mutual inclinations the number of terms in the perturbation function needed for a given accuracy
grows very fast, while our formulae have the same accuracy even for the largest mutual
inclinations. There is also a similar discomfort in the case of high eccentricities, in which
case the general formulae of \citet{nesvorny09} are more sensitive for these
parameters than in the hierarchical case. For example, we could reproduce satisfactory 
accuracy even for $e_1=0.9$, in which case the classical formulae are divergent. 
(Note, that although in this paper we concentrate only on the long period perturbations,
the same is valid for the apse-node time-scale variations, as will be illustrated
in the next paper.)
As a conclusion, the greater generality of the method used by \citet{nesvornymorbidelli08} and \citet{nesvorny09} 
beyond the evident non-hierarchical configurations is well (or even better) applicable also
in the nearly coplanar case, especially when the inner orbit is nearly circular
(in which strict case the first order hierarchical approximation becomes
insufficient, \citealp[see e.g][]{fordetal00}), but in other cases, as far as the
hierarchical assumption is satisfied, this latter could give a faster and simpler method.
Furthermore, since in the hierarchical approximation, the principal small parameter in
the perturbation equations is the ratio of the separations instead of the mass-ratio,
these formulae are valid for stellar mass objects as well, and can also be applicable for
planets orbiting an S-type orbit in binary stars, or even for hierarchical triple stellar systems.}

We analysed the above-mentioned dimensionless amplitudes for different arbitrary
initial parameters, as well as for two concrete systems \object{CoRoT-9b} and
\object{HD 80606b}. We found in general, that while the shape of the $O-C$
strongly varies with the angular orbital elements, the net amplitude (departing from
some specific configurations) depends only weakly on these elements, but strongly
on the eccentricities. Nevertheless, we found some situations around $i_\mathrm{m}=45$\degr
for the specific case of \object{CoRoT-9b}, where the $O-C$ almost disappeared.

We used \object{CoRoT-9b} and \object{HD 80606b} for case studies. Both giant planets
revolve on several month-period orbits. The former has an almost
circular orbit, while the latter has a comet-like, extremely eccentric orbit.
These large period systems are ideal for searching for further perturbing components,
as the amplitude of the $O-C$ is multiplied by $P_1^2/P_2$ and consequently, as
the magnitude of the perturbations determined by $P_1/P_2$, the same amplitude
perturbations cause a better detectable effect, if the characteristic size of
the system (i.e. $P_1$) is larger. 

We considered also the question of detection, as well as the correct identification of such perturbations. 
We emphasize again, that the $O-C$
curve is a very effective tool for detection of any period variations, due to
its cumulative nature. Nevertheless, some care is necessary. First, it has to be
kept in mind, that the detectability of a period variation depends on the curvature
of the $O-C$ curve. (If the plotted $O-C$ is simply a straight line, with any
slope, it means that the period is constant, which is known with an error equal
to that slope.) This implies that the interval which is necessary to detect
the period variation coming from a periodic phenomenon depends more strong
on which phase is observed than on the amplitude of the total variation.
(We illustrated this possibility for both system sampled.) We illustrated also,
that in the case of a very eccentric third companion the fastest rate period variation
lasts a very short interval, which consists of only a few transit events. This
emphasizes the importance of observing all possible transits (and, of course occultation)
events, with great accuracy. A further question is the possible misinterpretation
of the $O-C$ diagram. For example, as was shown, in the \object{HD 80606b} system
we can expect a secular period change due to tidal dissipation, which has the
same order of magnitude than might have been measured due to the periodic
perturbations of some hypothetical third body as our sample. These two types of
perturbations could be separated in two different ways. One way is
simply a question of time. In the case of a sufficiently long observing window,
a periodic pertubation would separate from a secular one, which would produce
a parabola-like $O-C$ continuously. But, there is also a faster possibility.
In the case that not only primary transits, but also secondary occultations are observed
with similar frequency and accuracy, then, on subtracting the two (primary and
secondary) $O-C$ curves from each-other, the secular change, i.e. the effect
of dissipation would disappear, since it is similar for primary transits and secondary
occultations. This fact also makes desirable the collection of as many as possible
occultation observations too.

\begin{acknowledgements} 
This research has made use of NASA's Astrophysics Data System Bibliographic Services.
We thank Drs. John Lee Greenfell and Imre Barna B\'\i r\'o for the linguistic corrections.
\end{acknowledgements}

\begin{appendix}

\section{Relation between the observable, and the dynamical orbital elements. \label{App:elemrel}}
The start of the paper remarked that some of the relations between the elements given in this section
are valid strictly in the presented form only for the situation shown in Fig.~\ref{fig:krsz1}.  
According to the actual orientations of the orbital
planes, the spherical triangle (with sides $u_\mathrm{m1}$, $u_\mathrm{m2}$, $\Omega_1-\Omega_2$)
could be oriented in different ways, and so, some precise discussion are necessary,
which is omitted in the present appendix.

The relation between the pericentrum arguments are as follows:
\begin{eqnarray}
\omega_1&=&g_1+u_\mathrm{m1}, \\
\omega_2&=&g_2+u_\mathrm{m2}+180\degr.
\end{eqnarray}
Consequently, the true longitudes measured from the sky ($u$) and from
the intersection of the two orbital planes ($w$) are:
\begin{eqnarray}
u_1&=&v_1+\omega_1, \\
u_2&=&v_2+\omega_2, \\
w_1&=&v_1+g_1, \\
&=&u_1-u_\mathrm{m1}, \\
w_2&=&v_2+g_2+180\degr, \\
&=&u_2-u_\mathrm{m2}.
\end{eqnarray}
There are two relations between the mutual inclination ($i_\mathrm{m}$)
and the two dynamical inclinations ($j_{1,2}$). One of these is trivial,
while the second comes from the relation of the inclinations to the
orbital angular momenta. The trivial case is
\begin{equation}
i_\mathrm{m}=j_1+j_2,
\end{equation}
while the non-trivial case comes from the fact, that
\begin{eqnarray}
\cos j_1&=&\frac{\vec{C}\vec{C}_1}{CC_1}, \\
\sin j_1&=&\frac{|\vec{C}\times\vec{C}_1|}{CC_1},
\end{eqnarray}
i.e.
\begin{eqnarray}
\cos j_1&=&\frac{C_1}{C}+\frac{C_2}{C}I, \\
\sin j_1&=&\frac{C_2}{C}\sin i_\mathrm{m},
\end{eqnarray}
where $\vec{C}_{1,2}$ represents the orbital angular momentum vector of the two
orbits, $\vec{C}$ the net orbital angular momentum vector, and
\begin{eqnarray}
C_1&=&\frac{m_1m_2}{m_{12}}\sqrt{Gm_{12}a_1\left(1-e_1^2\right)}, \\
C_2&=&\frac{m_{12}m_3}{m_{123}}\sqrt{Gm_{123}a_2\left(1-e_2^2\right)}, \\
C&=&C_1\cos j_1+C_2\cos j_2.
\end{eqnarray}
(The rotational angular momenta were neglected.)
At this point we remark that for hierarchical systems it is a very reasonable
to assume 
\begin{equation}
C_1\cos j_1=\mathrm{constant},
\end{equation}
(since the first order, doubly averaged Hamiltonian of such systems does not contain 
its conjugated variable, $h_1$), which property, together with the constancy of 
the semi-major axis connects the eccentricity ($e_1$) with the dynamical inclination ($j_1$).

Further relations can be written by the use of the several identities of spherical
triangles. For example:
\begin{eqnarray}
\cos i_\mathrm{m}&=&\cos i_1\cos i_2+\sin i_1\sin i_2\cos(\Omega_2-\Omega_1), \\
\sin i_\mathrm{m}\sin u_\mathrm{m2}&=&\sin i_1\sin(\Omega_2-\Omega_1), \\
\sin i_\mathrm{m}\sin u_\mathrm{m1}&=&\sin i_2\sin(\Omega_2-\Omega_1), \\
\sin i_\mathrm{m}\cos u_\mathrm{m2}&=&-\cos i_1\sin i_2+\sin i_1\cos i_2\cos(\Omega_2-\Omega_1), \\
\sin i_\mathrm{m}\cos u_\mathrm{m1}&=&\cos i_2\sin i_1-\sin i_2\cos i_1\cos(\Omega_2-\Omega_1), 
\end{eqnarray}
and similar ones can be written for the two smaller spherical triangles. 
Finally, in hierarchical systems usually $C_2>>C_1$, so we can take
the following as a first approximation:
\begin{eqnarray}
j_1&=&i_\mathrm{m}, \\
j_2&=&0, \\
i_0&=&i_2, \\
h_1&=&u_\mathrm{m2}.
\end{eqnarray}

We note that in Fig.~\ref{fig:krsz1}, for practical reasons, we chose the intersection of
the invariable plane and the sky for the arbitrary starting point of the observable nodes ($\Omega_1$, $\Omega_2$). 
Traditionally, in astrometry, these quantities are measured from north to east on the sky, nevertheless such an approximation is valid since the $\Omega$-s only appear in the equations in the form of their differences, and derivatives.
(Nodes cannot be determined from photometry and spectroscopy, as both the light curves and
the radial velocity curves are invariant for any orientation of the orbital planes
projected onto the sky.)

\onecolumn

\section{The derivatives used for calculating the long-period dynamical $O-C$ up
to sixth order in $e_1$ \label{App:longO-C}}
The derivatives of the indirect terms are as follows:
\begin{eqnarray}
\frac{\mathrm{d}{e_1\cos\omega_1}}{\mathrm{d}{v_2}}&=&A_\mathrm{L}\left(1+e_2\cos v_2\right)\left\{(1-e_1^2)^{1/2}\left\{-\frac{3}{5}e_1\sin\omega_1\left[\left(I^2-\frac{1}{3}\right)+\left(1-I^2\right)\cos(2v_2+2g_2)\right]\right.\right. \nonumber \\
&&-e_1\sin(\omega_1-2g_1)\left[\left(1-I^2\right)+\left(1+I^2\right)\cos(2v_2+2g_2)\right] \nonumber \\
&&\left.-e_1\cos(\omega_1-2g_1)2I\sin(2v_2+2g_2)\right\}\nonumber \\
&&+\frac{\sin i_\mathrm{m}\cot i_1}{\left(1-e_1^2\right)^{1/2}}\left\{\left[\frac{2}{5}\left(1+\frac{3}{2}e_1^2\right)\cos u_\mathrm{m1}-e_1^2\cos(2g_1+u_\mathrm{m1})\right]e_1\sin\omega_1I\left[1-\cos(2v_2+2g_2)\right]\right. \nonumber \\
&&\left.\left.+\left[\frac{2}{5}\left(1+\frac{3}{2}e_1^2\right)\sin u_\mathrm{m1}+e_1^2\sin(2g_1+u_\mathrm{m1})\right]e_1\sin\omega_1\sin(2v_2+2g_2)\right\}\right\},
\end{eqnarray}
\begin{eqnarray}
\frac{\mathrm{d}{e_1^2f_3(e_1)\sin2\omega_1}}{\mathrm{d}{v_2}}&=&A_\mathrm{L}\left(1+e_2\cos v_2\right)\left\{(1-e_1^2)^{1/2}\left\{\frac{6}{5}e_1^2f_3\cos2\omega_1\left[\left(I^2-\frac{1}{3}\right)+\left(1-I^2\right)\cos(2v_2+2g_2)\right]\right.\right. \nonumber \\
&&+\left[2e_1^2f_4\cos(2\omega_1-2g_1)-\frac{1}{6}e_1^4f_5\cos(2\omega_1+2g_1)\right]\left[\left(1-I^2\right)+\left(1+I^2\right)\cos(2v_2+2g_2)\right] \nonumber \\
&&\left.-\left[2e_1^2f_4\sin(2\omega_1-2g_1)+\frac{1}{6}e_1^4f_5\sin(2\omega_1+2g_1)\right]2I\sin(2v_2+2g_2)\right\}\nonumber \\
&&+\frac{\sin i_\mathrm{m}\cot i_1}{\left(1-e_1^2\right)^{1/2}}f_3\left\{\left[-\frac{4}{5}\left(1+\frac{3}{2}e_1^2\right)\cos u_\mathrm{m1}+2e_1^2\cos(2g_1+u_\mathrm{m1})\right]e_1^2\cos2\omega_1I\left[1-\cos(2v_2+2g_2)\right]\right. \nonumber \\
&&\left.\left.-\left[\frac{4}{5}\left(1+\frac{3}{2}e_1^2\right)\sin u_\mathrm{m1}+2e_1^2\sin(2g_1+u_\mathrm{m1})\right]e_1^2\cos2\omega_1\sin(2v_2+2g_2)\right\}\right\},
\end{eqnarray}
\begin{eqnarray}
\frac{\mathrm{d}{e_1^3f_6(e_1)\cos3\omega_1}}{\mathrm{d}{v_2}}&=&A_\mathrm{L}\left(1+e_2\cos v_2\right)\left\{(1-e_1^2)^{1/2}\left\{-\frac{9}{5}e_1^3f_6\sin3\omega_1\left[\left(I^2-\frac{1}{3}\right)+\left(1-I^2\right)\cos(2v_2+2g_2)\right]\right.\right. \nonumber \\
&&-\left[3e_1^3f_7\sin(3\omega_1-2g_1)-\frac{3}{8}e_1^5\sin(3\omega_1+2g_1)\right]\left[\left(1-I^2\right)+\left(1+I^2\right)\cos(2v_2+2g_2)\right] \nonumber \\
&&-\left.\left[3e_1^3f_7\cos(3\omega_1-2g_1)+\frac{3}{8}e_1^5\cos(3\omega_1+2g_1)\right]2I\sin(2v_2+2g_2)\right\}\nonumber \\
&&+\frac{\sin i_\mathrm{m}\cot i_1}{\left(1-e_1^2\right)^{1/2}}f_6\left\{\left[\frac{6}{5}\left(1+\frac{3}{2}e_1^2\right)\cos u_\mathrm{m1}-3e_1^2\cos(2g_1+u_\mathrm{m1})\right]e_1^3\sin3\omega_1I\left[1-\cos(2v_2+2g_2)\right]\right. \nonumber \\
&&\left.\left.+\left[\frac{6}{5}\left(1+\frac{3}{2}e_1^2\right)\sin u_\mathrm{m1}+3e_1^2\sin(2g_1+u_\mathrm{m1})\right]e_1^3\sin3\omega_1\sin(2v_2+2g_2)\right\}\right\},
\end{eqnarray}
\begin{eqnarray}
\frac{\mathrm{d}{e_1^4f_8(e_1)\sin4\omega_1}}{\mathrm{d}{v_2}}&=&A_\mathrm{L}\left(1+e_2\cos v_2\right)\left\{(1-e_1^2)^{1/2}\left\{\frac{12}{5}e_1^4f_8\cos4\omega_1\left[\left(I^2-\frac{1}{3}\right)+\left(1-I^2\right)\cos(2v_2+2g_2)\right]\right.\right. \nonumber \\
&&+\left[4e_1^4f_5\cos(4\omega_1-2g_1)-\frac{3}{5}e_1^6\cos(4\omega_1+2g_1)\right]\left[\left(1-I^2\right)+\left(1+I^2\right)\cos(2v_2+2g_2)\right] \nonumber \\
&&\left.-\left[4e_1^4f_5\sin(4\omega_1-2g_1)+\frac{3}{5}e_1^6\sin(4\omega_1+2g_1)\right]2I\sin(2v_2+2g_2)\right\}\nonumber \\
&&+\frac{\sin i_\mathrm{m}\cot i_1}{\left(1-e_1^2\right)^{1/2}}f_8\left\{\left[-\frac{8}{5}\left(1+\frac{3}{2}e_1^2\right)\cos u_\mathrm{m1}+4e_1^2\cos(2g_1+u_\mathrm{m1})\right]e_1^4\cos4\omega_1I\left[1-\cos(2v_2+2g_2)\right]\right. \nonumber \\
&&\left.\left.-\left[\frac{8}{5}\left(1+\frac{3}{2}e_1^2\right)\sin u_\mathrm{m1}+4e_1^2\sin(2g_1+u_\mathrm{m1})\right]e_1^4\cos4\omega_1\sin(2v_2+2g_2)\right\}\right\},
\end{eqnarray}
\begin{eqnarray}
\frac{\mathrm{d}{e_1^5\cos5\omega_1}}{\mathrm{d}{v_2}}&=&A_\mathrm{L}\left(1+e_2\cos v_2\right)\left\{(1-e_1^2)^{1/2}\left\{-3e_1^5\sin5\omega_1\left[\left(I^2-\frac{1}{3}\right)+\left(1-I^2\right)\cos(2v_2+2g_2)\right]\right.\right. \nonumber \\
&&-5e_1^5\sin(5\omega_1-2g_1)\left[\left(1-I^2\right)+\left(1+I^2\right)\cos(2v_2+2g_2)\right] \nonumber \\
&&\left.-5e_1^5\cos(5\omega_1-2g_1)2I\sin(2v_2+2g_2)\right\}\nonumber \\
&&+\frac{\sin i_\mathrm{m}\cot i_1}{\left(1-e_1^2\right)^{1/2}}\left\{\left[2\left(1+\frac{3}{2}e_1^2\right)\cos u_\mathrm{m1}-5e_1^2\cos(2g_1+u_\mathrm{m1})\right]e_1^5\sin5\omega_1I\left[1-\cos(2v_2+2g_2)\right]\right. \nonumber \\
&&\left.\left.+\left[2\left(1+\frac{3}{2}e_1^2\right)\sin u_\mathrm{m1}+5e_1^2\sin(2g_1+u_\mathrm{m1})\right]e_1^5\sin5\omega_1\sin(2v_2+2g_2)\right\}\right\},
\end{eqnarray}
\begin{eqnarray}
\frac{\mathrm{d}{e_1^6\sin6\omega_1}}{\mathrm{d}{v_2}}&=&A_\mathrm{L}\left(1+e_2\cos v_2\right)\left\{(1-e_1^2)^{1/2}\left\{\frac{18}{5}e_1^6\cos6\omega_1\left[\left(I^2-\frac{1}{3}\right)+\left(1-I^2\right)\cos(2v_2+2g_2)\right]\right.\right. \nonumber \\
&&+6e_1^6\cos(6\omega_1-2g_1)\left[\left(1-I^2\right)+\left(1+I^2\right)\cos(2v_2+2g_2)\right] \nonumber \\
&&\left.-6e_1^6\sin(6\omega_1-2g_1)2I\sin(2v_2+2g_2)\right\}\nonumber \\
&&+\frac{\sin i_\mathrm{m}\cot i_1}{\left(1-e_1^2\right)^{1/2}}f_8\left\{\left[-\frac{12}{5}\left(1+\frac{3}{2}e_1^2\right)\cos u_\mathrm{m1}+6e_1^2\cos(2g_1+u_\mathrm{m1})\right]e_1^4\cos6\omega_1I\left[1-\cos(2v_2+2g_2)\right]\right. \nonumber \\
&&\left.\left.-\left[\frac{12}{5}\left(1+\frac{3}{2}e_1^2\right)\sin u_\mathrm{m1}+6e_1^2\sin(2g_1+u_\mathrm{m1})\right]e_1^6\cos6\omega_1\sin(2v_2+2g_2)\right\}\right\}.
\end{eqnarray}
The direct terms are coming from ($a^{3/2}e^m\cos(nv)$ and $\dot\Omega\cos i_1$) are as follows:
\begin{eqnarray}
\left(\frac{\mathrm{d}e_1\cos v}{\mathrm{d}v_2}\right)_\mathrm{dir}&=&A_\mathrm{L}\left(1+e_2\cos v_2\right)(1-e_1^2)^{1/2}\left\{-\frac{2}{5}\left(1+\frac{1}{2}e_1^2\right)\left[\left(I^2-\frac{1}{3}\right)+\left(1-I^2\right)\cos(2v_2+2g_2)\right]\right. \nonumber \\
&&-e_1^2\cos2g_1\left[\left(1-I^2\right)+\left(1+I^2\right)\cos(2v_2+2g_2)\right] \nonumber \\
&&\left.-e_1^2\sin2g_12I\sin(2v_2+2g_2)\right\}+...,
\end{eqnarray}
\begin{eqnarray}
\left(\frac{\mathrm{d}e_1^2f_3(e_1)\cos2v}{\mathrm{d}v_2}\right)_\mathrm{dir}&=&A_\mathrm{L}\left(1+e_2\cos v_2\right)(1-e_1^2)^{1/2}\left\{\frac{3}{5}e_1^2\left(1+\frac{1}{4}e_1^2+\frac{1}{18}e_1^4\right)\left[\left(I^2-\frac{1}{3}\right)+\left(1-I^2\right)\cos(2v_2+2g_2)\right]\right. \nonumber \\
&&-\frac{1}{5}e_1^2\left(1-\frac{53}{12}e_1^2-\frac{3}{2}e_1^4\right)\cos2g_1\left[\left(1-I^2\right)+\left(1+I^2\right)\cos(2v_2+2g_2)\right] \nonumber \\
&&\left.-\frac{1}{5}e_1^2\left(1-\frac{53}{12}e_1^2-\frac{3}{2}e_1^4\right)\sin2g_12I\sin(2v_2+2g_2)\right\}+...,
\end{eqnarray}
\begin{eqnarray}
\left(\frac{\mathrm{d}e_1^3f_6(e_1)\cos3v}{\mathrm{d}v_2}\right)_\mathrm{dir}&=&A_\mathrm{L}\left(1+e_2\cos v_2\right)(1-e_1^2)^{1/2}\left\{-\frac{3}{5}e_1^4\left(1+\frac{1}{2}e_1^2\right)\left[\left(I^2-\frac{1}{3}\right)+\left(1-I^2\right)\cos(2v_2+2g_2)\right]\right. \nonumber \\
&&+\frac{1}{5}e_1^2\left(1+e_1^2-\frac{49}{16}e_1^4\right)\cos2g_1\left[\left(1-I^2\right)+\left(1+I^2\right)\cos(2v_2+2g_2)\right] \nonumber \\
&&\left.+\frac{1}{5}e_1^2\left(1+e_1^2-\frac{49}{16}e_1^4\right)\sin2g_12I\sin(2v_2+2g_2)\right\}+...,
\end{eqnarray}
\begin{eqnarray}
\left(\frac{\mathrm{d}e_1^4f_8(e_1)\cos4v}{\mathrm{d}v_2}\right)_\mathrm{dir}&=&A_\mathrm{L}\left(1+e_2\cos v_2\right)(1-e_1^2)^{1/2}\left\{\frac{1}{2}e_1^6\left[\left(I^2-\frac{1}{3}\right)+\left(1-I^2\right)\cos(2v_2+2g_2)\right]\right. \nonumber \\
&&-\frac{2}{5}e_1^4\left(1+\frac{3}{4}e_1^2\right)\cos2g_1\left[\left(1-I^2\right)+\left(1+I^2\right)\cos(2v_2+2g_2)\right] \nonumber \\
&&\left.-\frac{2}{5}e_1^4\left(1+\frac{3}{4}e_1^2\right)\sin2g_12I\sin(2v_2+2g_2)\right\}+...,
\end{eqnarray}
\begin{eqnarray}
\left(\frac{\mathrm{d}e_1^5\cos5v}{\mathrm{d}v_2}\right)_\mathrm{dir}&=&A_\mathrm{L}\left(1+e_2\cos v_2\right)(1-e_1^2)^{1/2}\left\{\frac{1}{2}e_1^6\cos2g_1\left[\left(1-I^2\right)+\left(1+I^2\right)\cos(2v_2+2g_2)\right]\right. \nonumber \\
&&\left.+\frac{1}{2}e_1^6\sin2g_12I\sin(2v_2+2g_2)\right\}+...,
\end{eqnarray}
\begin{equation}
\left(\frac{\mathrm{d}e_1^6\cos6v}{\mathrm{d}v_2}\right)_\mathrm{dir}=0,
\end{equation}
where $+...$ refer to those terms which come from the normal force component, and will be
cancelled by equal but opposite in sign direct $\dot\Omega$-terms.
Furthermore, from
\begin{eqnarray}
\left(\mu^{-1/2}\frac{\mathrm{d}a^{3/2}}{\mathrm{d}v_2}\frac{\left(1-e_1^2\right)^{3/2}}{(1+e_1\cos v)^2}\right)_\mathrm{dir}&=&\frac{P_1}{2\pi}A_\mathrm{L}\left(1+e_2\cos v_2\right)(1-e_1^2)^{1/2}\left\{\frac{6}{5}e_1^2\left(1+\frac{9}{16}e_1^2+\frac{7}{16}e_1^4\right)\left[\left(I^2-\frac{1}{3}\right)+\left(1-I^2\right)\cos(2v_2+2g_2)\right]\right. \nonumber \\
&&+\frac{21}{20}e_1^2\left(1+\frac{9}{14}e_1^2+\frac{57}{112}e_1^4\right)\cos2g_1\left[\left(1-I^2\right)+\left(1+I^2\right)\cos(2v_2+2g_2)\right] \nonumber \\
&&\left.+\frac{21}{20}e_1^2\left(1+\frac{9}{14}e_1^2+\frac{57}{112}e_1^4\right)\sin2g_12I\sin(2v_2+2g_2)\right\},
\end{eqnarray}
and from $\dot\Omega$-term:
\begin{eqnarray}
\left(\frac{\mathrm{d}\Omega}{\mathrm{d}v_2}\cos{i_1}\frac{\left(1-e_1^2\right)^{3/2}}{(1+e_1\cos v)^2}\right)_\mathrm{dir}&=&A_\mathrm{L}\left(1+e_2\cos v_2\right)\frac{\sin i_\mathrm{m}\cot i_1}{(1-e_1^2)^{1/2}}\left\{\left[-\frac{2}{5}\left(1+\frac{3}{2}e_1^2\right)\cos u_\mathrm{m_1}+e_1^2\cos(u_\mathrm{m1}+2g_1)\right]I\left[1-\cos(2v_2+2g_2)\right]\right. \nonumber \\
&&\left.\left[-\frac{2}{5}\left(1+\frac{3}{2}e_1^2\right)\sin u_\mathrm{m_1}+e_1^2\sin(u_\mathrm{m1}+2g_1)\right]\sin(2v_2+2g_2)\right\}-...
\end{eqnarray}
By the use of equations above, and integrating for $v_2$, finally we get:
\begin{eqnarray}
(O-C)_{v_2}&=&\frac{P_1}{2\pi}A_\mathrm{L}\left\{\left(1-e_1^2\right)^{1/2}\left\{\left[\frac{4}{5}f_1(e_1)+\frac{6}{5}K_1(e_1,\omega_1)\right]\left[\left(I^2-\frac{1}{3}\right){\cal{M}}+\frac{1}{2}\left(1-I^2\right){\cal{S}}(2v_2+2g_2)\right]\right.\right. \nonumber \\
&&+\left[\frac{51}{20}e_1^2f_2(e_1)\cos2g_1+2K_2(e_1,\omega_1,g_1)+\frac{1}{8}e_1^2K_4(e_1,\omega_1,g_1)\right]\left[\left(1-I^2\right){\cal{M}}+\frac{1}{2}\left(1+I^2\right){\cal{S}}(2v_2+2g_2)\right] \nonumber \\
&&\left.-\frac{1}{2}\left[\frac{51}{20}e_1^2f_2(e_1)\sin2g_1+2K_3(e_1,\omega_1,g_1)+\frac{1}{8}e_1^2K_5(e_1,\omega_1,g_1)\right]2I{\cal{C}}(2v_2+2g_2)\right\} \nonumber \\
\nonumber \\
&&+\frac{\sin i_\mathrm{m}\cot i_1}{\left(1-e_1^2\right)^{1/2}}\left\{\left[-\frac{2}{5}\left(1+\frac{3}{2}e_1^2\right)\cos u_\mathrm{m1}+e_1^2\cos(2g_1+u_\mathrm{m1})\right]\left[1+2K_1(e_1,\omega_1)\right]I\left[{\cal{M}}-\frac{1}{2}{\cal{S}}(2v_2+2g_2)\right]\right. \nonumber \\
&&\left.\left.+\frac{1}{2}\left[\frac{2}{5}\left(1+\frac{3}{2}e_1^2\right)\sin u_\mathrm{m1}+e_1^2\sin(2g_1+u_\mathrm{m1})\right]\left[1+2K_1(e_1,\omega_1)\right]{\cal{C}}(2v_2+2g_2)\right\}\right\},
\end{eqnarray}
where
\begin{eqnarray}
K_1(e_1,\omega_1)&=&\mp e_1\sin\omega_1+\frac{3}{4}e_1^2f_3\cos2\omega_1\pm\frac{1}{2}e_1^3f_6\sin3\omega_1-\frac{5}{16}e_1^4f_8\cos4\omega_1\mp\frac{3}{16}e_1^5\sin5\omega_1+\frac{7}{64}e_1^6\cos6\omega_1, \\
K_2(e_1,\omega_1,g_1)&=&\mp e_1\sin(\omega_1-2g_1)+\frac{3}{4}e_1^2f_4\cos(2\omega_1-2g_1)\pm\frac{1}{2}e_1^3f_7\sin(3\omega_1-2g_1)-\frac{5}{16}e_1^4f_5\cos(4\omega_1-2g_1)\nonumber \\
&&\mp\frac{3}{16}e_1^5\sin(5\omega_1-2g_1)+\frac{7}{64}e_1^6\cos(6\omega_1-2g_1), \\
K_3(e_1,\omega_1,g_1)&=&\mp e_1\cos(\omega_1-2g_1)-\frac{3}{4}e_1^2f_4\sin(2\omega_1-2g_1)\pm\frac{1}{2}e_1^3f_7\cos(3\omega_1-2g_1)+\frac{5}{16}e_1^4f_5\sin(4\omega_1-2g_1)\nonumber \\
&&\mp\frac{3}{16}e_1^5\cos(5\omega_1-2g_1)-\frac{7}{64}e_1^6\sin(6\omega_1-2g_1), \\
K_4(e_1,\omega_1,g_1)&=&-e_1^2f_5\cos(2\omega_1+2g_1)\mp e_1^3\sin(3\omega_1+2g_1)+\frac{3}{4}e_1^4\cos(4\omega_1+2g_1),\\
K_5(e_1,\omega_1,g_1)&=&-e_1^2f_5\sin(2\omega_1+2g_1)\pm e_1^3\cos(3\omega_1+2g_1)+\frac{3}{4}e_1^4\sin(4\omega_1+2g_1),
\end{eqnarray}
and
\begin{eqnarray}
f_1&=&1+\frac{25}{8}e_1^2+\frac{15}{8}e_1^4+\frac{95}{64}e_1^6, \\
f_2&=&1+\frac{31}{51}e_1^2+\frac{23}{48}e_1^4, \\
f_3&=&1+\frac{1}{6}e_1^2+\frac{1}{16}e_1^4, \\
f_4&=&1+\frac{1}{4}e_1^2+\frac{1}{8}e_1^4, \\
f_5&=&1+\frac{3}{4}e_1^2, \\
f_6&=&1+\frac{3}{8}e_1^2, \\
f_7&=&1+\frac{1}{2}e_1^2, \\
f_8&=&1+\frac{3}{5}e_1^2.
\end{eqnarray}

\end{appendix}

\end{document}